\documentclass[reprint,showpacs,twocolumn,superscriptaddress,aps]{revtex4-2}

\usepackage{amsmath,amsthm,amssymb}
\usepackage[pdftex]{graphicx}  
\usepackage{xcolor}
\usepackage{mathtools}
\usepackage{bbm}
\usepackage{subfigure}
\usepackage{dcolumn}
\usepackage{braket}
\usepackage{bm}
\usepackage{comment}

\usepackage[colorlinks,citecolor=blue,linkcolor=blue,urlcolor=blue,hyperindex]{hyperref}

\definecolor{purple}{rgb}{0.7,0,0.7}
\definecolor{dkgreen}{rgb}{0,0.6,0}
\definecolor{brown}{rgb}{0.8,0.4,0}
\definecolor{midnightblue}{rgb}{0.39,0.58,0.93}

\newcommand{\roger}[1]{{\ifmmode \text{\color{purple}(RM) #1} \else {\color{purple}(RM) #1} \fi}}

\usepackage[normalem]{ulem}
\newcommand*{\D}[2][]{\mathinner{\mathrm{d}^{#1}\mkern-1mu{#2}}}
\newcommand{\Pf}{\operatorname{Pf}}
\newcommand{\avg}[1]{\left\langle #1 \right\rangle}

\usepackage{dsfont}
\usepackage{physics}

\begin{document}
\title{Measurement-altered Ising quantum criticality}

\author{Sara Murciano}
\thanks{These authors contributed equally to this work.}
\affiliation{Department of Physics and Institute for Quantum Information and Matter, California Institute of Technology, Pasadena, CA 91125, USA}
\affiliation{Walter Burke Institute for Theoretical Physics, California Institute of Technology, Pasadena, CA 91125, USA}

\author{Pablo Sala}
\thanks{These authors contributed equally to this work.}
\affiliation{Department of Physics and Institute for Quantum Information and Matter, California Institute of Technology, Pasadena, CA 91125, USA}
\affiliation{Walter Burke Institute for Theoretical Physics, California Institute of Technology, Pasadena, CA 91125, USA}

\author{Yue Liu}
 \affiliation{Department of Physics and Institute for Quantum Information and Matter, California Institute of Technology, Pasadena, CA 91125, USA}
 
\author{Roger S. K. Mong}
 \affiliation{Department of Physics and Astronomy and Pittsburgh Quantum Institute,
University of Pittsburgh, Pittsburgh, PA 15260, USA}

\author{Jason Alicea}
 \affiliation{Department of Physics and Institute for Quantum Information and Matter, California Institute of Technology, Pasadena, CA 91125, USA}
 \affiliation{Walter Burke Institute for Theoretical Physics, California Institute of Technology, Pasadena, CA 91125, USA}

\date{\today}

\begin{abstract}
   Quantum critical systems constitute appealing platforms for the exploration of novel measurement-induced phenomena due to their innate sensitivity to perturbations.  We study the impact of measurement on paradigmatic Ising quantum critical chains using an explicit protocol, whereby correlated ancilla are entangled with the critical chain and then projectively measured.  Using a perturbative analytic framework supported by extensive numerical simulations, we demonstrate that measurements can qualitatively alter long-distance correlations in a manner dependent on the choice of entangling gate, ancilla measurement basis, measurement outcome, and nature of ancilla correlations.  Measurements can, for example, modify the Ising order-parameter scaling dimension and catalyze order parameter condensation.  We derive numerous quantitative predictions for the behavior of correlations in select measurement outcomes, and also identify two strategies for detecting measurement-altered Ising criticality in measurement-averaged quantities. 
 First, averaging the square of the order-parameter expectation value over measurement outcomes retains memory of order parameter condensation germinated in fixed measurement outcomes---even though on average the order parameter itself vanishes.  Second, we show that, in certain cases, observables can be averaged separately over measurement outcomes residing in distinct symmetry sectors, and that these `symmetry-resolved averages' reveal measurement effects even when considering standard linearly averaged observables.   We identify complementary regimes in which symmetry-resolved averages and post-selection can be pursued reasonably efficiently in experiment, with the former generically outperforming the latter in the limit of sufficiently weak ancilla-critical chain entanglement. 
 Our framework naturally adapts to more exotic quantum critical points and highlights opportunities for potential experimental realization in NISQ hardware and in Rydberg arrays.  
 \end{abstract} 

\maketitle

\section{Introduction}

Measurements are increasingly viewed as not only a means of probing quantum matter, but also as a resource for generating novel quantum phenomena that may be difficult or impossible to realize solely with unitary evolution.
For instance, local measurements that tend to suppress entanglement can compete with entanglement-promoting dynamics---leading to entanglement transitions when these effects compete to a draw~\cite{Li2018,skinner2019,Li2019}.  Well-studied examples include the volume-to-area law entanglement transition in random Clifford circuits~\cite{Chan2019,Gullans2020,choi2020,Vasseur2020} and the transition from a critical phase with  logarithmic scaling to an area-law phase, e.g., in monitored free fermions~\cite{Alberton2021,Biella_2021,Turkeshi2022} (see also Refs.~\onlinecite{deluca,Gullans_2020,bao2020,Boorman,fan2021,swingle,Li2021,Friedman,Turkeshi1,Turkeshi2,muller,Lavasani,Hsieh3,BAO2021,Regemortel,ippoliti_2021,xhek1,xhek2,federica,mario2,shane1}).  Measurements additionally provide shortcuts to preparing certain long-range entangled quantum states~\cite{piroli2021} including wavefunctions associated with topological order \cite{Verresen,nat21,nat22,Lu22,bravyi_2022} and quantum criticality \cite{Zhu22,leeji2022}, and can also induce spontaneous symmetry breaking via quantum monitoring of a system \cite{delcampo}.  With the advent of analog quantum simulators and noisy intermediate scale quantum hardware, these directions are becoming increasingly experimentally relevant.  Indeed, recent experiments have reported signatures of measurement-induced entanglement transitions \cite{exp1,exp2} as well as measurement-assisted preparation of the toric code with a finite-depth quantum circuit \cite{Iqbal22}.

Despite the impressive progress in this arena, dealing with inherent randomness associated with quantum measurements poses a nontrivial ongoing challenge.  Measurement-induced quantum phenomena of interest commonly occur within particular measurement-outcome sectors.  Moreover, applying conventional averages of observables over measurement outcomes tends to erase measurement effects altogether.  Verification is therefore subtle and can proceed along several possible avenues---e.g., brute-force post-selection \cite{exp2}; decoding to `undo' randomness injected by measurement using classical post-processing \cite{leeji2022,shane2}, machine learning \cite{mach_learning,xhek_machine}, or active feedback and conditional control \cite{Gullans2020,exp1,Iqbal22}; considering non-unitary circuits that are space-time duals to unitary evolution \cite{ippolitiPRL,fractal,grover2021}; or  via cross-entropy benchmarking \cite{CrossEntropy}.

Quantum critical systems offer promising venues for exploring nontrivial measurement-induced behavior.  First, gaplessness renders such systems inherently sensitive to small perturbations---suggesting that even weak disturbances generated by measurements can yield profound consequences.  Second, quantum criticality traditionally manifests in long-distance correlations among \emph{local} observables; one might then anticipate that developing verification protocols here poses a gentler challenge relative to, say, identifying more nuanced entanglement modifications. Pioneering work by Garratt et al.~\cite{AltmanMeasurementLL} demonstrated that even arbitrarily weak measurements can indeed qualitatively impact long-distance correlations in a one-dimensional gapless Luttinger liquid, opening up a new frontier of `measurement-altered quantum criticality'. 
\onecolumngrid

\begin{figure}[h!]
\centering
\includegraphics[width=\textwidth]{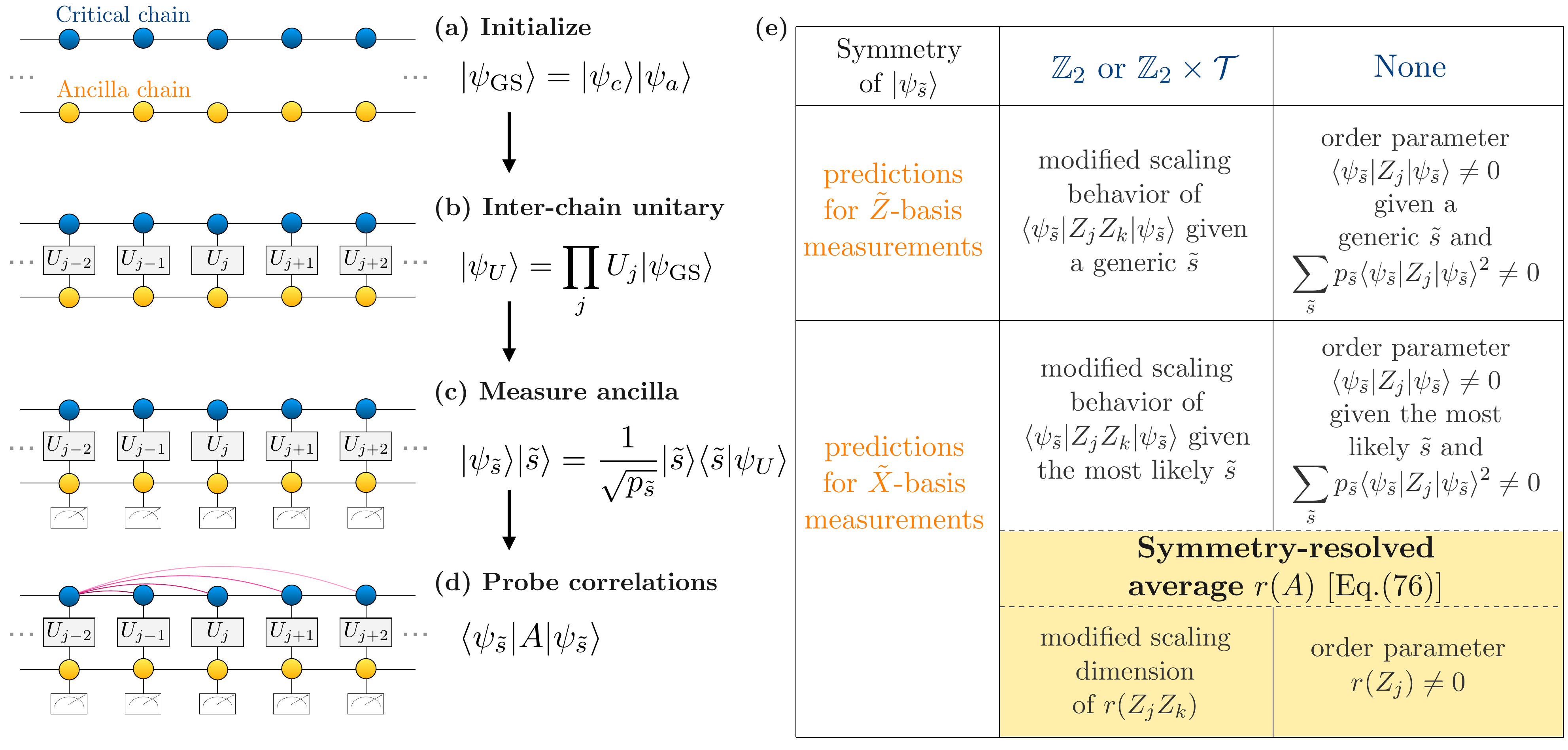}
\caption{\textbf{\bf Protocol used to explore measurement-altered Ising criticality (left) and summary of the main results (right).}  
(a) The upper chain is always prepared in the ground state $\ket{\psi_c}$ of the critical transverse-field Ising model.  The ancilla chain is initialized into the ground state $\ket{\psi_a}$ of the Ising model either in the paramagnetic phase or at criticality. 
 (b) After a unitary that entangles the two chains followed by (c) ancilla measurements, the ancilla chain enters a product state $\ket{\tilde s}$ while the upper chain enters a state $\ket{\psi_{\tilde s}}$ dependent on the measurement outcome $\tilde s$.  (d) Physical operators $A$ for the top chain are then probed in the state $\ket{\psi_{\tilde s}}$.  The table (e) summarizes our predictions for the four cases that we explore, distinguished by the ancilla measurement basis and symmetry of the post-measurement wavefunction $\ket{\psi_{\tilde s}}$.  }
    \label{fig:protocol}
\end{figure}

\twocolumngrid

More precisely, Ref.~\onlinecite{AltmanMeasurementLL}
showed that, in close analogy with the classic Kane-Fisher impurity problem \cite{KaneFisher}, measurement effects can be turned `on' or `off' by varying the Luttinger parameter that characterizes the interaction strength.  Reference~\onlinecite{AltmanMeasurementLL} additionally proposed detection protocols both for post-selected and (unconventionally) measurement-averaged correlators.  Earlier works~\cite{mario1,vicari,Hsieh1} also showed that measurements can nontrivially impact entanglement in quantum critical states, albeit with quite different protocols.  Subsequently, Ref.~\onlinecite{entropyLuttinger} associated certain effects of measurement on Luttinger liquids with an entanglement transition.  Measurements have since been further investigated in the context of $(2+1)$-dimensional quantum critical points~\cite{lee2023}; see also Refs.~\onlinecite{topological1,topological2,Hsieh2}.

In this paper we develop a theory of measurement-altered criticality in paradigmatic one-dimensional Ising quantum critical chains.  Ising quantum critical points arise in myriad physical contexts---ranging from Mott insulating spin systems to Rydberg atom arrays---and can also arise in non-interacting model Hamiltonians, thus greatly facilitating analytical and numerical progress. We consider the explicit protocol summarized in Fig.~\ref{fig:protocol}; to retain nontrivial correlations in the critical chain's wavefunction, the protocol entangles the critical degrees of freedom with a second chain of correlated ancilla and then projectively measures the latter.  Our use of ancilla not only provides a practical tool for weakly measuring the critical chain, but further opens a large phase space in which to explore measurement effects.   
Numerous questions naturally arise here: How are critical correlations modified in specific post-selected measurement outcomes?  How do such modifications depend on the choice of entangling gate and ancilla measurement basis used in the protocol?  What role do correlations among the ancilla play?  And how can one extract nontrivial effects of measurement in practice?

On a technical level, the Ising conformal field theory governing the quantum critical chains we study does not admit any marginal operators that can serve to tune the impact of measurements, unlike the Luttinger liquid setting \cite{AltmanMeasurementLL}, naively suggesting mundane behavior.  On the contrary, we find that measurements can wield exceptionally rich and experimentally accessible consequences on Ising quantum-critical spin correlations: $(i)$ Scaling dimensions that are otherwise `non-negotiable' in the pristine Ising conformal field theory can become continuously variable in translationally invariant measurement post-selection sectors.  $(ii)$ For certain unitary entangling gates in our protocol, coarse-grained spin-spin correlations can be formally obtained from a perturbed Ising conformal field theory for arbitrary measurement outcomes.  $(iii)$ Measurements can catalyze order-parameter condensation with a spatial profile dependent on the measurement outcome.  Averaging the \emph{square} of the order parameter over measurement outcomes returns a nontrivial result that appears to survive in the thermodynamic limit.   
$(iv)$ When the ancilla are also initialized into a critical state, we argue that measurements can alter power-law spin-spin correlations in a manner qualitatively different from modifications generated with paramagnetic ancilla.  
$(v)$ For certain ancilla measurement bases, measurement outcomes can be partitioned into distinct symmetry sectors.  We show that correlations linearly averaged over \emph{particular} symmetry sectors retain nontrivial signatures of measurements.  Interestingly, appropriately normalized differences in such symmetry-resolved averages closely mimic correlations evaluated with post-selected uniform measurement outcomes, yet as we show can generically be extracted more efficiently compared to post-selection when the ancilla entangle sufficiently weakly with the critical chain.   $(vi)$ By assessing the order of magnitude of the experimental trials required to probe measurement-altered criticality via symmetry-resolved averages and post-selection, we identify complementary regimes in which each technique remains viable even for large systems containing O$(100)$ spins.

Figure~\ref{fig:protocol} summarizes our main findings, all of which we substantiate using a perturbative analytic framework supplemented by extensive numerical simulations. 
Our results collectively shed new light on the interplay between measurements and quantum criticality, and can be potentially tested experimentally in Rydberg arrays~\cite{Slagle2021,scholl2023} and presently available NISQ devices. In the latter realm, a hybrid of classical algorithms for representing correlated quantum states, and physical qubits that can exploit these algorithms, was recently used to create the ground state of the critical transverse-field Ising chain and measure order-parameter power-law correlations~\cite{Haghshenas23}.  Such experimental developments bode well for future realization of measurement-altered Ising quantum criticality.

We proceed in Sec.~\ref{sec:isingprotocol} by first reviewing the microscopic model and continuum theory used throughout, and then detailing our protocol.  Section~\ref{sec:action} derives an effective action formalism that incorporates measurement effects into a perturbation to the Ising conformal field action.  
We critically assess the conditions under which our perturbative action formalism is expected to be valid in Sec.~\ref{sec:properties}. Sections~\ref{Ztilde_basis} and~\ref{Xtilde_basis} then examine consequences of our protocol with different ancilla measurement bases.  In Sec.~\ref{sec:evenodd} we develop the formalism of symmetry-resolved measurement averages for detecting measurement-altered Ising criticality and critically compare with post-selection-based schemes.    
Finally, Sec.~\ref{sec:discussion} provides a summary and outlook.

\section{Setup and protocol}\label{sec:isingprotocol}

\subsection{Review of Ising criticality}

Throughout this paper we explore the effect of measurements on an Ising quantum critical point realized microscopically in the canonical transverse field Ising model,
\begin{equation}\label{eq:spins}
    H=\sum_j(-JZ_{j}Z_{j+1}-hX_{j}).
\end{equation}
Here $Z_j$ and $X_j$ are Pauli operators acting on site $j$ of a chain with periodic boundary conditions (unless specified otherwise).  Additionally, we assume ferromagnetic interactions, $J>0$, and a positive transverse field, $h>0$.  Equation~\eqref{eq:spins} preserves both time reversal symmetry $\mathcal{T}$---which leaves $X_j$ and $Z_j$ invariant but enacts complex conjugation---and global $\mathbb{Z}_2$ spin flip symmetry generated by $G = \prod_j X_j$.  At $h>J$, the system realizes a symmetry-preserving paramagnetic phase.  For $h<J$, a ferromagnetic phase emerges, characterized by a non-zero order parameter $\langle Z_j\rangle \neq 0$ that indicates spontaneously broken $\mathbb{Z}_2$ symmetry.  The paramagnetic and ferromagnetic phases are related under a duality transformation that interchanges $J\leftrightarrow h$.  

Ising criticality appears at the self-dual point $J = h$---to which we specialize hereafter.  The low-energy critical theory is most easily accessed via a Jordan-Wigner transformation to Majorana fermion operators
\begin{align}
  \gamma_{Aj} = \left(\prod_{k<j}X_k\right) Z_j,~~~\gamma_{Bj} = \left(\prod_{k<j}X_j\right) i X_j Z_j.
\end{align}  
In this basis the $J = h$ Hamiltonian becomes
\begin{equation}\label{eq:majorana}
  H_c = iJ \sum_j(\gamma_{Aj+1}-\gamma_{Aj})\gamma_{Bj}.
\end{equation}
Focusing on long-wavelength Fourier components of $\gamma_{Aj}$ and $\gamma_{Bj}$, which comprise the important degrees of freedom at criticality, yields the continuum Hamiltonian $\mathcal{H}_c = i v \int_x (\partial_x \gamma_A) \gamma_B$ with $v \propto J$.  Upon changing basis to $\gamma_A = \gamma_R+\gamma_L$ and $\gamma_B = \gamma_R - \gamma_L$, we arrive at
\begin{equation}
  \mathcal{H}_c = -iv \int_x(\gamma_R \partial_x \gamma_R - \gamma_L\partial_x \gamma_L),
  \label{HCFT}
\end{equation}
which describes kinetic energy for right- and left-moving Majorana fermions $\gamma_R$ and $\gamma_L$.  

Equation~\eqref{HCFT} corresponds to an Ising conformal field theory (CFT) with central charge $c = 1/2$~\cite{yellowbook}.  The Ising CFT exhibits three primary fields: the identity $\mathbbm{1}$, the `spin field' $\sigma$ (scaling dimension $\Delta_\sigma = 1/8$), and the `energy field' $\varepsilon$ (dimension $\Delta_\varepsilon = 1$).  The spin field is odd under $\mathbb{Z}_2$ symmetry and represents the continuum limit of the ferromagnetic order parameter.  Consequently, only correlators containing an even number of spin fields can be non-zero at criticality. 
For instance, one- and two-point spin-field correlators read
\begin{equation}
  \langle \sigma(x) \rangle = 0,~~~  \langle \sigma(x)\sigma(x')\rangle \sim \frac{1}{|x-x'|^{1/4}}.
  \label{sigma_standard}
\end{equation}
The energy field is a composite of right- and left-movers, $\varepsilon = i \gamma_R \gamma_L$; this field is odd under duality and hence represents a perturbation that moves the system off of criticality.  
The operator product expansion for two fields at different points determine the following fusion rules:
\begin{equation}
    \begin{cases}
    \sigma \times \sigma = \mathbbm{1}+\varepsilon,\\[1ex]
    \varepsilon \times \varepsilon = \mathbbm{1},\\[1ex]
    \sigma \times \varepsilon = \sigma.
    \end{cases}
    \label{fusionrules}
\end{equation}

Local microscopic spin operators admit straightforward expansions in terms of the above CFT fields and their descendants.  In particular, we have 
\begin{align}
  Z_j &\sim \sigma + \cdots
  \label{Zexpansion}
  \\
  X_j -\langle X \rangle &\sim \varepsilon + \cdots,
  \label{Xexpansion}
\end{align}
where the ellipses denote fields with subleading scaling dimension.
Equation~\eqref{Zexpansion} follows from $\mathbb{Z}_2$ symmetry together with translation invariance of the ferromagnetic phase (for the antiferromagnetic case, $J<0$, an additional $(-1)^j$ factor would appear on the right side).  In Eq.~\eqref{Xexpansion}, $\langle X\rangle$ denotes the (position-independent) ground-state expectation value of $X_j$---which is generically non-zero due to the transverse field.  In the thermodynamic limit at criticality one finds $\langle X \rangle = 2/\pi$. 
Subtracting off this expectation value from the left-hand side removes terms proportional to the identity field on the right-hand side.  One can understand the appearance of $\varepsilon$ in Eq.~\eqref{Xexpansion} by noting that perturbing the microscopic Hamiltonian with a term $\propto\sum_j X_j$ moves the system off of criticality, corresponding to the generation of the energy field in the continuum theory.  (The same conclusion follows by expressing $X_j$ in terms of Majorana fermions and taking the continuum limit~\footnote{Notice that $\avg{i\gamma_R\gamma_L}=0$ when evaluated in the critical Ising CFT}.)

With the aid of relations like Eqs.~\eqref{Zexpansion} and \eqref{Xexpansion}, standard techniques relate ground-state expectation values of microscopic operators to averages of CFT fields expressed in path-integral language.  We illustrate the approach in a way that will be useful for exploring the influence of measurements in Sec.~\ref{lowenergy}.  The ground-state expectation value of a microscopic operator $A$ in the critical chain's ground state $\ket{\psi_c}$,
\begin{equation}
  \langle A\rangle = \bra{\psi_c} A \ket{\psi_c},
  \label{Adef}
\end{equation}
can always be expressed as
\begin{equation}
  \langle A\rangle = \lim_{\beta \rightarrow \infty} \frac{1}{Z} \Tr{e^{-\beta H_c/2} A e^{-\beta H_c/2}}.
  \label{A2}
\end{equation}
The $e^{-\beta H_c/2}$ factors in the numerator project away excited state components from the bra and ket in each element of the trace, and the partition function $Z =\lim_{\beta\to\infty} \Tr{e^{-\beta H_c}}$ in the denominator ensures proper normalization.  
Next we take the continuum limit of both sides:
\begin{align}
  \langle A\rangle &\sim \langle \mathcal{A}\rangle
  = \lim_{\beta \rightarrow \infty}\frac{1}{\mathcal{Z}}\Tr{e^{-\beta \mathcal{H}_c/2}\mathcal{A}e^{-\beta \mathcal{H}_c/2}}.
  \label{A3}
\end{align}  
Calligraphic fonts indicate low-energy expansions of the corresponding quantities in Eq.~\eqref{A2}.  For instance, if $A = Z_j Z_{j'}$ then Eq.~\eqref{Zexpansion} gives $\mathcal{A} = \sigma(x_j)\sigma(x_{j'})$ for $x_{i}$ a continuum coordinate corresponding to site $i$.     
Finally, Trotterizing the exponentials and inserting resolutions of identity in the fermionic coherent state basis yields
\begin{align}
  \langle \mathcal{A} \rangle = \lim_{\beta \rightarrow \infty}\frac{1}{\mathcal{Z}} \int \mathcal{D}\gamma_R \mathcal{D}\gamma_L e^{-\mathcal{S}_c}\mathcal{A}(\tau = 0)
  \label{Acontinuum}
\end{align}
with the Euclidean Ising CFT action
\begin{equation}
  \mathcal{S}_c = \int_x \int_{-\beta/2}^{\beta/2} \D{\tau}[\gamma_R (\partial_\tau -i v \partial_x)\gamma_R + \gamma_L( \partial_\tau+iv \partial_x) \gamma_L ].
  \label{S_c}
\end{equation}
Note that in our convention imaginary time $\tau$ runs from $-\beta/2$ to $+\beta/2$ with $\beta \rightarrow \infty$; the operator ordering in Eq.~\eqref{A3} then naturally gives $\mathcal{A}$ evaluated at $\tau = 0$ in Eq.~\eqref{Acontinuum}.

\subsection{Protocol}\label{sec:protocol}

Performing local projective measurements to all sites of the critical chain reviewed above would simply collapse the corresponding wave function into a trivial product state, thereby destroying all existent correlations. Hence, we introduce a second, ancillary transverse-field Ising chain that enables us to enact different types of generalized measurements on the critical Ising chain and characterize their non-trivial effect on its entanglement structure. Specifically, the ancilla Hamiltonian is
\begin{equation}\label{ancilla}
    H_{\rm anc}=\sum_j(-J_{\rm anc} \tilde Z_{j} \tilde Z_{j+1}-h_{\rm anc}\tilde X_{j}),
\end{equation}
where $\tilde Z_j$ and $\tilde X_j$ are Pauli operators for the ancilla spins and we assume $J_{\rm anc},h_{\rm anc} >0$.   Throughout we work in the regime $h_{\rm anc} \geq J_{\rm anc}$---i.e., the ancilla are either critical or realize the gapped paramagnetic phase, so that the 
spin-flip symmetry generated by $\tilde G = \prod_j \tilde X_j$ is always preserved in the ground state.  

Inspired by Ref.~\onlinecite{AltmanMeasurementLL}, we consider the following protocol  (see Fig.~\ref{fig:protocol} for a summary):

$(a)$ Initialize the system into the wavefunction
\begin{equation}
  \ket{\psi_{\rm GS}} = \ket{\psi_c}  \ket{\psi_a},
\end{equation}
where $\ket{\psi_c}$ is the ground state of the top (critical) chain and $\ket{\psi_a}$ is the ground state of the bottom (ancilla)  chain.

$(b)$ Apply a unitary gate $U_j$ to each pair of adjacent sites from the critical and ancilla chains, sending 
\begin{equation}
  \ket{\psi_{\rm GS}} \rightarrow \ket{\psi_U}  = \left(\prod_j U_j\right)\ket{\psi_c}  \ket{\psi_a}.    
  \label{psi_U}
\end{equation}
The unitaries we apply generally consist of single-spin ancilla rotations followed by a two-spin entangling gate, and are always implemented in a translationally invariant manner.  

$(c)$ Projectively measure all ancilla spins in some fixed basis, e.g., $\tilde Z$ or $\tilde X$, yielding measurement outcome 
\begin{equation}
  \tilde s \equiv \{\tilde s_j\}
\end{equation}
with $\tilde s_j \in \pm 1$.
The wavefunction correspondingly collapses to 
$\ket{\psi_{\tilde s}} \ket{\tilde s}$; here  $\ket{\psi_{\tilde s}}$ denotes the post-measurement state for the top chain, which depends on the outcome $\tilde s$.
More precisely, this step sends
\begin{align} 
  \ket{\psi_U} & \xrightarrow{\text{measure}} \ket{\psi_{\tilde s}} \ket{\tilde s}
\notag\\ &\qquad\qquad
  = \frac{1}{\sqrt{p_{\tilde s}}}\left(\prod_j \ket{\tilde s_j}\bra{\tilde s_j}U_j\right)\ket{\psi_c}\ket{\psi_a},
  \label{psi_m_def}
\end{align}
where the normalization 
\begin{equation}
 p_{\tilde s} = \bra{\psi_a}\bra{\psi_c}\left(\prod_j U_j^\dagger \ket{\tilde s_j}\bra{\tilde s_j}U_j\right)\ket{\psi_c}\ket{\psi_a}
 \label{p_def}
\end{equation}
specifies the probability for obtaining measurement outcome $\tilde s$.

$(d)$ Probe correlations on the top chain for the post-measurement state $\ket{\psi_{\tilde s}}$.  

Inspection of Eq.~\eqref{psi_m_def} reveals that the ancilla measurements can nontrivially impact correlations in the critical chain only if the measurement basis and unitaries are chosen such that $[\ket{\tilde s_j}\bra{\tilde s_j},U_j] \neq 0$~\footnote{Otherwise the measurement translates into a control unitary on the top chain depending on the measurement outcome of the bottom chain.}.  Even in this case, however, extracting measurement-induced changes in correlations poses a subtle problem.  For an arbitrary critical-chain observable $A$, performing a standard average of $\bra{ \psi_{\tilde s}}A \ket{\psi_{\tilde s}}$ over ancilla measurement outcomes simply recovers the expectation value $\bra{ \psi_U}A \ket{\psi_U}$ taken in the pre-measurement state.  Indeed, using Eqs.~\eqref{psi_U} and \eqref{psi_m_def} yields
\begin{align}
  &\sum_{\tilde s} p_{\tilde s}\bra{ \psi_{\tilde s}}A \ket{\psi_{\tilde s}} =  \sum_{\tilde s}\bra{\psi_U}\left(\prod_j \ket{\tilde s_j}\bra{\tilde s_j}\right)
  \nonumber \\
  &~~~~~\times A\left(\prod_j \ket{\tilde s_j}\bra{\tilde s_j}\right)\ket{\psi_U}
  \nonumber \\
  &~~~~~= \sum_{\tilde s}\bra{\psi_U}\left(\prod_j \ket{\tilde s_j}\bra{\tilde s_j}\right)
  A\ket{\psi_U}
  = \bra{\psi_U} A\ket{\psi_U}.
  \label{trivial}
\end{align}
The second equality follows from the fact that the projectors $\ket{\tilde s_j}\bra{\tilde s_j}$ commute with $A$ (because they act on different chains) and square to themselves, while the third follows upon removing a resolution of the identity for the ancilla chain. 

Our protocol performs a particular physical implementation of generalized measurements that combines additional degrees of freedom provided by the ancilla chain, a unitary entangling transformation, and projective measurements (see, e.g., Ref.~\onlinecite{nielsen2002quantum}). More importantly, it allows us to assess how quantum correlations among the ancilla, tunable via the ratio $h_{\rm anc}/J_{\rm anc}$, impact the measurement-induced changes in the critical chain's properties.  Let us make this connection more explicit. The post-measured state $\ket{\psi_{\tilde{s}}}$ given in Eq.~\eqref{psi_m_def} can be written as 
\begin{equation} \label{eq:psi_genmeas}
    \ket{\psi_{\tilde{s}}}=\frac{1}{\sqrt{p_{\tilde{s}}}}M_{\tilde{s}}\ket{\psi_c},
\end{equation}
where, given measurement outcome $\tilde s$, $M_{\tilde{s}}\equiv \bra{\tilde{s}}\prod_j U_j\ket{\psi_a}$ denotes the measurement operator acting on the critical chain. The full set of measurement operators satisfy the completeness relation $\sum_{\tilde{s}} M_{\tilde{s}}^\dagger M_{\tilde{s}}=\mathds{1}_c$. If $\ket{\psi_a}$ was a product state, then we could factorize $M_{\tilde{s}}=\prod_j M_{\tilde{s}_j}$ in terms of on-site measurement operators $M_{s_j}$. However, for nontrivially entangled $\ket{\psi_a}$---which we always consider below---this exact factorization no longer holds, though an approximate factorization can nevertheless suffice to capture the essential influence of measurement, depending on the precise form of $M_{\tilde s}$ and the range of ancilla correlations.  We return to this point in Sec.~\ref{sec:discussion}. 

Subsequent sections investigate the protocol with different classes of unitaries and measurement bases using a combination of field-theoretic and numerical tools. The latter combine covariant-matrix techniques for Gaussian states (explained in Appendix~\ref{app:Gaussian}) when characterizing a single chain; exact diagonalization (ED) to exactly evaluate averages over measurement outcomes; and tensor network methods~\cite{tenpy}, using the density matrix renormalization group (DMRG) method~\cite{dmrg} and its infinite variant (iDMRG)~\cite{iDMRG} to evaluate correlations on specific measurement outcomes. As a prerequisite, next we develop a perturbative formalism that we use extensively to distill measurement effects into a perturbation to the Ising CFT action.

\section{Effective action formalism}\label{sec:action}

Table~\ref{tab.unitaries} lists the four classes of unitaries $U_j$ that we examine, together with the corresponding ancilla measurement basis taken for each case.  In the second column, $\langle X\rangle = \bra{\psi_c}X_j\ket{\psi_c}$ as defined previously, $C$ is a constant, 
and $u$ characterizes the strength of the unitary---i.e., how far $U_j$ is from the identity. As we will see later, the symmetry of $U_j$ depends on whether $C = 0$ or $C \neq 0$ in a manner that qualitatively affects critical-chain correlations after measurement.  Throughout our analytical treatment, we assume small $u \ll 1$ that does not scale with system size.  Our goal in Sec.~\ref{perturbative} is to recast the post-measurement state in the form
\begin{equation}
    \ket{\psi_{\tilde s}} = \frac{1}{\sqrt{\mathcal{N}}}U' e^{-H_m/2}\ket{\psi_c}.
    \label{psi_m_goal}
\end{equation}
Here $U'$ is a unitary operator acting solely on the critical chain, while $H_m$ is a Hermitian operator, organized systematically in powers of $u$, that encodes the non-unitary change in $\ket{\psi_U}$ imposed by the measurement. One can view this representation of $\ket{\psi_{\tilde{s}}}$ as arising from a polar decomposition of the measurement operator $M_{\tilde{s}}$ in Eq.~\eqref{eq:psi_genmeas}, up to an overall constant (dependent on $\tilde{s}$) that is absorbed into the normalization factor $\mathcal{N}$. 
Section~\ref{lowenergy} uses the form in Eq.~\eqref{psi_m_goal} to develop a continuum-limit CFT framework for characterizing  observables given a fixed ancilla measurement outcome. 

\begin{table}
\centering
\setlength{\extrarowheight}{4pt} 
\begin{tabular}{|c|c|c|}
\hline
 Case & Unitary $U_j$ & Ancilla measurement basis \\
 \hline
 I & $\exp[ iu(X_j-\langle X \rangle)\tilde X_j ]$ & $\tilde Z$ \\
 \hline
 II & $\exp[ iu(Z_j-C)\tilde X_j ]$ & $\tilde Z$ \\
 \hline
 III & $\exp[ iu(X_j-\langle X \rangle)\tilde Z_j ]$ & $\tilde X$ \\
 \hline
 IV & $\exp[ iu(Z_j-C)\tilde Z_j ]$ & $\tilde X$ \\
 \hline
\end{tabular}
\caption{Four classes of unitaries used in our protocols, along with the corresponding ancilla measurement basis.  In cases II and IV, $C$ is a constant that controls whether or not the post-measurement state $\ket{\psi_{\tilde s}}$ preserves global spin-flip symmetry for the critical chain. 
}\label{tab.unitaries}
\end{table}

\subsection{Perturbative framework}
\label{perturbative}

All four unitaries in Table~\ref{tab.unitaries} take the form  
\begin{equation}
  U_j = e^{i u (O_j-\theta)\tilde O_j} . 
  \label{Uj_general}
\end{equation} 
The constant $\theta$ is either $\langle X \rangle$ (cases I, III) or $C$ (cases II, IV), while $O_j$ and $\tilde O_j$ denote Pauli matrices that respectively act on the critical chain and ancilla.   We refer to Appendix \ref{app:post} for a detailed derivation of the post-measurement state, providing here the final expression of $U'$ and $H_m$ appearing in Eq. \eqref{psi_m_goal}. Defining $U'=e^{iH'}$, we obtain to $O(u^2)$
\begin{align}
    H' &= u \sum_j a(j)(O_j-\langle O\rangle) ,
    \label{H_U}
    \\
    H_m &= u^2\sum_j m_j(O_j-\langle O\rangle)
    \nonumber \\ 
    &+ u^2\sum_{j \neq k}V_{jk}(O_j-\langle O\rangle)(O_k-\langle O\rangle).
    \label{H_m}
\end{align}
For later convenience we have organized the contributions in terms of $O_j-\langle O\rangle$, where $\langle O\rangle = \bra{\psi_c}O_j \ket{\psi_c}$ is the expectation value of $O_j$ in the initialized state, prior to applying the unitary and measuring.  Equation~\eqref{H_m} contains coefficients
\begin{align}\label{eq:Vjk}
    V_{jk} &= a(j,k)-a(j)a(k)
    \\
    m_j &= -2\theta [1-a(j)^2] + 2(\langle O\rangle -\theta) \sum_{k \neq j} V_{jk}.
    \label{mj}
\end{align}
where 
\begin{equation}\label{eq:ajk}
    a(j) = \frac{\langle \tilde{s} |\tilde{O}_j|\psi_a \rangle}{\langle \tilde{s}|\psi_a\rangle},\quad 
    a(j,k) = \frac{\langle \tilde{s} |\tilde{O}_j\tilde{O}_k\psi_a \rangle}{\langle \tilde{s}|\psi_a\rangle}.
\end{equation}
We stress that $U' = e^{i H'}$ factorizes into a product of operators acting on a single site $j$---i.e., one can always decompose $U' = \prod_j U'_j$ at order $O(u^2)$---whereas $e^{-H_m/2}$ admits no such factorization due to the $V_{jk}$ term.
The leading corrections to $H'$ and $H_m$ arise at $O(u^3)$ and $O(u^4)$, respectively.  

The derivation of the post-measurement state assumed $\bra{\tilde s}\psi_a \rangle \neq 0$, which holds provided measurement outcome $\tilde s$ can arise even at $u = 0$, where the unitary applied in our protocol reduces to the identity.  As we will see below, however, symmetry can constrain $\bra{\tilde s}\psi_a \rangle = 0$ for a class of $\tilde X$-basis measurement outcomes.  In the latter case our expansion for $H_m$ and $H'$ breaks down [as evidenced by the vanishing denominator in $a(i_1,\cdots, i_{N_f})$ from Eq.~\eqref{eq:aj}].  Nevertheless, even without an action-based framework, in Sec.~\ref{sec:evenodd} we will use a non-perturbative technique to constrain correlations resulting from such measurement outcomes.  For now we neglect this case and continue to assume $\bra{\tilde s}\psi_a \rangle \neq 0$ in the remainder of this section.

\subsection{Continuum limit}
\label{lowenergy}

Using Eq.~\eqref{psi_m_goal}, the expectation value of a general critical-chain observable $A$ in the state $\ket{\psi_{\tilde s}}$ associated with measurement outcome $\tilde s$ reads
\begin{equation}
  \langle A\rangle_{\tilde s} = \frac{1}{\mathcal{N}}\bra{\psi_c}e^{-H_m/2} A_{U'}e^{-H_m/2}\ket{\psi_c},
  \label{A1m}
\end{equation}
where $A_{U'} = U'^\dagger A U'$.  The numerator on the right-hand side of Eq.~\eqref{A1m} has the same form as the right side of Eq.~\eqref{Adef}, but with $A\rightarrow e^{-H_m/2} A_{U'}e^{-H_m/2}$ as a consequence of the unitary and measurement applied in our protocol. Furthermore, the normalization constant $\mathcal{N}=\bra{\psi_c}e^{-H_m}\ket{\psi_c}$ also has the form of Eq.~\eqref{Adef} with $A\to e^{-H_m}$. Following exactly the logic below Eq.~\eqref{Adef} for both the numerator and demoninator leads to the continuum expansion
\begin{align}
  \langle A\rangle_{\tilde s} &\sim \langle \mathcal{A}\rangle_{\tilde s} 
  \nonumber \\
  &= \lim_{\beta \rightarrow \infty}\frac{1}{\mathcal{Z'}} \int \mathcal{D}\gamma_R \mathcal{D}\gamma_L e^{-(\mathcal{S}_c+\mathcal{S}_m)}\mathcal{A}_{U'}(\tau = 0)
  \label{Acontinuum_m}
\end{align}
that generalizes Eq.~\eqref{Acontinuum}.  The new partition function is 
\begin{align}
  \mathcal{Z'} = \int \mathcal{D}\gamma_R \mathcal{D}\gamma_L e^{-(\mathcal{S}_c+\mathcal{S}_m)}.
\end{align}
Most crucially, the Ising CFT action $\mathcal{S}_c$ from Eq.~\eqref{S_c} has been appended with a `defect line' acting at all positions $x$ but only at imaginary time $\tau = 0$, encoded through
\begin{equation}\label{eq:action_m}
  \mathcal{S}_m = \int_{x}\mathcal{H}_m(\tau = 0)
\end{equation}
with $\mathcal{H}_m$ the continuum expansion of $H_m$. 
The explicit form of the defect line action $\mathcal{S}_m$ depends on the unitary $U$ and measurement basis, and will be explored in depth in Secs.~\ref{Ztilde_basis} and \ref{Xtilde_basis} for cases I through IV in Table~\ref{tab.unitaries}.  Specifically, we seek to understand its impact on observables $\mathcal{A}_{U'}$---which are also evaluated at $\tau = 0$ in the path integral description, thereby potentially altering critical properties of the original Ising CFT in a dramatic manner.  (Technically, $\mathcal{A}_{\bar U}$ is sandwiched between two factors of $e^{-\mathcal{S}_m/2}$ evaluated at slightly different imaginary times.  We approximated this combination as $\mathcal{A}_{\bar U}e^{-\mathcal{S}_m}$ since the leading scaling behavior is unchanged by this rewriting.) 
It is important, however, to first understand the conditions under which the perturbative expansion developed above is expected to be controlled.  To this end we now study the properties of the $V_{jk}$ and $m_j$ couplings in Eq.~\eqref{H_m}.

\section{Properties of $V_{jk}$ and $m_j$ couplings}\label{sec:properties}

In general, both $V_{jk}$ and $m_j$ vary nontrivially with the site indices in a manner dependent on the measurement outcome.  The couplings $V_{jk}$ control the interaction range in $H_m$, and exhibit a structure reminiscent of a connected correlator.  That is, $V_{jk}$ specifies how the overlap between the initial ancilla wavefunction and the measured state changes under a \emph{correlated} flip of spins at sites $j,k$.  It is thus natural to expect that $V_{jk}$ statistically averaged over measurement outcomes, denoted $\bar{V}_{jk}$, decays with $|j-k|$---either exponentially if the ancilla are initialized into the ground state of the gapped paramagnetic phase, or as a power-law if the ancilla are critical.  We confirm this expectation below.
The statistically averaged $m_j$ coefficients, denoted $\bar{m}_j$, would then not suffer from a divergence in the presence of the $\sum_{k \neq j} V_{jk}$ term in Eq.~\eqref{mj}, so long as $\bar{V}_{jk}$ decays faster than $1/|j-k|$ (which does not always hold as we will see later). 

For a \emph{particular} measurement outcome $\tilde s$, control of the expansion leading to $H_m$ in the previous subsection requires, at a minimum, that $V_{jk}$ and $m_j$ for this outcome are similarly well-behaved.  
For example, if the amplitude of $m_j$ for a particular $\tilde s$ grows with system size $N$, then $u \ll 1$ does not suffice to control the expansion (assuming that $u$ does not also scale with system size).  Additionally, if $V_{jk}$ for a given $\tilde s$ does not decay to zero with $|j-k|$, then the correspondingly infinite-range interaction in Eq.~\eqref{H_m} makes the expansion suspect.  Thus it is crucial to quantify not only the mean but also the variances ${\rm Var}(V_{jk})$ and ${\rm Var}(m_j)$ of the couplings in $H_m$---which will inform which set of measurement outcomes we analyze later on.  Next, we address this problem for the four classes of unitaries/ancilla measurement bases listed in Table~\ref{tab.unitaries}.

In all four cases we statistically average using the $u = 0$ distribution for the ancilla measurement outcomes, 
\begin{equation}
  p_{\tilde s}=\langle \tilde s|\psi_a\rangle^2\avg{\psi_c|e^{-H_m}|\psi_c} \sim \langle \tilde s|\psi_a\rangle^2\equiv p^{(0)}_{\tilde s} ,
  \label{p0}
\end{equation}
since $m_j$ and $V_{jk}$ already come with $u^2$ prefactors.  In Eq.~\eqref{p0} and many places below, we take advantage of the fact that the overlaps $\avg{\Tilde{s}|\psi_a}$ are non-negative, which follows because the transverse-field Ising model [Eq.~\eqref{ancilla}] is \emph{stoquastic} \cite{Bravyi-08}.  That is, on a given computational basis state $\ket{v}$---in this case the $\tilde Z$ or $\tilde X$ local basis---$\bra{v}H_{\mathrm{anc}}\ket{w}\leq 0$ for $v\neq w$, from which it follows that $\bra{v}\ket{\psi_a}\geq 0$ for arbitrary basis states $v$. Hence, all  elements $a(i_1,\cdots,i_{N_F})$ from Eq.~\eqref{eq:aj} are also non-negative. Finally, since the fermionized $H_{\mathrm{anc}}$ is quadratic in Majorana fermion fields, we  exploit the Gaussianity of both $\ket{\psi_a}$ and $\ket{\Tilde{s}}$ to evaluate the elements $a(i_1,\cdots,i_{N_F})$ 
using covariance-matrix techniques; see Appendix~\ref{app:Gaussian}.  These tools, along with standard results for transverse-field Ising chain correlators, allow us to compute the probabilities $p_{\tilde s}^{(0)}$ as well as the mean and variance of $m_j$ and $V_{jk}$ below.

\subsection{$\tilde Z$ measurement basis}

\begin{figure}[t!]
\centering
 \includegraphics[width = \columnwidth]{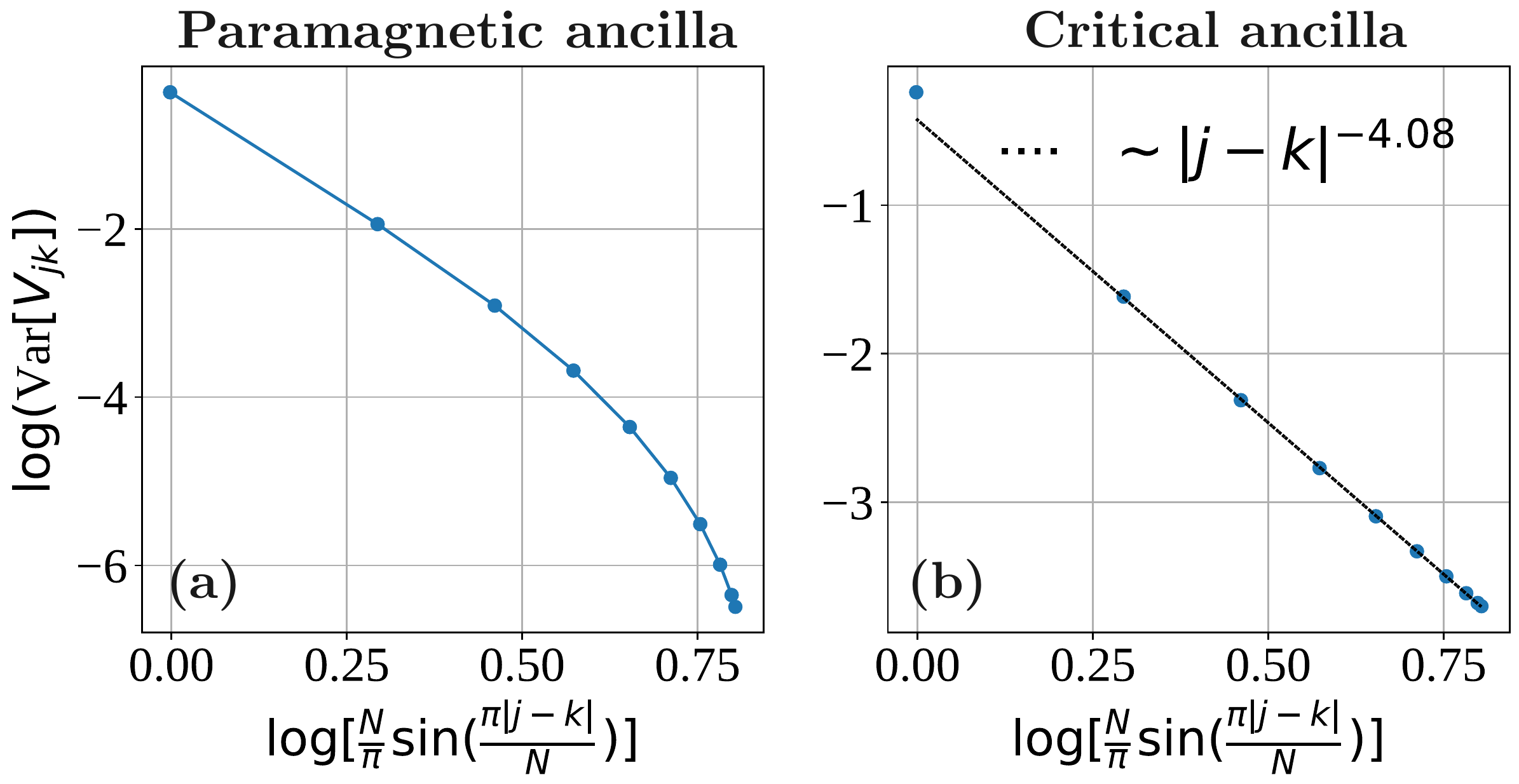}
\caption{(Color online). \textbf{Variance of $V_{jk}$ for $\tilde{Z}$-basis measurements.} Panels (a) and (b) respectively correspond to paramagnetic and critical ancilla.  The variance decays exponentially with $|j-k|$ in the former but decays approximately as $|j-k|^{-4}$ in the latter.  Data were obtained using the methods in Appendix~\ref{app:Gaussian} with a system size $N = 20$. }
 \label{fig_VarVjk}
\end{figure}

When the ancilla are measured in the $\tilde Z$ basis, the mean and variance of $V_{jk}$ evaluate to
\begin{align}
  \bar{V}_{jk} & = \sum_{\tilde s} p^{(0)}_{\tilde s} V_{jk}= 0
  \label{Vjkmean}
  \\
  {\rm Var}(V_{jk}) &= \sum_{\tilde s} p^{(0)}_{\tilde s} V_{jk}^2 - (\bar{V}_{jk})^2 
  \nonumber \\
  & = 1 + \sum_{\tilde s}p^{(0)}_{\tilde s}a(j)a(k)[a(j)a(k)-2a(j,k)]. 
  \label{Vjkvariance}
\end{align}
Figure~\ref{fig_VarVjk} illustrates ${\rm Var}(V_{jk})$ versus $|j-k|$ determined numerically at $N =20$, for both (a) paramagnetic ancilla 
and (b) critical ancilla ($h_{\rm anc}/J_{\rm anc} = 1$). In Fig.~\ref{fig_VarVjk}(a) and all subsequent simulations that use paramagnetic ancilla, we take $h_{\rm anc}/J_{\rm anc} = 1.5$. Additionally, when using periodic boundary conditions we present numerical results for correlations as a function of $\frac{N}{\pi}\sin(\frac{\pi|j-k|}{N})$ to reduce finite-size effects \cite{Slagle2021}. The variances in Fig.~\ref{fig_VarVjk} clearly tend to zero at large $|j-k|$, exponentially with paramagnetic ancilla and as a power-law (with decay exponent $\approx 4$) for critical ancilla.  This decay suggests that typical $\tilde Z$-basis measurement outcomes yield well-behaved, decaying interactions in the second line of Eq.~\eqref{H_m}.  

\begin{figure}[t!]
\centering
 \includegraphics[width = \columnwidth]{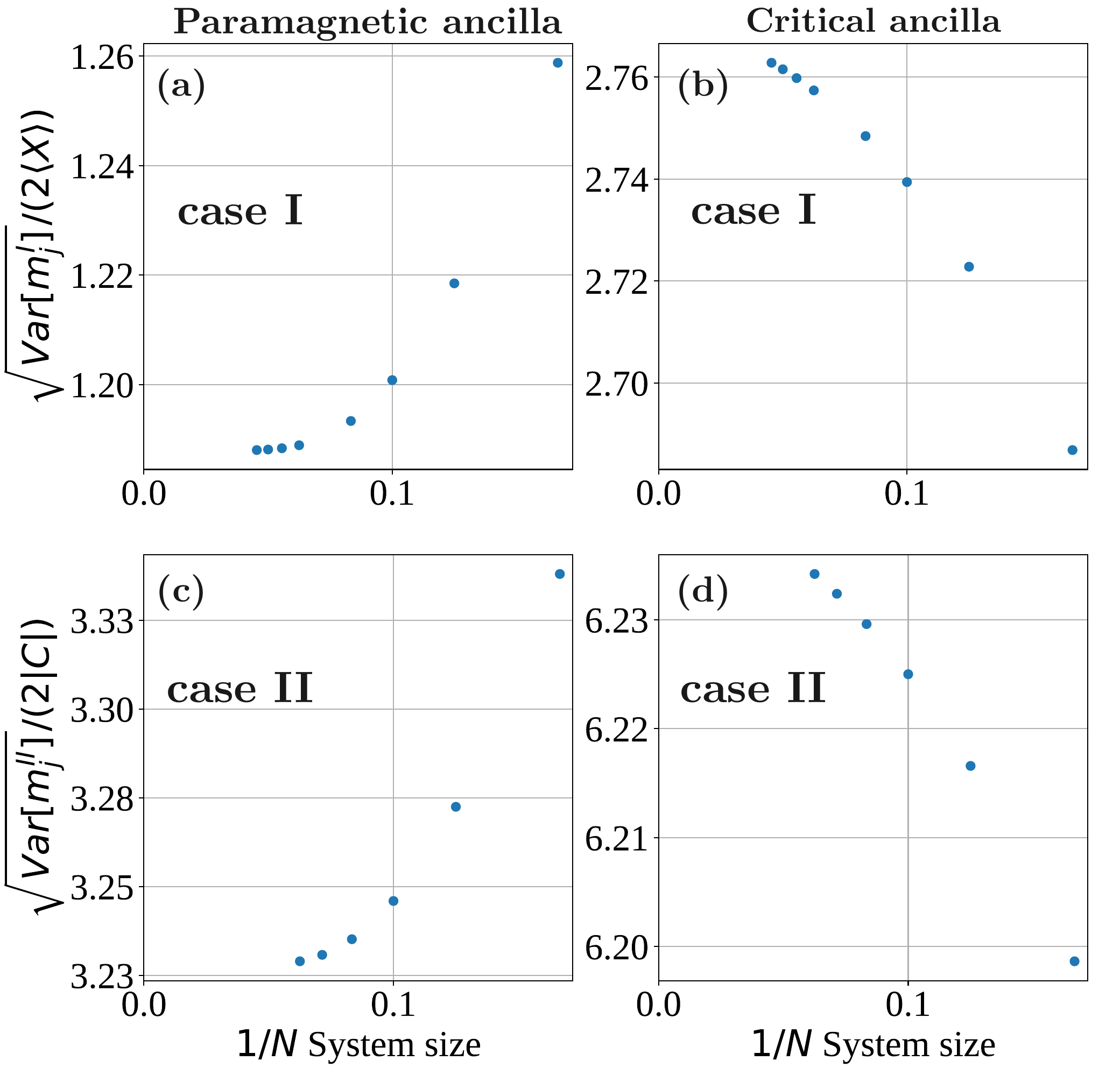}
\caption{(Color online). \textbf{Standard deviation of $m_j$ for $\tilde{Z}$-basis measurements.} 
For paramagnetic ancilla (panels $(a,c)$) in either case I or case II, the standard deviation decreases with $N$, with the trend suggesting saturation to a finite value in the thermodynamic limit.  With critical ancilla (panels $(b,d)$) the standard deviation increases extremely slowly with $N$ in both cases but remains comparable to the values with paramagnetic ancilla. 
Data were obtained using results from Appendix~\ref{app:Gaussian}. }
 \label{fig_var_mj}
\end{figure}

Due to the dependence on $\theta$ and $\langle O \rangle$ in Eq.~\eqref{mj}, the mean and variance of $m_j$ depend on the unitary applied in the protocol.  For case I in Table~\ref{tab.unitaries} we have $\theta = \langle O \rangle = \langle X \rangle = 2/\pi$ while case II corresponds to $\theta = C, \langle O \rangle = 0$.  For these cases we find
\begin{align}
    \bar{m}^{\rm I}_j &= \bar{m}^{\rm II}_j = 0
    \\
    {\rm Var}(m_j^{\rm I}) &= 4 \langle X \rangle^2 \left[-1 + \sum_{\tilde s} p^{(0)}_{\tilde s}a(j)^4\right]
    \\
    {\rm Var}(m^{\rm II}_j) &=4C^2 \left\{-1 + \sum_{\tilde s} p_{\tilde s}^{(0)}\left[a(j)^2 -  \sum_{k \neq j}V_{jk}\right]^2\right\}.
\end{align}
Figure~\ref{fig_var_mj} illustrates the numerically evaluated standard deviation $\sqrt{{\rm Var}(m^{\rm I,II}_j)}$ versus inverse system size $1/N$, again for both paramagnetic and critical ancilla.  (For case II we assume $C \neq 0$ here, since otherwise $m_j$ simply vanishes.) With paramagnetic ancilla, the standard deviation clearly converges at large $N$ to a finite value for both case I and case II.  
With critical ancilla, in both cases the standard deviation is modestly larger for the system sizes shown, albeit showing \emph{very} slow, potentially saturating, growth with $N$.  Although here we can not ascertain the trend for the thermodynamic limit, we expect that for experimentally relevant $N$ values the variance of $m_j$ remains of the same order of magnitude as for the paramagnetic case. 

\begin{figure}[t!]
\centering
 \includegraphics[width = \columnwidth]{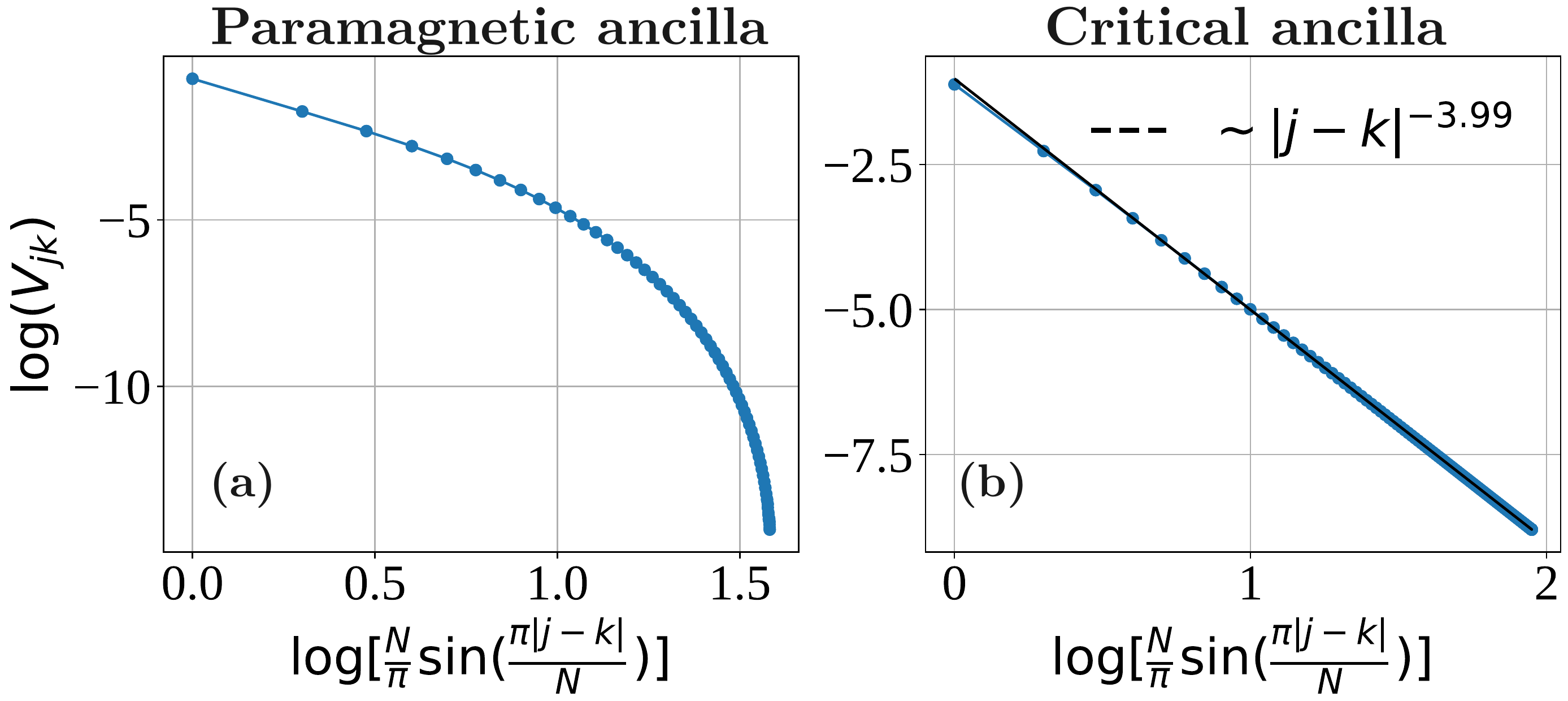}
\caption{(Color online). \textbf{$V_{jk}$ profiles for a uniform measurement outcome in case I of Table~\ref{tab.unitaries}.} Translation invariance of the uniform string outcome implies that $V_{jk}$ depends only on $|j-k|$.  Decay in $V_{jk}$ is exponential with paramagnetic ancilla (panel (a)) but power-law ($\sim |j-k|^{-4}$) with critical ancilla (panel (b)).  The data were obtained for $N = 280$ for critical ancilla using results from Appendix~\ref{app:Gaussian}. }   
 \label{fig_examples}
\end{figure}

The behavior of the variances discussed above suggests that, at least for paramagnetic ancilla, any typical string outcome yields a well-behaved defect-line action amenable to our perturbative formalism.   
To support this expectation, we illustrate $V_{jk}$ and $m_{j}$ for select measurement outcomes. 
First, Fig.~\ref{fig_examples} displays $V_{jk}$ for a uniform measurement outcome with $\ket{\tilde s} = \ket{\cdots \uparrow\uparrow\uparrow\cdots}$---which, along with its all-down partner, occurs with highest probability $p^{(0)}_{\tilde s}$ (as confirmed numerically for systems as large as $N = 26$).  Panels (a) and (b) correspond to paramagnetic and critical ancilla, respectively.  In the former, $V_{jk}$ decays exponentially with $|j-k|$, while in the latter it decays as $|j-k|^{-4}$. In both cases $m_j$ is translational invariant, as dictated by uniformity of the measurement outcome. Moreover, we have numerically verified that $m_j$ saturates to a constant value with increasing system size in the paramagnetic case, whereas it slowly grows (at the level of the third decimal digit) for critical ancilla.  The $V_{jk}$ and $m_j$ values discussed here can be combined to infer the (also uniform) $m_j$ profile for case II, which at $N \rightarrow \infty$ will simply differ from $m_j$ for case I by a finite value given the `fast' decay in $V_{jk}$.

\begin{figure}[t!]
 \includegraphics[width = \columnwidth]{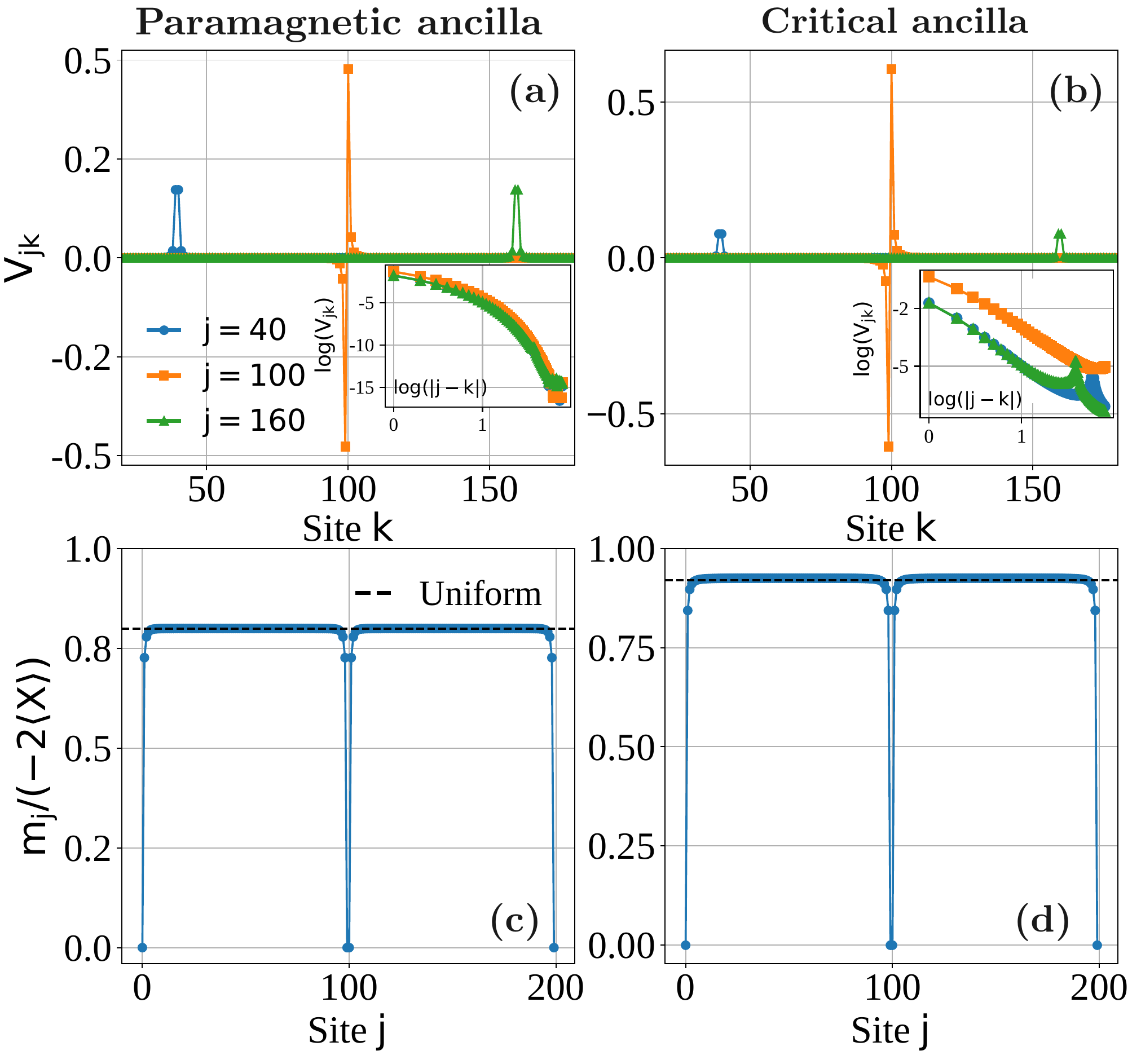}
\caption{(Color online). \textbf{$V_{jk}$ and $m_j$ profiles for a two-domain-wall measurement outcome in case I of Table~\ref{tab.unitaries}}.  Domain walls reside near sites $0$ and $100$ in a system with $N = 200$. Left and right columns show data for paramagnetic and critical ancilla, respectively.  Due to loss of translation symmetry, we show $V_{jk}$ versus $k$ for several $j$ values. As shown in the insets, $V_{jk}$ decays with the distance $|j-k|$, although we find a relative bump close to the domain walls.  The corresponding $m_j$ profiles (panels $(c,d)$) exhibit dips near zero in the immediate vicinity of the domain walls, but are otherwise roughly uniform matching the values obtained for a uniform string outcome (black dashed lines). Data obtained using results from Appendix~ \ref{app:Gaussian}.}  
 \label{fig_examplesDW}
\end{figure}

The next-most-probable set of measurement outcomes correspond to configurations with isolated spin flips introduced into the uniform $\tilde s$ string considered above.  Rather than consider such outcomes, we next examine a lower-probability configuration with two maximally separated domain walls: $\ket{\tilde s} = \ket{\cdots \uparrow \uparrow \uparrow \downarrow \downarrow \downarrow \cdots}$. 
(Due to periodic boundary conditions considered here, domain walls come in pairs.) Figure~\ref{fig_examplesDW} displays both $V_{jk}$ and $m_j$ for this measurement outcome with domain walls at $j=0,100$ for a system with $N=200$ assuming case I.  Since $V_{jk}$ now depends on $j$ and $k$ due to non-uniformity of the measurement outcome, in (a,b) we show $V_{jk}$ versus $k$ for three different $j$ values.  Overall decay with $|j-k|$ similar to that for the uniform measurement outcome persists here. For fixed $j$, a relative bump appears when $k$ sits close to a domain walls, but the height of the bump is nonetheless orders of magnitude smaller than when $k$ is close to $j$  (see insets).  In (c,d), the $m_j$ profiles resemble those for the uniform case, but with dips that tend to zero from below in the thermodynamic limit for $j$'s on either end of a given domain wall.  We have also verified that  still-lower-probability random $\tilde s$ strings also yield well-behaved $V_{jk}$ and $m_j$ couplings.

\subsection{$\tilde X$ measurement basis}

Switching the ancilla measurement basis from $\tilde Z$ to $\tilde X$ qualitatively changes the statistical properties of $m_j$ and $V_{jk}$.  By construction, the initialized ancilla wavefunction $\ket{\psi_a}$ is an eigenstate of the $\mathbb{Z}_2$-symmetry generator $\tilde G = \prod_j \tilde X_j$ with eigenvalue $+1$. For $\tilde X$-basis measurements, a given ancilla state $\ket{\tilde s}$ obtained after a measurement can also be classified by its $\tilde G$ eigenvalue; we refer to measurement outcomes with $\tilde G\ket{\tilde s} = + \ket{\tilde s}$ as `even strings' and outcomes with $\tilde G\ket{\tilde s} = - \ket{\tilde s}$ as `odd strings'.  (Due to the form of the unitary $U$ applied prior to measurement in this case, both sectors can still arise despite the initialization.)  Consider now an even-string measurement outcome with $\bra{\tilde s}\psi_a \rangle \neq 0$---as assumed in the perturbative expansion developed in Sec.~\ref{perturbative}.  Crucially, due to mismatch in $\tilde G$ eigenvalues, $\bra{\tilde s(i_1\cdots i_{N_f})}\psi_a\rangle$ then vanishes for \emph{any} odd number of flipped spins $N_f$.  It follows that $a(j) = 0$ in Eqs.~\eqref{eq:Vjk} and \eqref{mj}, leaving
\begin{align}\label{eq:iii}
    V_{jk} = a(j,k),~~~m_j = -2\theta + 2(\langle O\rangle -\theta) \sum_{k \neq j} V_{jk}.
\end{align}
Notice that $V_{jk}$ here is always non-negative [recall the discussion below Eq.~\eqref{p0}].

The mean and variance of $V_{jk}$ then reduce to simple ground-state ancilla correlation functions:
\begin{align}
    \bar{V}_{jk} = \bra{\psi_a} \tilde Z_j \tilde Z_k \ket{\psi_a},~~{\rm Var}(V_{jk}) = 1- \bra{\psi_a} \tilde Z_j \tilde Z_k \ket{\psi_a}^2.
    \label{Vmeanvariance}
\end{align}
At large $|j-k|$, the mean always decays to zero: for ancilla initialized in the paramagnetic phase with correlation length $\xi$ we have $\bar{V}_{jk}\sim e^{-|j-k|/\xi}$, while if the ancilla are critical $\bar{V}_{jk} \sim 1/|j-k|^{1/4}$.  The variance of $V_{jk}$, by contrast, grows towards unity at large $|j-k|$.   Correspondingly, the $V_{jk}$'s for particular measurement outcomes can differ wildly from the mean, and in particular need not decay with $|j-k|$.  

Remarkably, for case III in Table~\ref{tab.unitaries} $m_j$ takes on the same $j$-independent value for \emph{any} even-sector measurement outcome:
\begin{equation}
    m_j^{\rm III} = -2 \langle X \rangle.
\end{equation}
For case IV, however, $m_j$ depends nontrivially on $V_{jk}$ and hence the measurement outcome; here we find
\begin{align}
    \bar{m}_j^{\rm IV} &= -2C\left(1+ \sum_{k \neq j}\bar{V}_{jk}\right)
    \\
    {\rm Var}(m_j^{\rm IV}) &= (2C)^2\sum_{k,k' \neq j}(\bar{V}_{kk'} - \bar{V}_{jk}\bar{V}_{jk'})
\end{align}
with $\bar{V}_{jk}$ given in Eq.~\eqref{Vmeanvariance}.
Suppose that the ancilla are paramagnetic.  Exponential decay of $\bar{V}_{jk}$ with $|j-k|$ yields a finite mean $\bar{m}_j^{\rm IV}$, though the variance diverges linearly with system size, ${\rm Var}(m_j^{\rm IV}) \sim N$, due to contributions from the $\bar{V}_{kk'}$ term with $k'$ near $k$.  With critical ancilla, power-law decay of $\bar{V}_{jk}$ generates divergent mean and variance: $\bar{m}_j^{\rm IV} \sim N^{3/4}$ and ${\rm Var}(m_j^{\rm IV}) \sim N^{7/4}$.  In both scenarios the fluctuations of $m_j$ increase with system size faster than the average value.

We therefore can only apply the perturbative formulation developed in Sec.~\ref{perturbative} to a restricted set of $\tilde X$-basis measurement outcomes that lead to 
a well-behaved, decaying interaction term in $H_m$, and correspondingly well-behaved $m_j$ couplings.  Fortunately, the most probable measurement outcomes do indeed satisfy these criteria. 

\begin{figure}[t!]
 \includegraphics[width = 1.0\columnwidth]{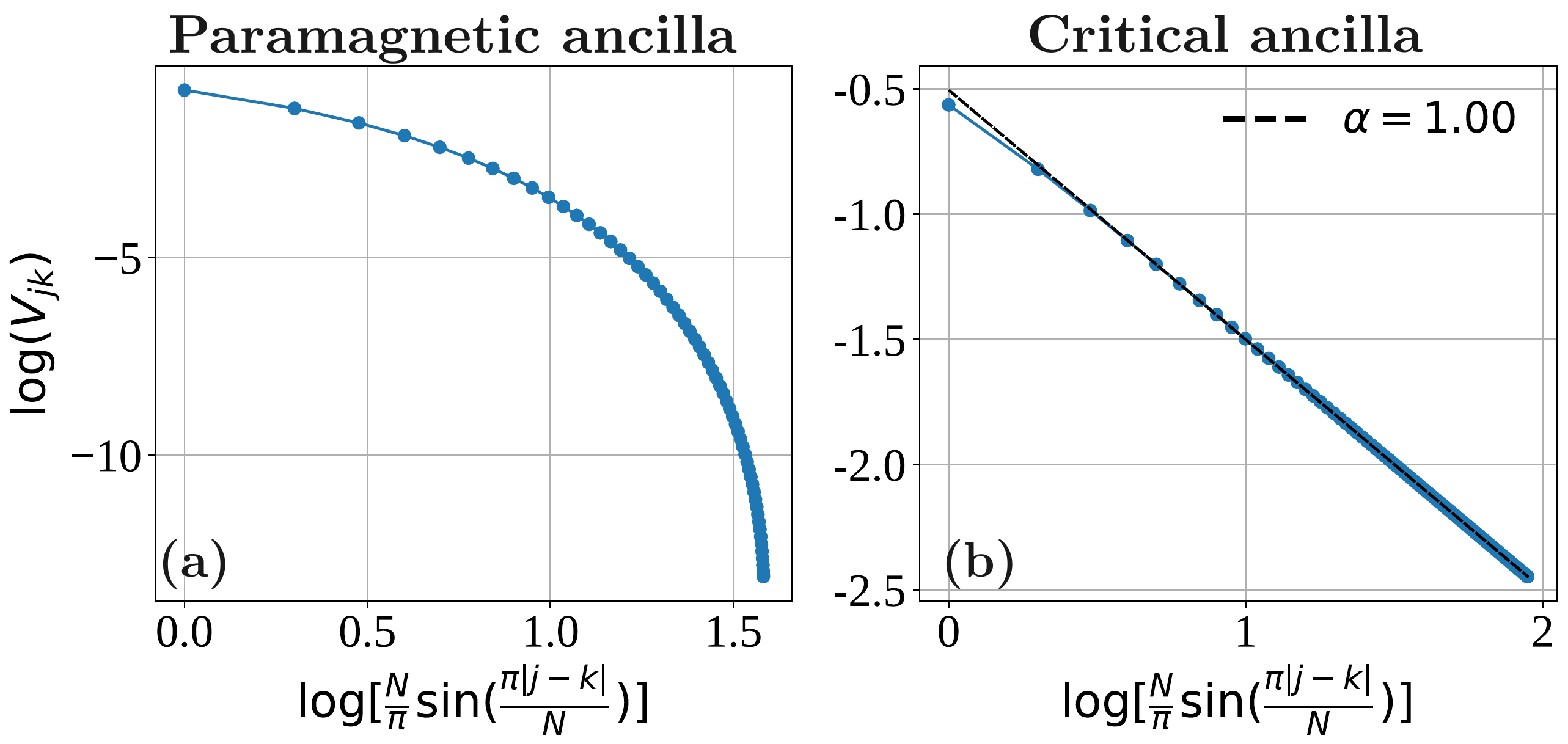}
\caption{(Color online). \textbf{$V_{jk}$ profile for a uniform $\tilde X$-basis measurement outcome.} Decay of $V_{jk}$ is exponential with paramagnetic ancilla and power-law ($\sim |j-k|^{-1}$) with critical ancilla.  In the critical case, note the significantly smaller exponent compared to Fig.~\ref{fig_examples}.  Data obtained using results from Appendix~ \ref{app:Gaussian} with system size $N = 280$ for critical ancilla.}  
 \label{fig_xbasis_uni}
\end{figure}

Figure~\ref{fig_xbasis_uni} plots  $V_{jk}$ for the highest-probability outcome,  corresponding to the uniform string $\ket{\tilde s} = \ket{\cdots \rightarrow\rightarrow\rightarrow\cdots}$~\footnote{The high probability of this measurement outcome becomes intuitive in the $h_{\rm anc}/J_{\rm anc} \gg 1$ regime.}. For (a) paramagnetic ancilla $V_{jk}$ decays to zero with $|j-k|$ exponentially, while for (b) critical ancilla it decays as $\sim|j-k|^{-1}$.  In case IV with $C \neq 0$, Eq.~\eqref{eq:iii} implies that the associated $m_j$ converges to a finite value as $N$ increases for paramagnetic ancilla, but diverges as $\ln N$ with critical ancilla.  (For $C = 0$, $m_j$ again simply vanishes.)  Therefore, modulo this possible logarithmic factor, the uniform string presents a `good' $\tilde X$-basis measurement outcome.

\begin{figure}[t!]
 \includegraphics[width = \columnwidth]{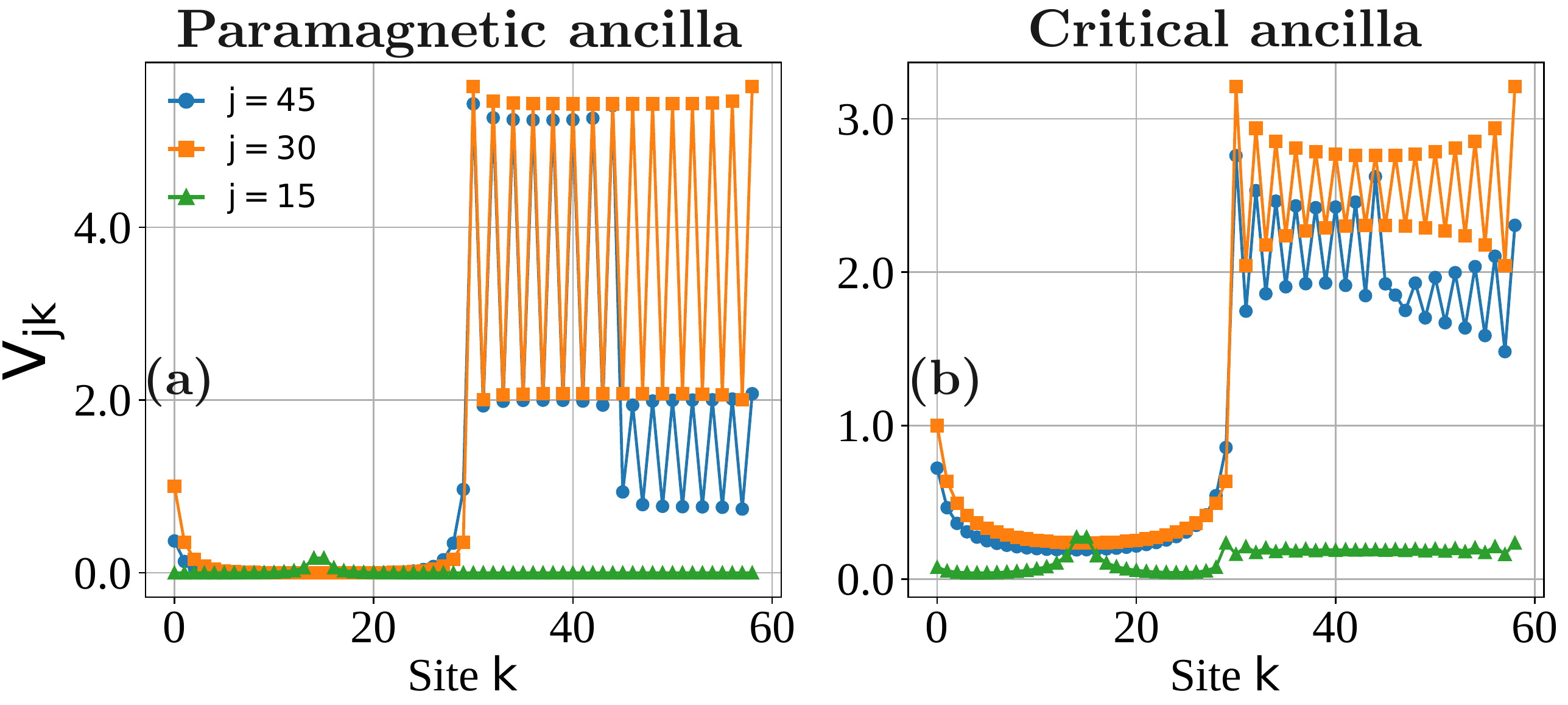}
\caption{(Color online). \textbf{$V_{jk}$ profile for a two-domain-wall $\tilde X$-basis measurement outcome.} The first 30 sites point in the energetically favorable $\rightarrow$ direction, while the remaining 30 sites point in the unfavorable $\leftarrow$ direction.  Non-decaying behavior of $V_{jk}$ occurs when $j,k$ both reside in the unfavorable domain. Data obtained using results from Appendix~\ref{app:Gaussian} for system size $N=60$. }
 \label{fig_xbasis_DW}
\end{figure}

As an example of a `bad' measurement outcome, consider next the domain-wall configuration $\ket{\tilde s} = \ket{\cdots \rightarrow\rightarrow\rightarrow \leftarrow\leftarrow\leftarrow\cdots}$. Figure~\ref{fig_xbasis_DW} shows that here $V_{jk}$ becomes highly non-local.  More precisely, $V_{jk}$ takes on sizable values whenever both $j$ and $k$ reside in the `$\leftarrow$' domain, regardless of their separation.  One can gain intuition for this observation by considering the ancilla ground state deep in the paramagnetic regime.  Here the ground state takes the form $\ket{\psi_a} = \ket{\rightarrow\rightarrow\cdots \rightarrow} + \cdots$, where the ellipsis denotes perturbative corrections induced by small $J_{\rm anc}/h_{\rm anc}$.  To leading order, these corrections involve spin flips on nearest-neighbor sites induced by $J_{\rm anc}$, i.e., admixture of $\ket{\cdots \rightarrow \rightarrow \leftarrow \leftarrow \rightarrow \rightarrow \cdots}$ components into the wavefunction.  Now consider the domain-wall outcome $\ket{\tilde s}$.  Flipping two spins at sites $j,k$ in the energetically unfavorable $\leftarrow$ domain tends to increase the overlap with the ground state---naturally leading to $V_{jk} = \frac{\bra{\tilde s(j,k)}\psi_a\rangle}{\bra{\tilde s}\psi_a \rangle}$ that can exceed unity even for distant $j,k$ as seen in our simulations.   Flipping one spin within each of the two domains takes the energetically favorable `$\rightarrow$' domain and introduces a \emph{single} $\leftarrow$ spin.  That domain then no longer resembles the ground state, which always harbors an even number of flipped spins.  The coupling $V_{jk}$ is therefore generically small with $j$ and $k$ in opposite domains, also as borne out in our numerics.  Finally, flipping two spins in the $\rightarrow$ domain again decreases the resemblance with the ground state---more so as the separation between the flipped sites $j$ and $k$ increases.  The corresponding $V_{jk}$ diminishes with $|j-k|$ in line with simulations yet again.  

More generally, `good' measurement outcomes are those for which flipping two far away spins invariably decreases overlap with the ground state such that $V_{jk} \rightarrow 0$ as $|j-k|$ increases.  In addition to the highest-probability uniform string, the next-highest-probability set of strings---which contain dilute sets of nearest-neighbor flipped spins relative to the uniform background---also satisfy this property.  Indeed, starting from such configurations, flipping spins at well-separated sites always locally produces regions with an odd number of flipped spins in a background of energetically favorable $\rightarrow$ spins, thereby obliterating the overlap with the ancilla ground state and hence $V_{jk}$.

\section{Protocol with $\tilde Z$-basis measurements}
\label{Ztilde_basis}

We now use our perturbative formalism to examine how correlations in the critical chain are modified by particular outcomes of $\tilde Z$-basis ancilla measurements in our protocol.  In Sec.~\ref{sec:properties} we saw that for this measurement basis both the mean and variance of $V_{jk}$ vanish as $|j-k|\rightarrow \infty$, suggesting that generic measurement outcomes yield well-behaved decaying interactions in $H_m$ [Eq.~\eqref{H_m}].  Moreover, with paramagnetic ancilla the variance of $m_j$ trended to a finite value at large system sizes, suggesting that the single-body piece in $H_m$ is also well-behaved for generic measurement outcomes.  Thus for paramagnetic ancilla, below we proceed with confidence considering unrestricted measurement outcomes from the lens of the continuum defect-line action obtained in Sec.~\ref{lowenergy}.  
For critical ancilla we saw that the variance of $m_j$ grew slowly with system size, warranting more caution in this scenario.  

Let us illustrate an example for case I where $m_j \propto 1-a(j)^2$ with $a(j) = \frac{\bra{\tilde s(j)}\psi_a\rangle}{\bra{\tilde s}\psi_a\rangle}$.  After the uniform strings, the next most likely measurement outcomes are those containing a single spin flip.  Consider one such state $\ket{\tilde s}$ with a single flipped spin at site $j_{\rm flip}$.  Subsequently flipping the spin at $j_{\rm flip}$ converts $\ket{\tilde s}$ back into the most probable, uniform string.  We thereby obtain $a(j_{\rm flip}) > 1$ and hence $m_{j_{\rm flip}} < 0$ for this measurement outcome.  With paramagnetic ancilla, this negative $m_{j_{\rm flip}}$ value saturates to a small constant as the system size increases.  With critical ancilla, by contrast, we find that the magnitude of this negative value continues to increase over accessible system sizes---but \emph{very} slowly similar to the standard deviation shown in the right panels of Fig.~\ref{fig_var_mj}.
We thus expect our formalism to apply also to general $\tilde Z$-basis measurement outcomes even for critical ancilla, at least over system sizes relevant for experiments.

We now consider the unitaries in cases I and II from Table~\ref{tab.unitaries} in turn.

\subsection{Case I}

We start with case I where the unitary reads $U_j=e^{iu(X_j-\langle X \rangle )\Tilde{X}_j}$.  This form of $U_j$ preserves the $\mathbb{Z}_2$ symmetries $G$ and $\tilde G$ for the critical and ancilla chains, but does not preserve time reversal symmetry $\mathcal{T}$ (which sends $U_j \rightarrow U_j^\dagger$).  Thus although the $\tilde Z$-basis measurements break $\tilde G$ symmetry, the post-measurement state $\ket{\psi_{\tilde{s}}}$ remains invariant under $G$.  These considerations tell us that, for case I, $U' = e^{i H'}$ in Eq.~\eqref{psi_m_goal} is generically nontrivial [as one can indeed see from Eq.~\eqref{H_U}] while $H_m$ and hence $\mathcal{S}_m$ must preserve $G$ symmetry.  
Indeed, Eq.~\eqref{H_m} now takes the manifestly $G$-invariant form
\begin{align}
    H_m &= u^2\sum_j m_j(X_j-\langle X\rangle)
    \nonumber \\ 
    &+ u^2\sum_{j \neq k}V_{jk}(X_j-\langle X\rangle)(X_k-\langle X\rangle)
    \label{H_mI}
\end{align}
with
\begin{align}
    m_j &= -2\langle X \rangle [1-a(j)^2].
    \label{mjI}
\end{align}
and $V_{jk}$ given (as for all cases) by Eq.~\eqref{eq:Vjk}.  The defect-line action, using the low-energy expansion from Eq.~\eqref{Xexpansion}, then reads
\begin{align}
    \mathcal{S}_m &= u^2\int_x m(x) \varepsilon(x,\tau = 0) 
    \nonumber \\
    &+ u^2 \int_{x,y}V(x,y) \varepsilon(x,\tau = 0)\varepsilon(y,\tau = 0).
    \label{SmI}
\end{align}
Here $m(x)$ and $V(x,y)$ represent the coarse-grained, continuum-limit counterparts of $m_j$ and $V_{jk}$.  

Provided $V(x,y)$ scales to zero faster than $1/|x-y|$---which is indeed generally the case both for paramagnetic and critical ancilla---we can approximate the second line of Eq.~\eqref{SmI} as a local interaction obtained upon fusing the two $\varepsilon$ fields according to the fusion rules summarized in Eq.~\eqref{fusionrules}.
The leading nontrivial fusion product is $-i\gamma_R \partial_x \gamma_R + i \gamma_L \partial_x \gamma_L$, which is a descendent of the identity that, crucially, has a larger scaling dimension compared to the $\varepsilon$ field appearing in the first line of Eq.~\eqref{SmI}~\cite{yellowbook}.  It follows that for capturing long-distance physics we can neglect the
$V(x,y)$ term altogether and simply take
\begin{equation}
    \mathcal{S}_m \approx u^2\int_x m(x) \varepsilon(x,\tau = 0).
    \label{SmIb}
\end{equation}
We are primarily interested in computing the two-point correlator
\begin{equation}
    \langle Z_j Z_{j'} \rangle_{\tilde s} \sim \langle \sigma(x_j) \sigma(x_{j'}) \rangle_{\tilde s}
    \label{ZZI}
\end{equation}
in the presence of Eq.~\eqref{SmIb}.  Technically, according to Eq.~\eqref{A1m} we need to conjugate the $Z_j$ operators with the unitary $U'$, which in case I rotates $Z$ about the $X$ direction.  Such a rotation only mixes in operators in the low-energy theory with (much) larger scaling dimension compared to $\sigma$ \footnote{Explicitly, after dropping terms that vanish by time-reversal symmetry, one finds $\avg{(U^{\prime})^{ \dagger}Z_jZ_{j^\prime}U^\prime}_{\tilde{s}}=\cos(2ua(j))\cos(2ua(j^\prime))\avg{Z_jZ_{j^\prime}}_{\tilde{s}}+\sin(2ua(j))\sin(2ua(j^\prime))\avg{Y_jY_{j^\prime}}_{\tilde{s}}$.  Given that  $(i)$ $Y_j$ maps to a CFT operator with larger scaling dimension than that for $Z_j$ ($Y_j\sim i\partial_{\tau}\sigma$, $\Delta_{\partial_{\tau}\sigma}=9/8$~\cite{PFEUTY}) and $(ii)$ our perturbative expansion focuses on the $u \ll 1$ regime, the $U'$ unitary can be safely neglected.}. Hence Eq.~\eqref{ZZI}---which is the same as what one would obtain by ignoring $U'$ altogether---continues to provide the leading decomposition for the correlator.

When $m(x)$ is independent of $x$, as arises for uniform measurement outcomes, the above defect line action is marginal, though for general measurement outcomes $m(x)$ retains nontrivial $x$ dependence.  References \onlinecite{mccoy_1980,CABRA_1994,NT2011} employed non-perturbative field-theory methods to study the effects of this type of defect line on spin-spin correlation functions in the two dimensional Ising model.  In particular, Ref.~\onlinecite{NT2011} derived the spin-spin correlation function for an $\varepsilon$ line defect whose coupling is an arbitrary function of position.  We report here their main result: 
\begin{align}\label{eq:spin_inh}
    &\langle \sigma(x)\sigma(x')\rangle_{\tilde{s}}\sim |x-x'|^{-\frac{1}{4}-\frac{\kappa u^2}{8} [m(x)+m(x')]}e^{\frac{\kappa u^2}{16}\mathcal{F}(x,x')},
\end{align}
where 
\begin{align}
\mathcal{F}(x,x') &= \int_x^{x'} \!\! \D{y} m(y)\frac{\D{}}{\D{y}}\ln\frac{[(y-x')^2+a^2]^{1+\kappa u^2m(x')}}{[(y-x)^2+a^2]^{1+\kappa u^2 m(x)}} \nonumber\\
&- \int_x^{x'} \!\! \D{y} \D{y'} [1+\kappa u^2m(y')]\left[\frac{\D{}}{\D{y}}m(y)\right]
\nonumber \\
&\times \frac{\D{}}{\D{y'}}\ln[(y-y')^2+a^2].
\label{Fdef}
\end{align}
Above, $a$ is a short-distance cutoff, and $\kappa$ is a dimensionless parameter that captures an overall constant neglected on the right side of Eq.~\eqref{Xexpansion} as well as difference in normalization conventions between our work and Ref.~\onlinecite{NT2011}.  We simply view $\kappa$ as a fitting parameter in our analysis.   Since $\mathcal{F}(x,x')$ in Eq.~\eqref{eq:spin_inh} already contains an $O(u^2)$ prefactor, to the order we are working it suffices to simply set $u = 0$ in Eq.~\eqref{Fdef}.  Some algebra then gives the far simpler expression
\begin{align}
    \langle \sigma(x)\sigma(x')\rangle_{\tilde{s}}\sim |x-x'|^{-\frac{1}{4}-\frac{\kappa u^2}{4} [m(x)+m(x')]}e^{\frac{\kappa u^2}{8}f(x,x')}
    \label{sigma_nice}
\end{align}
with 
\begin{align}
f(x,x') =
\int_x^{x'} \!\! \D{y} \ln\left[\frac{(y-x)^2+a^2}{(y-x')^2+a^2}\right]\frac{\D{}}{\D{y}}m(y). 
\label{fsimple}
\end{align}
Eqs.~\eqref{sigma_nice} and \eqref{fsimple} capture coarse-grained spin-spin correlations for general measurement outcomes, though for deeper insight we now explicitly examine some special cases.

\begin{figure}[t!]
\centering
\includegraphics[width=\linewidth]{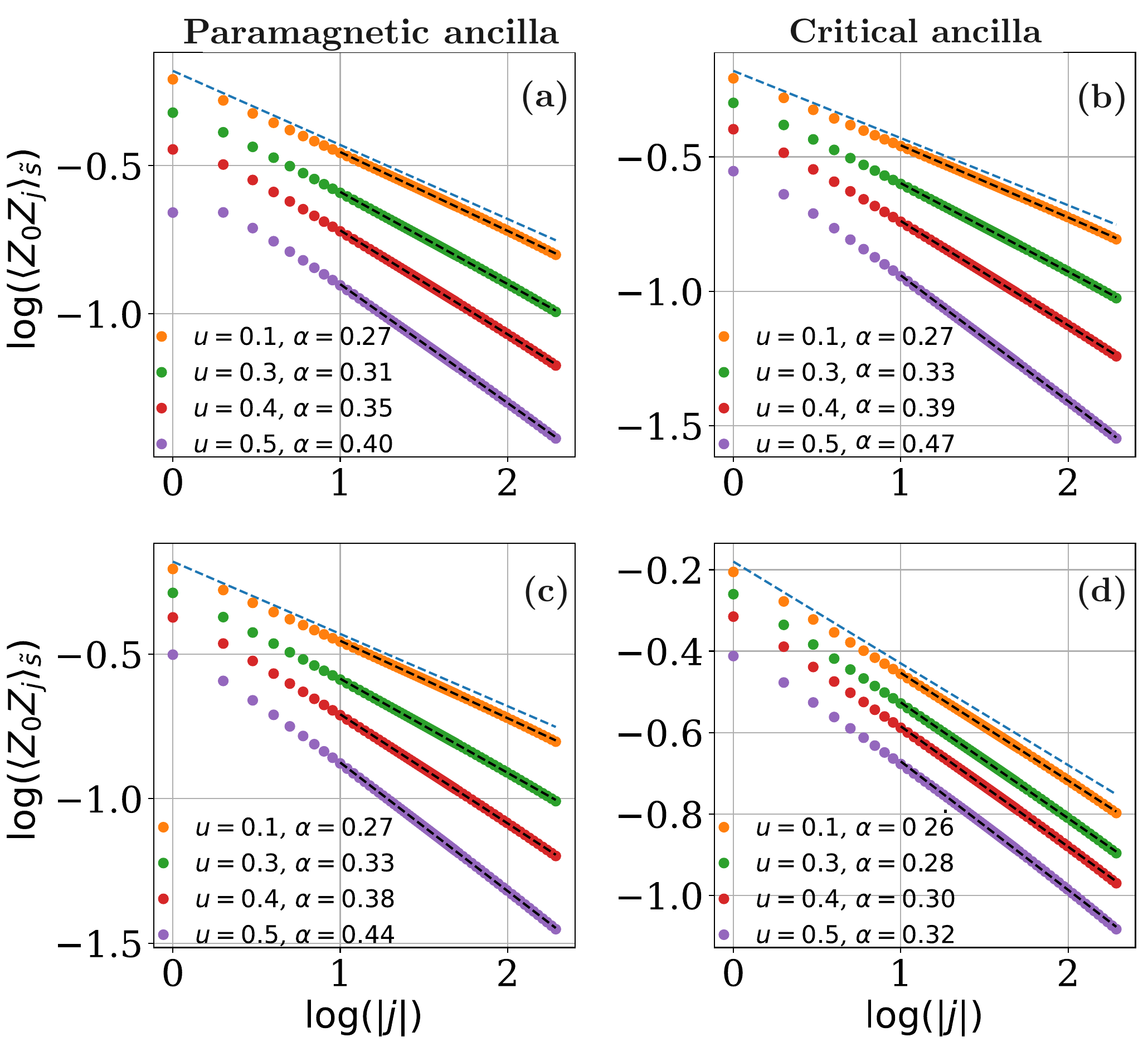}
\caption{(Color online). \textbf{Correlation function $\langle Z_0Z_j\rangle_{\tilde{s}}$ for uniform measurement outcomes.} The first row corresponds to case I from Table~\ref{tab.unitaries} while the second corresponds to case III. At $u = 0$ the curves exhibit an exponent $1/4$ that follows from the pristine Ising CFT.  Turning on $u\neq 0$ yields a measurement-induced \emph{increase} in the scaling dimension in all panels---as predicted by Eq.~\eqref{sigma_dim} for case I and~\eqref{correlator_caseIII} for case III.  Data were obtained using infinite DMRG with bond dimension $1000$ for paramagnetic ancilla and $2000$ for critical.}
 \label{fig_uniform_dmrg}
\end{figure}

\begin{figure}[t!]
\centering
\includegraphics[width=0.75\linewidth]{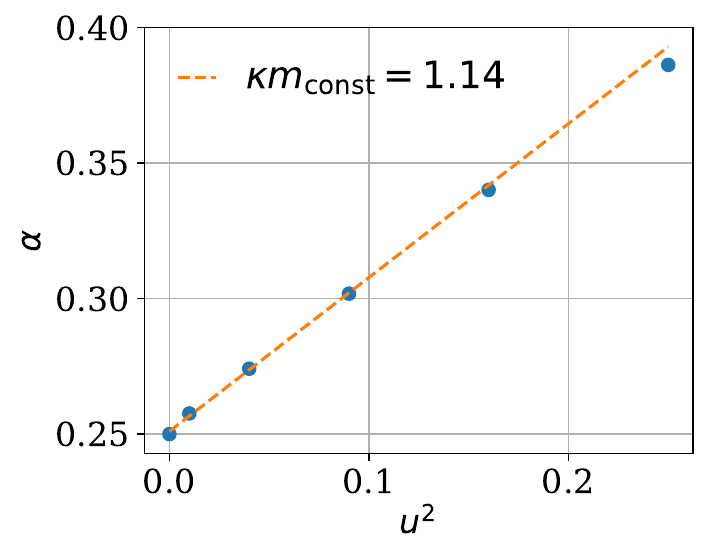}
\caption{(Color online). \textbf{Scaling of the power-law exponent for $\langle Z_0 Z_j\rangle_{\tilde s}$ with a uniform measurement outcome.}  Data correspond to case I in Table~\ref{tab.unitaries} assuming paramagnetic ancilla, and were obtained using iDMRG.  The numerically extracted exponent $\alpha$ scales approximately linearly with $u^2$ at small $u$, in quantitative agreement with $2\Delta_\sigma(u)$ predicted by Eq.~\eqref{sigma_dim}. A linear fit to Eq.~\eqref{sigma_dim} yields $\kappa m_{\rm const} = 1.14$. } 
 \label{fig_fit_dmrg}
\end{figure}

For a uniform measurement outcome (e.g., $s_j = +1$ for all $j$) giving constant $m(x) \equiv m_{\rm const}$, Eq.~\eqref{sigma_nice} simplifies to
\begin{align}   \langle\sigma(x)\sigma(x')\rangle_{\tilde{s}} &\sim \frac{1}{|x-x'|^{2\Delta_\sigma(u)}},~~~~({\rm uniform}~\tilde s)
    \label{correlator_uniform}
    \\
    \Delta_{\sigma}(u) &= \frac{1}{8}(1+2 \kappa u^2 m_{\rm const}),
    \label{sigma_dim}
\end{align}
consistent with the result found in Refs.~\onlinecite{mccoy_1980} and \onlinecite{CABRA_1994} in the limit $|x-x'|\gg a$ and to $O(u^2)$. Remarkably, the defect line in this post-selection sector yields an $O(u^2)$ change in the scaling dimension of the $\sigma$ field compared to the canonical result in Eq.~\eqref{sigma_standard}.  We confirm this change using infinite DMRG simulation reported in Fig.~\ref{fig_uniform_dmrg}$(a,b)$: with both paramagnetic and critical ancilla, the scaling dimension of the $\sigma$ field, $\Delta_{\sigma}(u)$, exceeds $1/8$ when $u \neq 0$, as is particularly clear at $u \geq 0.3$. The fact that the scaling dimension increases (rather than decreases) with $u$ is consistent with the fact that in this case the $e^{-H_m/2}$ non-unitary implements weak measurement in the $X$ basis---thereby naturally suppressing $Z$ correlations.
The scaling-dimension enhancement is quite similar for the paramagnetic and critical cases, as expected given that $m_j$ is only slightly larger in the latter [see black dashed line in Fig.~\ref{fig_examplesDW}$(c,d)$].  
Figure~\ref{fig_fit_dmrg} shows the dependence of the numerically extracted power-law exponent $\alpha$ as a function of $u^2$, revealing a linear dependence in agreement with Eq.~\eqref{sigma_dim}. The linear fit also allows us to extract a value $\kappa=-1.12$; note that $\kappa m_{\rm const}>0$---ensuring that $\Delta_{\sigma}(u)$ increases with $u$ in the presence of the defect line as observed in our numerical simulations. 

Next we examine a measurement outcome $\ket{\tilde s} = \ket{\cdots \uparrow \uparrow \uparrow \downarrow \downarrow \downarrow \cdots}$ with a domain wall.  This outcome yields nearly uniform $m_j$---see black dashed lines in Figs.~\ref{fig_examplesDW}$(c,d)$---except for a window around the domain wall where it approximately vanishes. We model the associated continuum $m(x)$ profile as
\begin{equation}
    m(x) = m_{\rm const} \{\Theta[(x_0-d)-x]+\Theta[x-(x_0+d)]\},
    \label{m_domain}
\end{equation}
where $x_0$ is the domain-wall location, $d$ is the spatial extent of suppressed $m(x)$ region on either side and $\Theta$ is the Heaviside function. 
Adequately capturing detailed behavior near the domain wall likely requires incorporating short-distance physics, though we expect that our low-energy framework can describe correlations among operators sufficiently far from $x_0$.  With this restriction in mind, we consider the two-point correlator $\langle\sigma(x)\sigma(x')\rangle_{\tilde{s}}$ with $x$ far to the left of the domain wall ($x \ll x_0$) and $x'>x$.  If $x'$ also sits to the left of the domain wall, then $f(x,x') = 0$ and the correlator retains---within our approximation---exactly the same form as in Eq.~\eqref{correlator_uniform}.  If, however, $x'$ sits to the right of the domain wall with $x' \gg x_0$, then we obtain 
\begin{align}
    f(x,x') \approx 4  d m_{\rm const}\left(\frac{1}{x_{0}-x} + \frac{1}{x'-x_{0}}\right),
\end{align}
resulting in a modest enhancement of the correlator amplitude compared to the domain-wall-free case. 
Summarizing, for the single-domain-wall measurement outcome we get
\begin{align}\label{eq:DW2}
  \langle\sigma(x)\sigma(x')\rangle_{\tilde{s}} &\sim \begin{cases}
      \frac{1}{|x-x'|^{2\Delta_\sigma(u)}}, &x' \ll x_0
      \\
      \frac{e^{\frac{1}{2}\kappa u^2 d m_{\rm const}\left(\frac{1}{x_0-x} + \frac{1}{x'-x_0}\right)}}{|x-x'|^{2\Delta_\sigma(u)}}, &x' \gg x_0
  \end{cases}.
\end{align}

\begin{figure}[t!]
\centering
\includegraphics[width = 1.0\columnwidth]{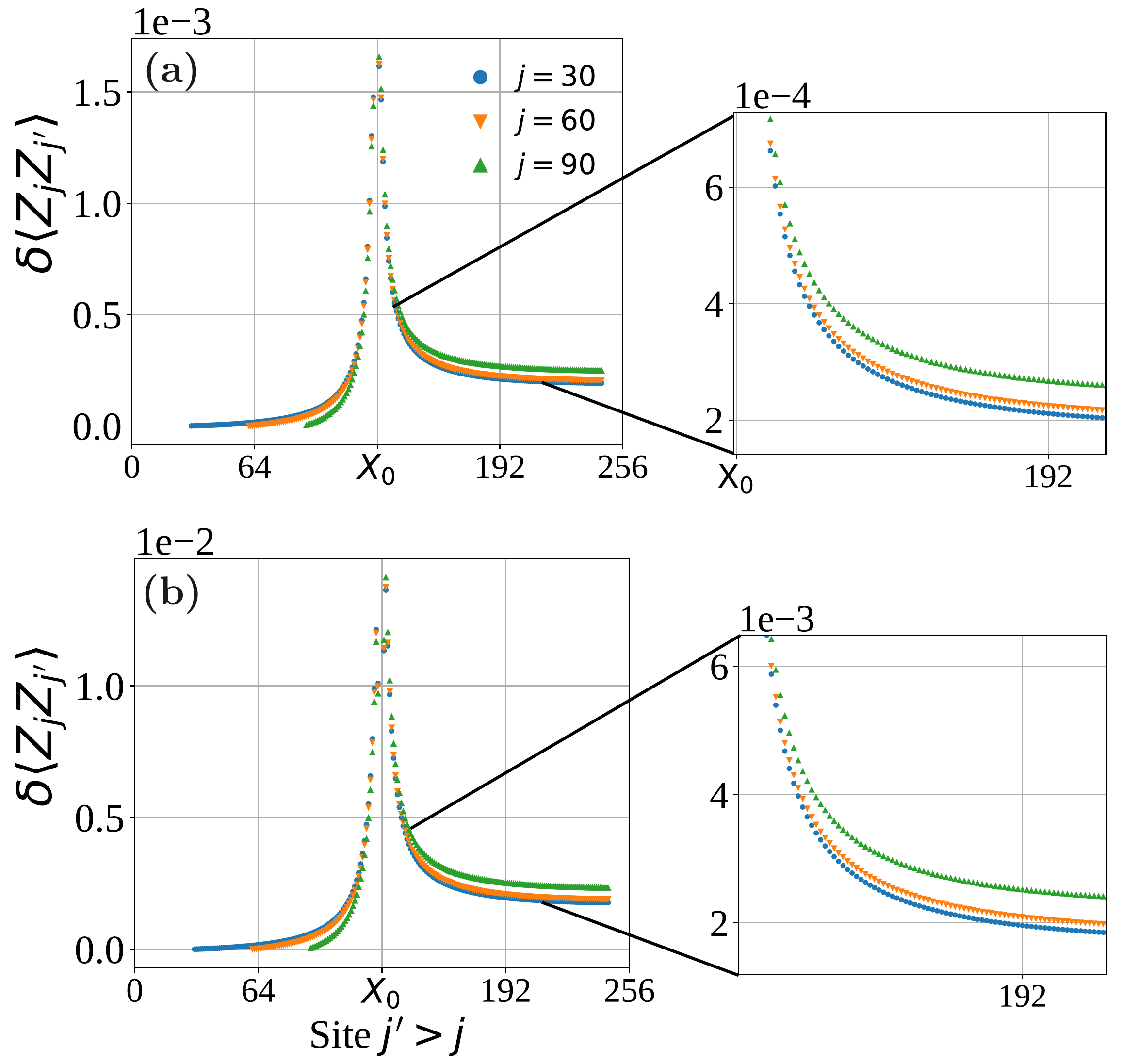}
\caption{(Color online). \textbf{Relative correlation function $\delta \langle Z_{j}Z_{j'}\rangle$ [Eq.~\eqref{relative_corr}] for a domain-wall measurement outcome in case I from Table~\ref{tab.unitaries}.}  Main panels illustrate the relative change in the two-point function resulting from insertion of a domain wall. Panel (a) corresponds to $u=0.1$, and panel (b) to $u=0.3$. The domain wall resides near site $x_0 = 128$ in an $N = 256$ system with open boundary conditions.  When $j$ and $j'$ both sit on one side of the domain wall, the change in correlations is negligible.  When they sit on opposite sides, however, the correlations \emph{increase} relative to the uniform measurement outcome.  As discussed in the main text, the behavior captured here reproduces the main qualitative features of the analytical prediction in Eq.~\eqref{eq:DW2}.  Data were obtained using DMRG with paramagnetic ancilla.}
 \label{fig_DW_dmrg}
\end{figure}

To test Eq.~\eqref{eq:DW2}, we performed DMRG simulations for a system of size $N = 256$ system with open boundary conditions, so that we can accommodate a single-domain-wall measurement outcome.
Figure~\ref{fig_DW_dmrg} plots the numerically determined function 
\begin{equation}
    \delta  \langle Z_j Z_{j'}\rangle \equiv \frac{ \langle Z_j Z_{j'}\rangle_{\tilde{s},\mathrm{DW}}- \langle Z_j Z_{j'}\rangle_{\tilde{s},\mathrm{unif}}}{ \langle Z_j Z_{j'}\rangle_{\tilde{s},\mathrm{unif}}},
    \label{relative_corr}
\end{equation}
i.e., the difference in the microscopic two-point correlator with and without a domain wall, normalized by the correlator for the uniform measurement outcome.  
Quite remarkably, the figure  reveals the main qualitative features predicted by our result in Eq.~\eqref{eq:DW2}: When both $j$ and $j'$ sit to the left of the domain wall, the difference in correlators approaches zero, while the correlator in the presence of a domain wall exhibits a small enhancement when $j$ and $j'$ sit on opposite sides of the domain wall.  Moreover, the enhancement factor modestly increases as $j$ approaches the domain wall---also in harmony with Eq.~\eqref{eq:DW2}.  The agreement between numerics and analytics here provides a very nontrivial check on our formalism.

In the presence of multiple well-separated dilute domain walls, the behavior of the correlator $\langle\sigma(x)\sigma(x')\rangle_{\tilde{s}}$ follows from a straightforward generalization of Eq.~\eqref{eq:DW2}.  For $x$ and $x'$ within the same domain, the correlator again reproduces that in a uniform measurement outcome, whereas moving $x'$ rightward leads to a relative uptick in the correlator upon passing  successive domain walls.  For dense domain walls, the $m_j$ pattern changes significantly, necessitating a separate analysis.

\subsection{Case II}

The unitary in case II, $U_j = e^{i u(Z_j-C)\tilde X_j}$, is invariant under $\tilde G$ but preserves neither $G$ nor $\mathcal{T}$.  Thus the post-measurement state generically breaks all microscopic symmetries.  A special case arises, however, when $C = 0$: here $U_j$ and hence the post-measurement state preserve the composite operation $G\mathcal{T}$.  In line with these symmetry considerations, case II yields
\begin{align}
    H_m = u^2 \sum_j m_j Z_j + u^2 \sum_{ j\neq k} V_{jk}Z_j Z_k
\end{align}
with 
\begin{equation}
    m_j = -2C\left[1-a(j)^2+\sum_{k\neq j}V_{jk}\right].
\end{equation}
Indeed, the $m_j$ term---which is odd under $G$---appears as long as $C \neq 0$.  The unitary $U' = e^{i H'}$, by contrast, is generically nontrivial even for $C = 0$ and always preserves $G \mathcal{T}$: $H'$ in Eq.~\eqref{H_U} is odd under $G$ in case II, but in $U'$ the minus sign is undone by $i\rightarrow -i$ from time reversal.  Nevertheless, we only consider $Z_j$ correlators below, which here are invariant under conjugation by $U'$.

The associated continuum defect-line action, now using Eq.~\eqref{Zexpansion}, is
\begin{align}
    \mathcal{S}_m &= u^2\int_x m(x) \sigma(x,\tau = 0) 
    \nonumber \\
    &+ u^2 \int_{x,y}V(x,y) \sigma(x,\tau = 0)\sigma(y,\tau = 0).
    \label{SmII}
\end{align}
As in case I, $V(x,y)$ decays fast enough that we can approximate the second line with a local interaction,  obtained here by fusing the pair of $\sigma$ fields.  The leading nontrivial fusion product is $\varepsilon$ [see Eq.~\eqref{fusionrules}].  For $C = 0$, where the $m(x)$ term drops out by symmetry, Eq.~\eqref{SmII} then reduces to the form 
\begin{equation}
  \mathcal{S}_m = u^2 \int_x M(x) \varepsilon(x,\tau = 0),~~~~(C = 0)
  \label{SmIIC0}
\end{equation} 
studied in the previous subsection with $M(x) = \int_y V(x,y)$.  One-point correlators $\langle Z_j \rangle_{\tilde s}$ vanish by symmetry (to all orders in $u$ due to preservation of $G\mathcal{T}$ symmetry) while two-point correlators  $\langle Z_j Z_{j'}\rangle_{\tilde s}$ can be computed using the methods deployed above.  For the remainder of this subsection we therefore take $C \neq 0$.  In this regime the $\sigma$ field arising from first line of Eq.~\eqref{SmII} has a smaller scaling dimension compared to the $\varepsilon$ field emerging from the second line.  We can therefore neglect the latter term, yielding
\begin{equation}
    \mathcal{S}_m \approx u^2\int_x m(x) \sigma(x,\tau = 0),~~~~(C\neq 0).
    \label{SmIIb}
\end{equation}
Physically, $m(x)$ plays the role of a longitudinal magnetic field that acts only at $\tau = 0$.  

\begin{figure}[t!]
\centering
\includegraphics[width=0.97\linewidth]{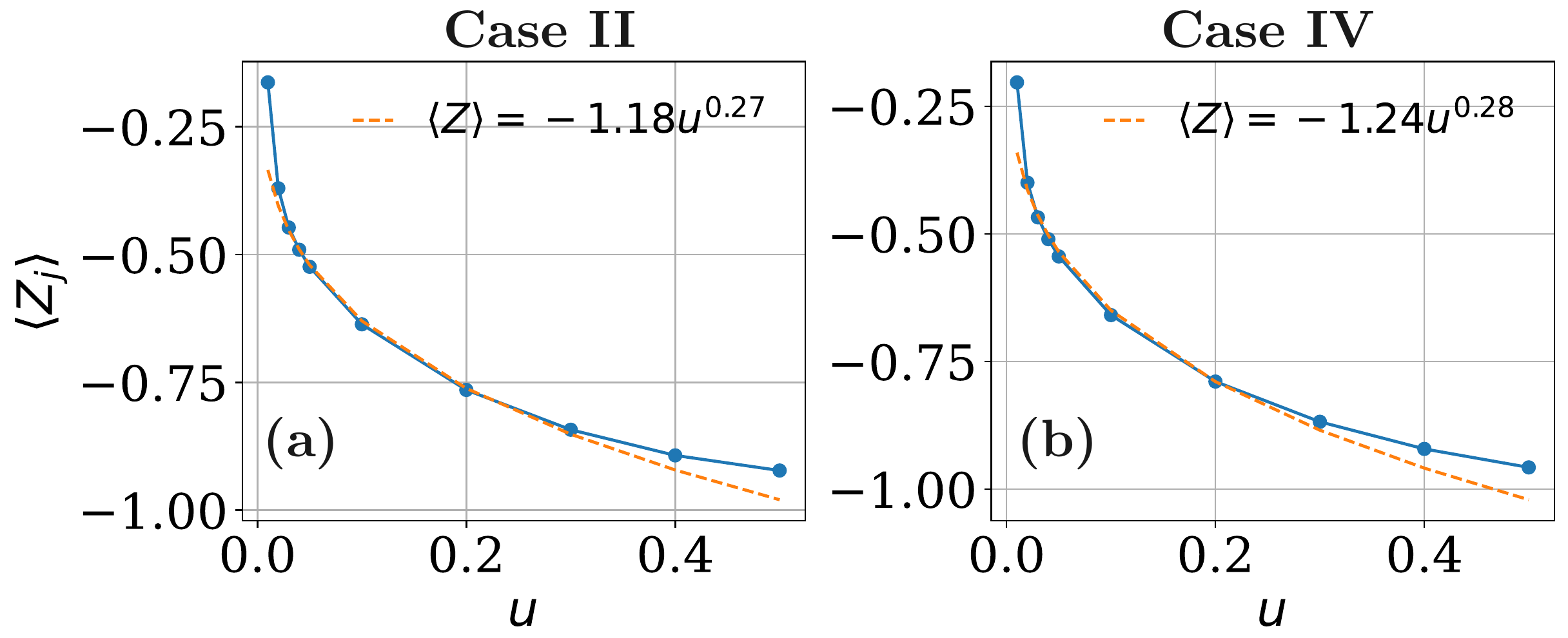}
\caption{(Color online). \textbf{One-point function $\langle Z_j\rangle$ with a uniform measurement outcome.} Panels (a) and (b) respectively correspond to cases II and IV from Table~\ref{tab.unitaries}.  Results were obtained using infinite DMRG with $C = -1$, assuming paramagnetic ancilla.  The non-zero value generated by measurement at $u \neq 0$ validates analytic predictions, e.g., Eq.~\eqref{onepoint}. 
 Moreover, the fits shown by the dashed lines exhibit excellent agreement with the $u$ dependence extracted using renormalization group arguments.  } 
 \label{fig_fit_dmrg2}
\end{figure}

For uniform strings where $m(x) = m_{\rm const}$, the defect-line action in Eq.~\eqref{SmIIb} constitutes a strongly relevant perturbation.  Clearly then $\sigma(x,\tau = 0)$ and hence $Z_j$ take on uniform, non-zero expectation values in this post-selection sector:
\begin{equation}
  \langle Z_j \rangle_{\tilde s} \sim \langle \sigma(x_j)\rangle_{\tilde s} = g(u),~~~~({\rm uniform}~\tilde s),
  \label{onepoint}
\end{equation}
where the function $g(u)$ vanishes as $u\to 0$ but tends to a non-zero constant at $u \neq 0$ in the thermodynamic limit.  In sharp contrast,  prior to measurements we have $\avg{\psi_U|Z_j|\psi_U}=0$ for all $u$. Infinite DMRG results for $\langle Z_j\rangle_{\tilde s}$ presented in  Fig.~\ref{fig_fit_dmrg2} confirm the qualitative behavior predicted by Eq.~\eqref{onepoint}.  

For a more quantitative treatment, we apply the renormalization group (RG) technique to obtain the dependence of $\avg{\sigma(x_j)}_{\Tilde{s}}$ on $u$.
Rescaling the spatial coordinate $x$ by a factor $b$ and defining $x'=x/b$, the $\sigma$ field transforms as $\sigma(x) = b^{-1/8} \sigma'(x')$.
The defect-line action $S_m = u^2 m_\text{const} \int \D{x} \sigma(x)$ is then rewritten as $S_m = u^2 m_\text{const} b^{1-1/8} \int \D{x'}\sigma'(x')$ and in particular exhibits a renormalized coupling strength $u^2 b^{7/8}$.
Suppose now that at some coupling strength $u_\text{ref}^2$, the magnetization is a fixed constant $\avg{\sigma'(0)}_{\tilde s} = M_\text{ref}$.
We can back out the observables at arbitrary $u$ by finding the RG map that takes $u \to u_\text{ref}$.
First, choose the scaling parameter $b$ such that $u^2 b^{7/8} = u_\text{ref}^2$, i.e., $b = (u/u_\text{ref})^{-16/7}$.
We then obtain $\avg{\sigma(0)}_{\Tilde{s}} = b^{-1/8}M_{\mathrm{ref}}\propto u^{2/7}$.
Despite the simplicity of this argument, a fit of $\avg{Z_j}$ in Fig.~\ref{fig_fit_dmrg2} yields a scaling $\sim u^{0.27}$ with an exponent that agrees well with our prediction of $2/7 \approx 0.29$.   

\begin{figure}[t!]
\centering
\includegraphics[width = \columnwidth]{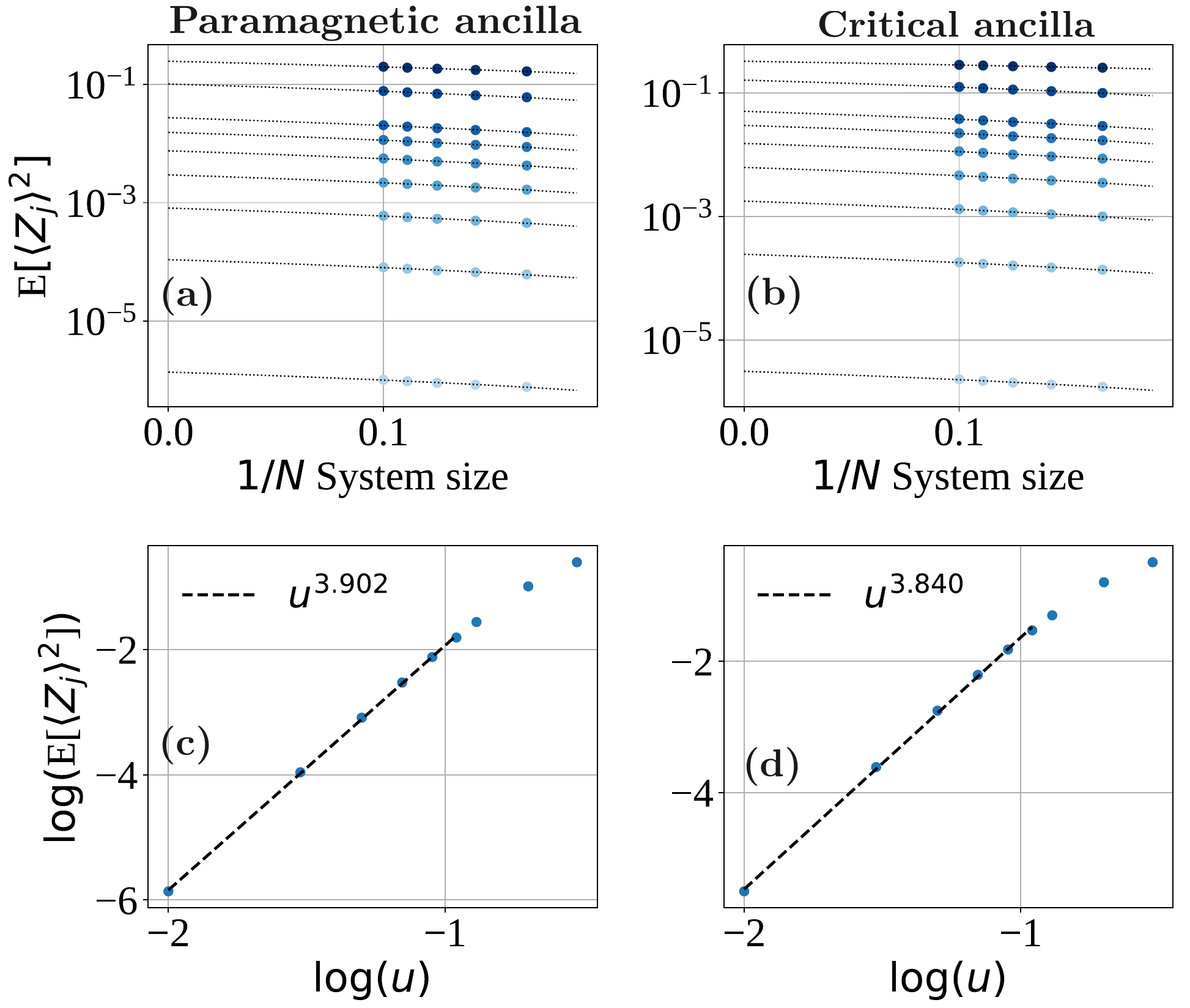}
\caption{(Color online). \textbf{Average of $\langle Z_i \rangle^2_{\tilde s}$ over measurement outcomes in case II of Table~\ref{tab.unitaries}.} Data were obtained using exact diagonalization for two chains of length $N = 6, 7, 8, 9, 10$ with $C = -1$.  Panels $(a,b)$ reveal well-behaved scaling with system size; larger values of $u$ correspond to darker blue, with the darkest color corresponding to $u = 0.13$. Panels $(c,d)$ show the extrapolated dependence of $\textrm{E}(\avg{Z_j}_{\tilde{s}}^2)$ with $u$. For small $u$, we find $\sim u^4$ scaling, consistent with the crude expectation that $\avg{Z_j}_{\tilde s}\sim u^2 m_j$ for a particular measurement outcome.}
 \label{fig_av_Z2}
\end{figure}

We are not aware of works that compute the one-point function in the presence of arbitrary position-dependent $m(x)$'s that arise with generic measurement outcomes.  Nevertheless, we expect that, at least for smoothly varying $m(x)$ profiles,  $\langle \sigma(x)\rangle_{\tilde s}$ polarizes for each $x$ with an orientation determined by the sign of $m(x)$.  Since $m_j$ averages to zero as shown in Sec.~\ref{sec:properties}, averaging $\langle Z_j \rangle_{\tilde s}$ over measurement outcomes then naturally erases the effects of measurements as must be the case on general grounds.  In contrast, such cancellation need not arise when averaging $\langle Z_j \rangle_{\tilde s}^2$ over measurement outcomes.  We thus anticipate that
\begin{equation}\label{eq:avg_zsquare}
    \sum_{\tilde s}p_{\tilde s}\langle Z_j \rangle_{\tilde s}^2 \neq 0.
\end{equation}
Very crudely, if for a random measurement outcome $\langle Z_j\rangle_{\tilde s} \sim u^2 m_j$, then the nonlinear average above would be proportional to $ u^4 {\rm Var}(m_j)$.  Our exact diagonalization results presented in Fig.~\ref{fig_av_Z2} support these predictions.  The top panels show $\langle Z_j\rangle_{\tilde s}^2$ averaged over all $\tilde Z$-basis measurement outcomes versus $1/N$ for several values of $u$.  For both paramagnetic and critical ancilla, extrapolation to $N \rightarrow \infty$ yields nonzero values for all $u \neq 0$ cases; additionally, the lower panels show that the extrapolated values indeed scale very nearly as $u^4$ for small $u$.

\section{Protocol with $\tilde X$-basis measurements}
\label{Xtilde_basis}

Recall that for $\tilde X$-basis measurements, our perturbative formalism applies only to the highest probability subset of even-string ($\tilde G \ket{\tilde s} = + \ket{\tilde s}$) measurement outcomes.  These outcomes, to which we exclusively focus in this section, include the uniform state with $\tilde s_j = +1$ on every site, and descendant states containing a dilute set of adjacent spin flips.  Interestingly, even in this restricted space of measurement outcomes, we will encounter qualitative differences between paramagnetic versus critical ancilla in our protocol.

For both cases III and IV, the unitary $U' = e^{i H'}$ is trivial in the even-string measurement sector.  This result immediately follows from Eq.~\eqref{H_U} using the fact that $a(j) = 0$ for any even-string $\tilde s$.  Hence in the ensuing analysis we need only consider $H_m$ and the associated defect-line action $\mathcal{S}_m$.  Note also that the $U_j$ unitaries for cases III and IV explicitly violate $\tilde G$ symmetry; nevertheless, ancilla measurements project onto an even-string $\tilde s$ (by assumption), so that both the initial and post-measurement states are $\tilde G$ eigenstates with eigenvalue $+1$.  The situation is reversed compared to the protocol with $\tilde Z$-basis measurements, where the $U_j$ unitaries preserved $\tilde G$ while measurements produced a wavefunction that was not a $\tilde G$ eigenstate.

\subsection{Case III}

The case-III unitary $U_j=e^{iu(X_j-\langle X \rangle )\Tilde{Z}_j}$ yields a defect-line action
\begin{align}
    \mathcal{S}_m &= u^2m\int_x \varepsilon(x,\tau = 0) 
    \nonumber \\
    &+ u^2 \int_{x,y}V(x,y) \varepsilon(x,\tau = 0)\varepsilon(y,\tau = 0).
    \label{SmIII}
\end{align}
Equation~\eqref{SmIII} has the same form as Eq.~\eqref{SmI} from case I---with the crucial difference that here $m(x)$ is replaced with a \emph{constant} $m = -2 \langle X \rangle$ that is the same for all of the (restricted) strings that we consider.  For paramagnetic ancilla, results from Sec.~\ref{sec:properties} imply that $V(x,y)$ decays exponentially with $|x-y|$, allowing us to once again fuse the $\varepsilon$'s in the second line into a subleading term compared to the first.  More care is needed for critical ancilla, since for the uniform string outcome $V(x,y)$ decays like $1/|x-y|$.  The second line then represents an inherently long-range, power-law-decaying  interaction.  Such a term is, however, still less relevant by power counting compared to the first line.  Thus, similar to case I, we can approximate the defect-line action as simply
\begin{align}
    \mathcal{S}_m &\approx u^2m\int_x \varepsilon(x,\tau = 0).
    \label{SmIIIb}
\end{align}

In this case one-point $\avg{Z_j}$ correlators again vanish by symmetry, while two-point correlators correspondingly behave as
\begin{align}
    \langle\sigma(x)\sigma(x')\rangle_{\tilde{s}} \sim \frac{1}{|x-x'|^{2\Delta_\sigma(u)}},~
    \Delta_{\sigma}(u) = \frac{1}{8}(1+2 \kappa u^2 m).
    \label{correlator_caseIII}
\end{align}
At least within the approximations used here, a pristine power-law with $O(u^2)$ enhanced scaling dimension occurs for any even-string measurement outcome conforming to our perturbative formalism, even if the outcome is not translationally invariant.  Surely additional ingredients beyond those considered here would restore dependence on the measurement outcome; such terms, however, reflect subleading contributions, e.g., the neglected $V(x,y)$ term above.  By contrast, for case I the dependence on measurement outcome was already encoded in the leading $m(x) \varepsilon$ term in the defect-line action.

The lower panels of Fig.~\ref{fig_uniform_dmrg} confirm the modified power-law behavior for the uniform measurement outcome, which again is especially clear at $u \geq 0.3$. 
Notice that for $u = 0.1$, the fitted scaling dimension is nearly the same for cases I and III, and for both paramagnetic and critical ancilla.  This similarity is expected from our perturbative framework given that the leading defect-line actions [Eqs.~\eqref{SmIb} and~\eqref{SmIIIb}] take the same form with similar coupling strengths in the uniform measurement outcome sector.  At the larger value of $u = 0.3$, the extracted scaling dimensions in case III differ for paramagnetic and critical ancilla---even though our $O(u^2)$ theory predicts precisely the same exponent in both scenarios.  Such a correction is not surprising, given that at higher orders in $u$, even the leading term in the defect-line action can discriminate between paramagnetic and critical ancilla.  Indeed, we have checked that in the paramagnetic case $\Delta_{\sigma}(u)$ scales like $u^2$ over a wider range of $u$ compared to the case with critical ancilla.

\subsection{Case IV}
\label{sec:caseIV}

For case IV, with unitary $U_j = e^{iu (Z-C)\tilde Z}$, the defect-line action reads
\begin{align}
    \mathcal{S}_m &= u^2\int_x m(x) \sigma(x,\tau = 0) 
    \nonumber \\
    &+ u^2 \int_{x,y}V(x,y) \sigma(x,\tau = 0)\sigma(y,\tau = 0),
    \label{SmIV}
\end{align}
which has identical structure to that of case II but with modified couplings $m(x)$ and $V(x,y)$ due to the shift in ancilla measurement basis.  As in case II, the special limit $C = 0$ still yields $m(x) = 0$, as required by symmetry.  

\begin{figure}[t!]
\centering
\includegraphics[width = \columnwidth]{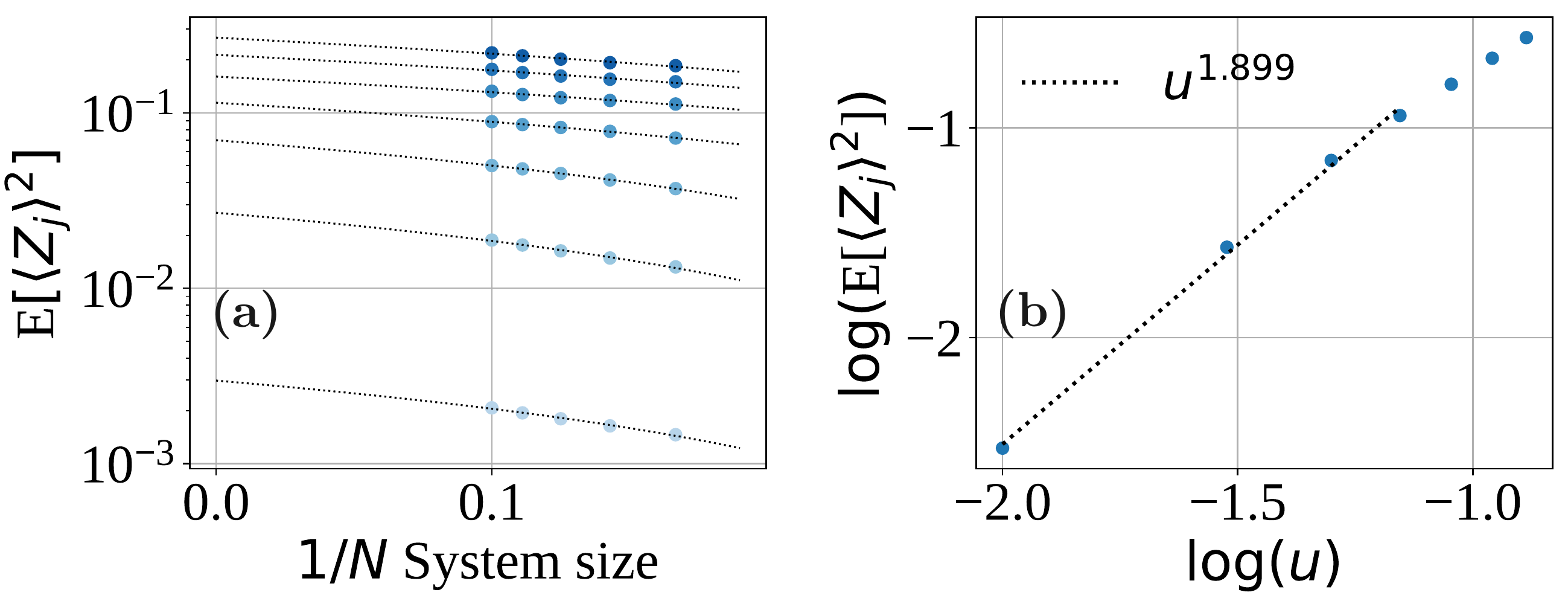}
\caption{(Color online). \textbf{Average of $\langle Z_i \rangle^2_{\tilde s}$ over measurement outcomes in case IV of Table~\ref{tab.unitaries}.} All parameters are the same as in Fig.~\ref{fig_av_Z2}, except here we consider only paramagnetic ancilla.  The scaling of $\textrm{E}(\avg{Z_j}_{\tilde{s}}^2)$ with $u$ in panel (b) is much steeper $(\sim u^2)$ compared to the scaling found in Fig.~\ref{fig_av_Z2}.  }
 \label{fig_av_Z2_xbasis}
\end{figure}

Let us first take $C \neq 0$.  Following the logic used for case III above, the second line is always subleading compared to the first, independent of whether the ancilla are paramagnetic or critical.   (Technically, however, $m(x)$ diverges logarithmically with system size when the ancilla are critical, so extra care is warranted when applying our perturbative formalism in this scenario.) 
For the uniform or nearly uniform measurement outcomes that we can treat here, the strongly relevant $m(x) \sigma$ perturbation leads once again to a non-zero one-point function $\langle \sigma \rangle_{\tilde s} \neq 0$ that scales as $u^{7/8}$, as reproduced in DMRG simulations [Figure~\ref{fig_fit_dmrg2}(b)]. Similar to our previous discussion in case II, we numerically find that $\langle Z_j\rangle_{\tilde s}^2$ averaged over all measurement outcomes also appears to yield a non-zero value at large $N$ (at least for paramagnetic ancilla), even though our perturbative formulation now only applies to a restricted set of measurement outcomes.  See Fig.~\ref{fig_av_Z2_xbasis} and notice the rather different scaling with $u$ compared to Fig.~\ref{fig_av_Z2}.  The results for critical ancilla, however, did not show an obvious trend and so we do not report them.  

When $C = 0$, the approximation invoked above no longer applies, and the defect-line action instead becomes
\begin{align}
    \mathcal{S}_m &= u^2 \int_{x,y}V(x,y) \sigma(x,\tau = 0)\sigma(y,\tau = 0),~~~~(C = 0).
    \label{SmIVC0}
\end{align}
Here the nature of the initial ancilla state becomes pivotal.  For gapped ancilla, exponential decay in $V(x,y)$ enables fusing the $\sigma$ fields into a single $\varepsilon$ field.  One then obtains the form in Eq.~\eqref{SmIIC0} that, for uniform or nearly uniform measurement outcomes, modifies the power-law correlations in $\langle \sigma(x)\sigma(x')\rangle_{\tilde s}$ as described previously.  For critical ancilla this prescription breaks down since $V(x,y)$ scales like $1/|x-y|$.  The resulting power-law-decaying interaction between $\sigma$'s in Eq.~\eqref{SmIVC0} is strongly relevant by power counting; the system's fate then depends on whether the power-law interaction is ferromagnetic or antiferromagnetic.  On one hand, ferromagnetic $\sigma(x,\tau = 0) \sigma(y,\tau= 0)$ interaction would promote order-parameter correlations at $\tau=0$---possibly replacing power-law decay in the spin-spin correlation function with true long-range order, i.e.,  turning the critical chain into a cat state~\footnote{Notice that for $C=0$, $\ket{\psi_{\tilde{s}}}$ is an eigenstate of $G$, even though $U$ breaks this symmetry explicitly.}.  On the other, antiferromagnetic interaction would produce frustration, leading to a subtle interplay with ferromagnetic order-parameter correlations built into the pre-measurement critical theory.  

\begin{figure}[t!]
\centering
\includegraphics[width =1.0 \columnwidth]{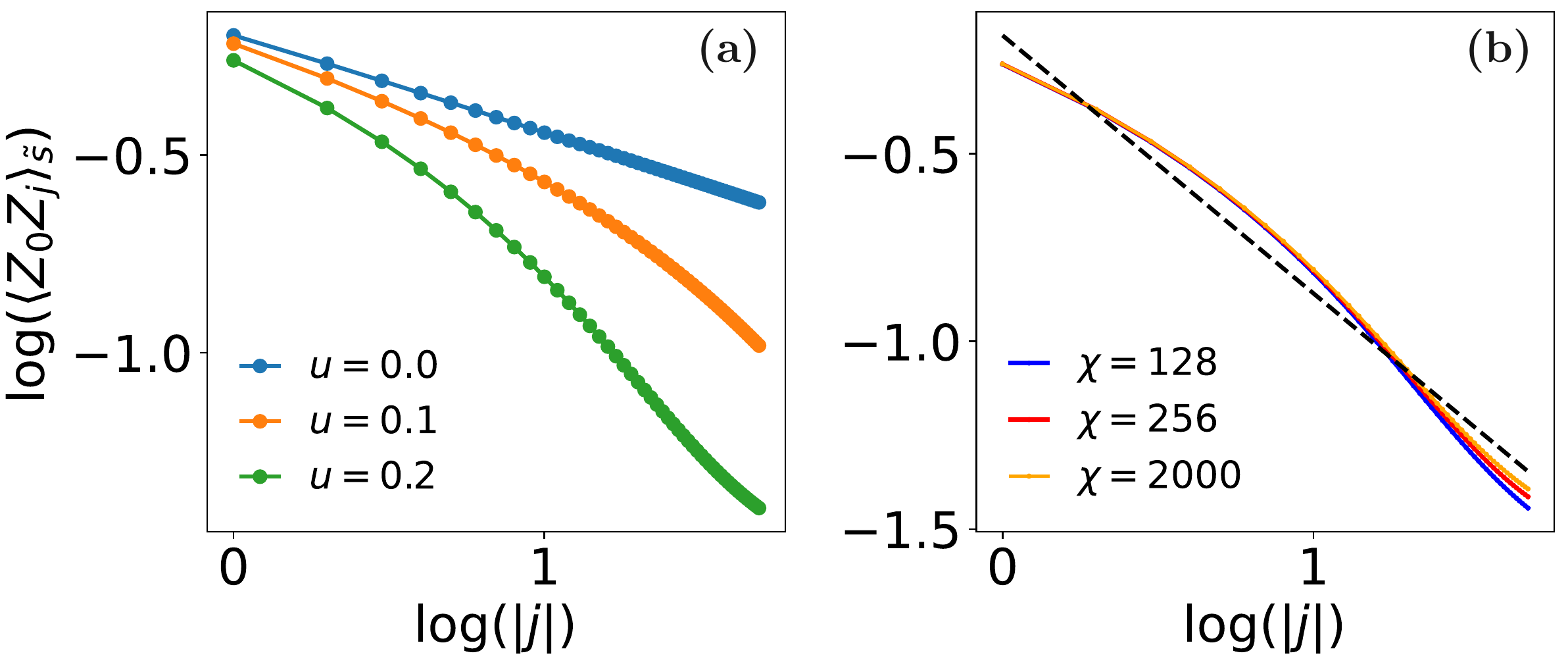}
\caption{(Color online).  \textbf{Correlation function $\langle Z_0Z_j\rangle_{\tilde{s}}$ for case IV in Table~\ref{tab.unitaries} with $C=0$ and critical ancilla.} (a) At $u >0$ the correlator appears to decay faster than a power-law for $|j|$ less than $O(10)$ in response to the measurement-induced long-range interaction in Eq.~\eqref{SmIVC0}. Data were obtained using iDMRG with bond dimension $2000$. (b) Two-point correlator at $u = 0.2$ for a range of bond dimensions $\chi$.  Over the separations shown in (a), the data are well-converged.  For larger separations, however, the data continue to evolve with bond dimension.  
}
 \label{fig_rZZC0}
\end{figure}

Since $V(x,y)$ is always non-negative in case IV, $\mathcal{S}_m$ in Eq.~\eqref{SmIVC0} realizes the antiferromagnetic scenario.  Figure~\ref{fig_rZZC0} presents infinite DMRG simulations of $\langle Z_0 Z_j \rangle_{\tilde s}$ for this case~\footnote{We show connected correlations, as for finite bond dimension $\langle Z\rangle_{\tilde{s}}$ gives non-zero values.}.  
For separations $|j|$ smaller than $O(10)$ we find signatures of faster-than-power-law decay induced by measurements.  For larger separations, however, we find a possible revival of correlations (though in this regime the DMRG data continue to evolve over the bond dimensions simulated).  
 While correlations might exhibit exponential decay at short distances, we expect algebraic decay to take over at long distances---which may be related to physics of antiferromagnetic 
 long-range Ising chains analyzed in Refs.~\onlinecite{Tagliacozzo,Vodola_2016}.  This expectation is consistent with the fact that, as we will show in an upcoming work~\cite{future_work}, when a short-range-correlated system entangles with critical ancilla, measuring the latter imprints long-range correlations into the former. Hence, it is natural to anticipate that tuning the short-range-correlated system to criticality only further enhances its long-range correlations.

With critical ancilla, the full two-chain system prior to measurement corresponds to a free-fermion problem with total central charge $c = 1/2 + 1/2$.  Thus here the setup resembles a single-channel Luttinger-liquid in the special case with Luttinger parameter $K = 1$.  For the Luttinger-liquid measurement protocol considered in Ref.~\onlinecite{AltmanMeasurementLL}, uniform measurement outcomes were shown to produce a marginal defect-line action.  Our protocol, by contrast, yields a \emph{relevant} defect-line action both for $C = 0$ and $C \neq 0$ in case IV---thereby qualitatively modifying correlations as discussed above.  Interestingly, it follows that the total central charge alone does not dictate the impact of measurements on long-distance correlations.  Additional factors including the allowed physical operators and details of the measurement protocol also play a role.  For example, the protocol from Ref.~\onlinecite{AltmanMeasurementLL} used an uncorrelated ancilla chain to mediate measurements on the Luttinger liquid, whereas in our effective $c = 1/2 + 1/2$ setup measurements are enacted `internally' without invoking an additional auxiliary chain.

\section{Exact averaging over even/odd strings in $\tilde X$-measurement protocol}
\label{sec:evenodd}

\subsection{Symmetry-resolved averages}\label{sec:evenoddA}

With $\tilde X$-basis measurements, outcomes $\tilde s$ can be divided into sectors according to whether $\tilde G \ket{\tilde s} = +\ket{\tilde s}$ or $-\ket{\tilde s}$.  
Here we exploit this neat even/odd-string dichomotomy to obtain illuminating, exact expressions for the average of critical-chain observables $A$ over measurement outcomes confined to a particular $\epsilon = \pm 1$ parity sector.  Using Eqs.~\eqref{psi_m_def} and \eqref{p_def}, the average in sector $\epsilon$ reads
\begin{align}
    \langle A\rangle_\epsilon &= \sum_{\tilde s \in \epsilon} p_{\tilde s}\bra{\psi_{\tilde s}}A\ket{\psi_{\tilde s}}
    \nonumber \\
    &= \sum_{\tilde s \in \epsilon}\bra{\psi_a}\bra{\psi_c} U^\dagger \left(\prod_j \ket{\tilde s_j}\bra{\tilde s_j} \right)U A_U\ket{\psi_c}\ket{\psi_a}
    \nonumber \\
    &= \bra{\psi_a}\bra{\psi_c} U^\dagger \tilde P_\epsilon U A_U\ket{\psi_c}\ket{\psi_a},
    \label{Aepsilon}
\end{align}
where $U = \prod_j U_j$ represents the unitary applied prior to measurement and $A_U = U^\dagger A U$.  In the last line 
\begin{equation}
    \tilde P_\epsilon = \sum_{\tilde{s}\in \epsilon}\ket{\tilde{s}}\bra{\tilde{s}} = \frac{1}{2}\left(1+\epsilon \tilde G\right)
    \label{Ptilde}
\end{equation}
projects onto the measurement-outcome sector with parity $\epsilon$.  For the unitaries in either case III or IV from Table~\ref{tab.unitaries}, the anticommutation relation $\{\tilde Z_j,\tilde G\} = 0$ implies that $U^\dagger \tilde G = \tilde G U$.  We can therefore express Eq.~\eqref{Aepsilon}, after also using $\tilde G \ket{\psi_a} = \ket{\psi_a}$, as
\begin{multline}
    \langle A\rangle_\epsilon = \frac{1}{2} \bra{\psi_a} \bra{\psi_c} A_U\ket{\psi_c} \ket{\psi_a} \\+ \frac{\epsilon}{2}\bra{\psi_a}\bra{\psi_c} U^2 A_U\ket{\psi_c}\ket{\psi_a}.
    \label{Aepsilon2}
\end{multline}
Notice that the second term is real for any Hermitian operator $A$, since any imaginary parts vanish by parity constraints.
Summing over the even and odd sectors yields $\langle A\rangle_+ + \langle A\rangle_- = \bra{\psi_a} \bra{\psi_c} A_U \ket{\psi_c}\ket{\psi_a}$---which, in agreement with Eq.~\eqref{trivial}, is simply the result one would obtain without performing any measurements.  The difference between the even- and odd-sector correlators, by contrast, isolates the second term in Eq.~\eqref{Aepsilon2},
\begin{align}
    \langle A\rangle_+ - \langle A\rangle_- = \bra{\psi_a}\bra{\psi_c} U^2 A_U\ket{\psi_c}\ket{\psi_a},
    \label{Adiff}
\end{align}
and \emph{does} retain nontrivial imprints of the measurements enacted in our protocol.  Equation~\eqref{Adiff} is equivalent to the expectation value of the \emph{non-local} operator $A \tilde G$ taken in the pre-measurement state $U \ket{\psi_c}\ket{\psi_a}$; crucially, measuring the ancilla in the $\tilde X$ basis provides access to such non-local information.  

There is, however, no free lunch here: On general grounds the right side of Eq.~\eqref{Adiff} should decay to zero with system size $N$ for any fixed $u \neq 0$.  To see why, let $p_\epsilon = \sum_{{\tilde s} \in \epsilon} p_{\tilde s}$ denote the probability for obtaining parity sector $\epsilon$ after a measurement, and consider the difference
\begin{equation}
    \Delta p\equiv p_+ - p_- = \bra{\psi_a}\bra{\psi_c} U^2 \ket{\psi_c}\ket{\psi_a}.
    \label{pdiff}
\end{equation}
Equation~\eqref{pdiff} simply corresponds to Eq.~\eqref{Adiff} with $A$ being the identity.  At $u = 0$, where $U$ also reduces to the identity, we obtain $\Delta p = 1$---reflecting the fact that the initial ancilla state $\ket{\psi_a}$ resides in the even-parity sector by construction.  Turning on $u \ll 1$, the state $U^2 \ket{\psi_c}\ket{\psi_a} = \prod_j U_j^2\ket{\psi_c}\ket{\psi_a}$ exhibits a small $O(u^2)$ probability for flipping a particular $\tilde X$-basis ancilla spin.  Yet the net effect over a macroscopic number of sites $N$ inevitably translates into a `large' change in the probability for remaining in the even-parity sector.  In terms of Eq.~\eqref{pdiff}, this logic implies that $U^2 \ket{\psi_c}\ket{\psi_a}$ becomes orthogonal to $\ket{\psi_c}\ket{\psi_a}$ at fixed $u\neq 0$ with $N \rightarrow \infty$, leading to $\Delta p \approx 0$.  The insertion of $A_U$ in Eq.~\eqref{Adiff}, assuming it represents physically relevant combinations of local operators, can not change this conclusion, implying that $\langle A\rangle_+ - \langle A\rangle_-$ vanishes with $N$ as well. In Appendix~\ref{app:exp_decay} we numerically show that, with paramagnetic ancilla, these quantities decay exponentially with system size.

We propose the ratio
\begin{align}
    r(A) \equiv \frac{\langle A\rangle_+ - \langle A\rangle_-}{\Delta p} = \frac{\bra{\psi_a}\bra{\psi_c} U^2 A_U\ket{\psi_c}\ket{\psi_a}}{\bra{\psi_a}\bra{\psi_c} U^2 \ket{\psi_c}\ket{\psi_a}}
    \label{ratio}
\end{align}
as an appealing diagnostic of $\tilde X$-basis measurement effects on Ising criticality.  Equation~\eqref{ratio} need not vanish in the thermodynamic limit.  Moreover, both the numerator and denominator comprise linear averages over experimentally accessible quantities, circumventing the need for post-selection.  (But again there is no free lunch---the individually small numerator and denominator would need to be obtained with sufficient accuracy to yield a meaningful ratio as quantified further below.)  

The formalism developed in Sec.~\ref{perturbative} and Appendix~\ref{app:post} allows us to rewrite $r(A)$ in a more illuminating form that directly connects with the results from Sec.~\ref{Xtilde_basis} \footnote{We are very grateful to Zack Weinstein for suggesting this approach.}.  For simplicity we focus for now on observables $A$ that commute with $U$ so that  $U^2 A_U=A U^2$ (see below for a comment on the generic case).
As detailed in Appendix~\ref{AppB}, we can express the ratio in Eq.~\eqref{ratio} as \begin{align}
 r(A)=\frac{\bra{\psi_c}A e^{-H^r_m}\ket{\psi_c}}{\bra{\psi_c} e^{-H^r_m}\ket{\psi_c}},
 \label{rA_nice}
\end{align}
where through our perturbative formalism we obtain at $O(u^2)$ 
\begin{align}
    H^r_m =& u^2\sum_j m^r(O_j-\langle O\rangle)\nonumber \\&+ u^2\sum_{j \neq k}V^r_{jk}(O_j-\langle O\rangle)(O_k-\langle O\rangle).
    \label{eq:H_m2}
\end{align}
Equation~\eqref{eq:H_m2} is analogous to Eq.~\eqref{H_m} but involves distinct couplings
\begin{align}\label{eq:Vjkr}
    V^r_{jk} &= 2\langle \psi_a|\tilde Z_{j} \tilde Z_{k}|\psi_a\rangle
    \\
    m^r &= -4\theta + 2(\langle O\rangle -\theta) \sum_{k \neq 0} V^r_{0k}.
    \label{mj2}
\end{align}
Most notably, compared to the $V_{jk}$ and $m_j$ couplings from cases III and IV of Table~\ref{tab.unitaries}, $V^r_{jk}$ here depends only on $|j-k|$ and follows from the expectation value of $\tilde Z_j\tilde Z_k$ in the initial ancilla ground state (rather than depending on some particular measurement outcome).  For similar reasons $m^r$ does not depend on position.  If $[A,U] \neq 0$, then Eq.~\eqref{rA_nice} holds together with an additional subleading term resulting from the commutator. For example, if we are interested in $A=Z_jZ_{j'}$ in case III from Table \ref{tab.unitaries}, then $[Z_j,U]=-2\,i \,Y_j \sin (u)[i\cos(u\,\theta)\tilde Z_{j}+\sin(u\,\theta)]\prod_{k\neq j}e^{iu(O_k-\theta)\tilde Z_k}$. Given that $Y_j$ maps to a CFT operator with larger scaling dimension than
that for $Z_j$, we can already deduce that the additional term coming from the commutator involves subleading contributions that we can safely neglect.  For further analysis see Appendix~\ref{AppB}.

Taking the continuum limit, the ratio in Eq.~\eqref{rA_nice} can be recast in terms of a path integral perturbed by a defect-line action akin to Eq.~\eqref{eq:action_m}; recall the steps below Eq.~\eqref{A1m}.  Let us now specialize to paramagnetic ancilla, where $V_{jk}^r$ decays exponentially leading to a purely local action and finite $m^r$.  We can then immediately import results from Sec.~\ref{Xtilde_basis} to obtain $r(A)$ for spin correlators of interest.  For case III we find
\begin{equation}
    r(Z_j Z_k) \sim |j-k|^{-2\Delta^r_\sigma(u)} ~~~{\rm (case~III)}
    \label{rIII}
\end{equation}
with nontrivially modified scaling dimension 
\begin{equation}
    \Delta^r_{\sigma}(u) = \frac{1}{8}(1+2 \kappa u^2 m^r),
\end{equation}
where $m^r=-4\langle X\rangle$.  This result is, remarkably, nearly identical to the prediction for post-selected uniform measurement outcomes in case III; comparing with Eq.~\eqref{correlator_caseIII}, the sole difference is that $m^r$ is twice as larger as $m$, leading to a more pronounced upward shift in scaling dimension.   
For case IV with $C \neq 0$ we similarly find that
\begin{equation}
    r(Z_j) \sim u^{2/7} ~~~~~~{\rm (case~IV,~C\neq 0)},
    \label{rIV}
\end{equation}
which also emulates predictions for the corresponding post-selected uniform measurement outcome.  

\begin{figure}[t!]
\centering
\includegraphics[width = \columnwidth]{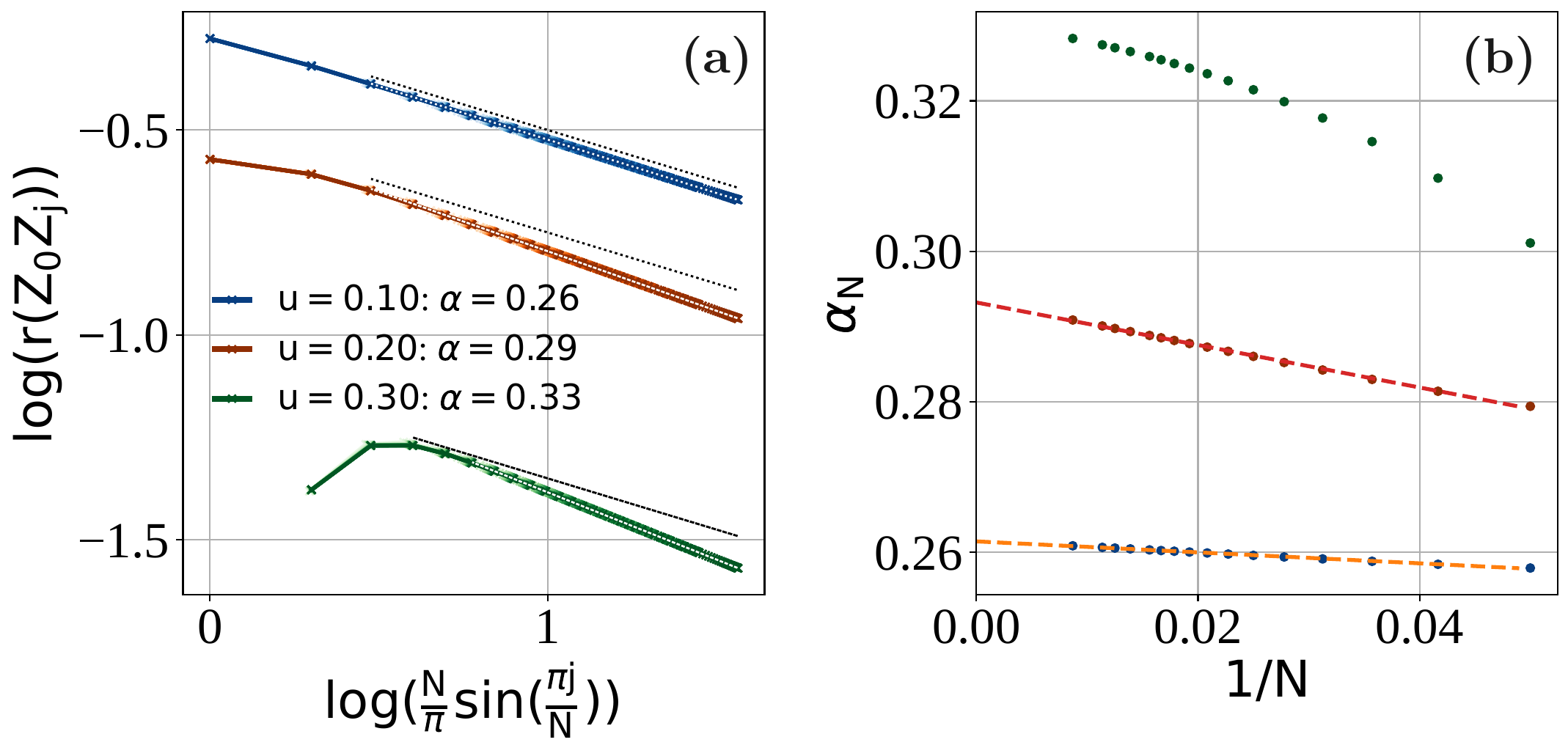}
\caption{(Color online).  \textbf{Ratio $r(Z_0Z_j)$ in Eq.~\eqref{ratio} involving symmetry-resolved measurement averages.} The data correspond to case III from Table~\ref{tab.unitaries} with paramagnetic ancilla, and different system sizes between $N=20$ (light colors in (a)) and $N=115$ (dark colors in (a)). At $u > 0$, the curves exhibit power-law decay with exponent exceeding $1/4$ in agreement with Eq.~\eqref{rIII}. The shift in scaling dimension becomes particularly clear for larger values of $u$ when compared with the black dotted lines, corresponding to a power-law decay exponent $0.25$. 
The tendency continues upon extrapolating to the thermodynamic limit, as shown in (b). The power-law exponents $\alpha$ displayed in panel (a) were obtained from fitting the data with the largest system size available.  Data were obtained using finite DMRG with periodic boundary conditions and bond dimension $1000$. }
 \label{fig_rZZj}
\end{figure}

\begin{figure}[t!]
\centering
\includegraphics[width=0.8\linewidth]{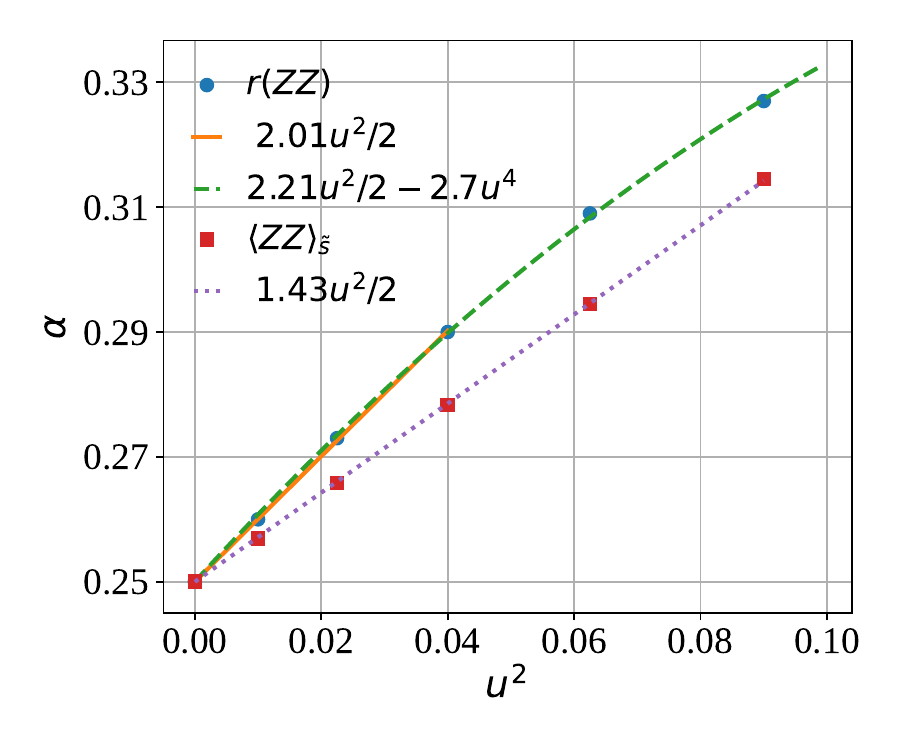}
\caption{(Color online). \textbf{Comparison between power-laws for the $r(Z_0Z_j)$ ratio and $\langle Z_0 Z_j\rangle_{\tilde s}$ correlator with a uniform measurement outcome in case III.}  The measurement-induced shift in power-law exponent for $r(Z_0Z_j)$ exceeds that of $\langle Z_0 Z_j\rangle_{\tilde s}$, in qualitative agreement with perturbative analytical predictions (though the enhancement is smaller than the predicted factor of 2).  The green dashed line represents a quartic fit ($1/4+\kappa m_{\textrm{const}}u^2/2 + bu^4$) in $u$ of the exponent for the $r$ ratio, while the orange and purple lines are the result of a quadratic fit ($1/4+\kappa m_{\textrm{const}}u^2/2$).
Data correspond to paramagnetic ancilla and were obtained with finite DMRG for a system of size $N=88$ with periodic boundary conditions and bond dimensions $800$ and $1000$.  } 
 \label{fig_fit_comparison}
\end{figure}

\begin{figure}[t!]
\centering
\includegraphics[width = \columnwidth]{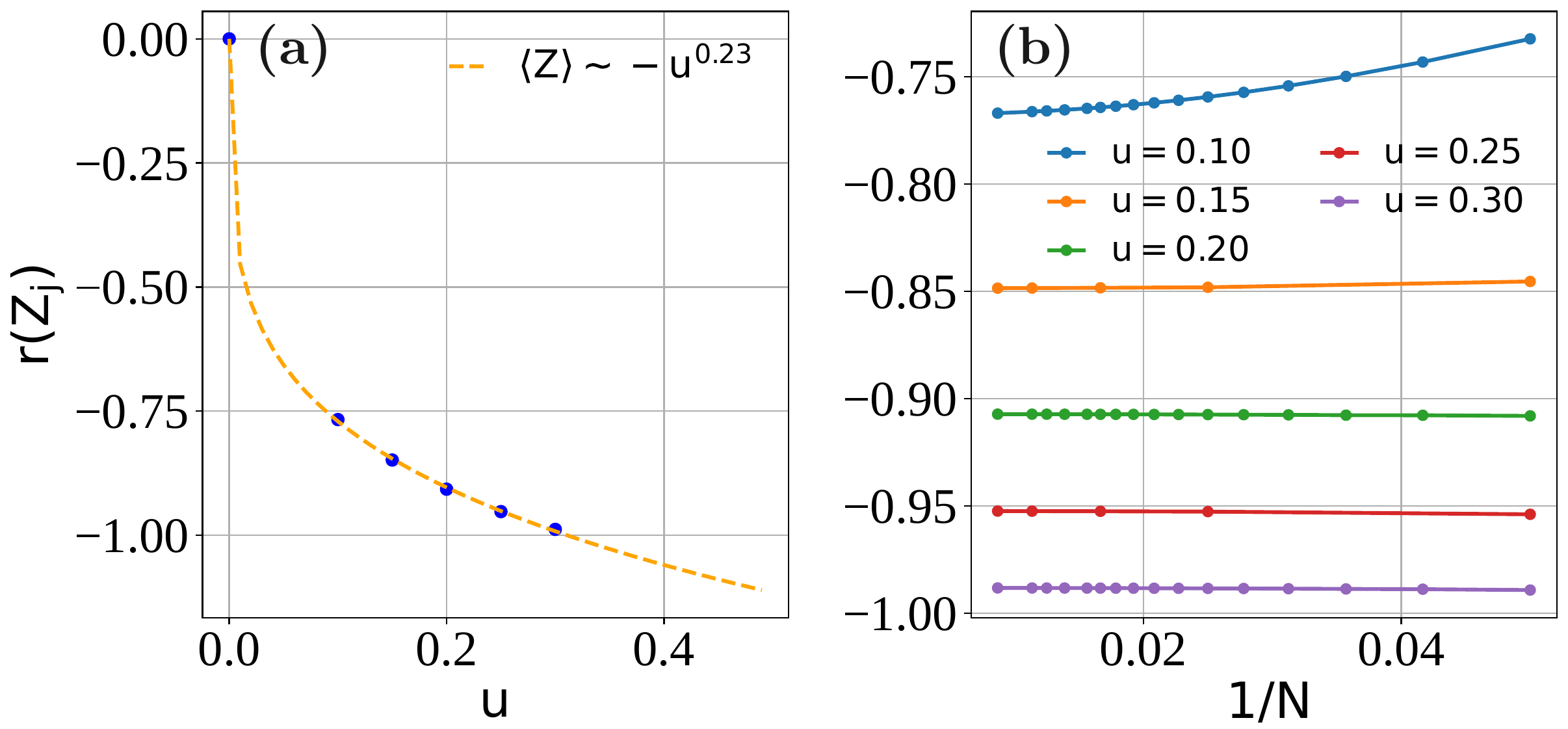}
\caption{(Color online). \textbf{Ratio $r(Z_j)$ in Eq.~\eqref{ratio} involving symmetry-resolved measurement averages.} The data correspond to case IV from Table~\ref{tab.unitaries} with paramagnetic ancilla, $C=-1$, and system sizes between $N=20$ (light colors in(a)) and $N=115$ (dark colors in (a)). Panel (a) shows that the small-$u$ data are well-fit by a $\sim u^{0.23}$ scaling form---consistent with the prediction from Eq.~\eqref{rIV}. As shown in (b), increasing $u$ suppresses the dependence on system size such that $r(Z_j)$ quickly saturates to a finite value. Data were obtained using finite DMRG with periodic boundary conditions and bond dimension $1000$.}
 \label{fig_rZj}
\end{figure}

We performed a numerical experiment of Eqs.~\eqref{rIII} and \eqref{rIV} using DMRG, focusing again on paramagnetic ancilla. Figure~\ref{fig_rZZj}
presents the results for $r(Z_0Z_j)$ in case III, which indeed reveals power-law decay with scaling dimension exceeding 1/8 at $u>0$. In Fig.~\ref{fig_fit_comparison}, we further contrast the data with the power-law exponents obtained for $\langle Z_0 Z_j \rangle_{\tilde s}$ with a uniform measurement outcome in case III. There we use finite DMRG for a system size of $N=88$ to treat both quantities with a common numerical method. Our perturbative formalism predicts that, at small $u$, the measurement-induced change in the power-law for $r(Z_0Z_j)$ should exceed that for $\langle Z_0 Z_j \rangle_{\tilde s}$ by a factor of 2. We indeed recover a more pronounced enhancement for the former---albeit by a factor smaller than 2. Note also that the $r(Z_0Z_j)$ power-law exponent clearly exhibits more dramatic higher-order-in-$u$ corrections that are beyond our leading perturbative treatment. Figure~\ref{fig_rZj}
reports the results for $r(Z_j)$ in case IV with non-zero $C$. Just as in Fig.~\ref{fig_fit_dmrg2} taken for uniform measurement outcomes in cases II and IV, we obtain good quantitative agreement with the prediction in Eq.~\eqref{rIV}. Moreover, panel (b) shows that the $r(Z_j)$ values quickly saturate as $N$ increases, at least for $u \geq 0.2$.

Despite the striking resemblance discussed above between $r(A)$ and correlators in post-selected uniform measurement outcomes, we stress that these quantities are not quite identical.  The distinction becomes particularly apparent with critical ancilla---for which $V_{jk}^r$ encodes a power-law interaction in the asssociated defect-line action with much slower decay (exponent $1/4$) compared to the decay found in cases III and IV (exponent $1$).  In case IV with $C = 0$, the inherently long-range $\sigma \sigma$ interaction mediated by $V_{jk}^r$ is much more strongly relevant compared to the (also strongly relevant) interaction encountered in Eq.~\eqref{SmIVC0}.  Moreover, in case IV with $C \neq 0$, $m^r$ correspondingly diverges rapidly with system size, signalling a clear breakdown of the perturbative expansion used above.   By contrast, in Sec.~\ref{sec:caseIV} we saw that critical ancilla yield only a mild logarithmic divergence in $m_j$.  We leave a detailed investigation of the properties of $r(A)$ with critical ancilla for future work.

\subsection{Comparison with post-selection}\label{sec:evenoddB}

We now critically assess the experimental feasibility of probing measurement-altered criticality via symmetry-resolved averages and contrast with the alternative strategy of post-selection.  For the latter, we focus in particular on post-selecting the uniform ancilla measurement string $\tilde{s}_{\text{uni}}$---which as we saw previously is the most likely measurement outcome and leads to clear measurement-induced changes of correlators that closely resemble symmetry-resolved averages.  Quite different challenges accompany these two approaches.  For the experimental extraction of symmetry-resolved averages, every protocol iteration---regardless of the specific ancilla measurement outcome---can in principle nontrivially inform evaluation of the ratio $r(A)$ in Eq.~\eqref{ratio}.  As stressed above, however,  the numerator and denominator both decay exponentially with system size, suggesting that obtaining sufficient statistics to reliably measure $r(A)$ requires a correspondingly large number of experimental trials.  With post-selection, nearly all protocol iterations reveal no information about the observable $A$ in the target ancilla measurement outcome $\tilde s_{\rm uni}$.  But within the rare instances in which the target outcome emerges, evaluating $\bra{\psi_{\tilde s_{\rm uni}}} A\ket{\psi_{\tilde s_{\rm uni}}}$ becomes relatively straightforward for two reasons.  First, this expectation value generally does not decay exponentially with system size [contrary to Eq.~\eqref{Adiff}].  Second, due to translation invariance of $\ket{\psi_{\tilde s_{\rm uni}}}$, one can interrogate all system spins in the post-measurement state to reduce the number of recurrences of $\tilde s_{\rm uni}$ needed to resolve correlations to a desired accuracy; even a single successful trial suffices to approximate the expectation value of both one- and two-point correlations with an error scaling as $1/\sqrt{N}$, with $N$ the system size.  In what follows we quantify the number of trials required for both approaches.

Let us first assess symmetry-resolved averages and respectively write the numerator and denominator of $r(A)$ as
\begin{align}
  \mathcal{N} = \sum_{\tilde s}p_{\tilde s} \epsilon_{\tilde s} A_{\tilde s},~~~~~
  \Delta p = \sum_{\tilde s}p_{\tilde s} \epsilon_{\tilde s},
  \label{ND}
\end{align}
where $A_{\tilde s} = \bra{\psi_{\tilde s}} A\ket{\psi_{\tilde s}}$ and $\epsilon_{\tilde s}$ denotes the parity for measurement outcome $\tilde s$. Both quantities decay exponentially with system size as 
\begin{equation}
    \mathcal{N},\Delta p\sim e^{-\eta_{\text{sra}}(u)N}.
    \label{scaling}
\end{equation}
The function $\eta_{\text{sra}}(u)$ vanishes as $u\rightarrow 0$, reflecting the fact that, in the $u = 0$ limit, we obtain $\Delta p = 1$ exactly while $\mathcal{N}$ reduces to the critical correlator $\bra{\psi_c} A \ket{\psi_c}$ that (at least for the few-body operators $A$ of interest) does not decay exponentially with system size. 
For simplicity we will assume that in a given protocol iteration yielding a  particular $\tilde s$, one can determine both $\epsilon_{\tilde s}$ and $A_{\tilde s}$ in a single shot.  (In practice, each iteration would yield an eigenvalue of $A$, and determining $A_{\tilde s}$ would require multiple iterations yielding the same outcome $\tilde s$. Our assumption mods out these standard repetitions;  
moreover, since averaging over measurement outcomes restores translation invariance, here too one can probe all system spins to reduce the required number of repetitions, similar to the situation noted above for post-selection.)  After $M$ experimental protocol iterations yielding a set of outcomes $\tilde s_{i = 1,\ldots,M}$ and associated parities $\epsilon_{\tilde s_{i}}$ and observables $A_{\tilde s_{i}}$, the quantities in Eq.~\eqref{ND} can be estimated by
\begin{align}
    \mathcal{N}_M = \frac{1}{M}\sum_{i = 1}^M \epsilon_{\tilde s_i} A_{\tilde s_i}, ~~~~~
    \Delta p_M= \frac{1}{M}\sum_{i = 1}^M \epsilon_{\tilde s_i}.
\end{align}
In the limit $M \rightarrow \infty$ one obtains the exact results $\mathcal{N}_M \rightarrow \mathcal{N}$ and $\Delta p_M \rightarrow \Delta p$.  

It is crucial to now understand the variance of the sampling distribution that quantifies the quality of these estimations at finite $M$.
Given an estimator $\theta_M$ the sample variance is
\begin{equation}
    \text{Var}_M\,({\theta})=\frac{\text{Var}(\theta)}{M},
\end{equation}
i.e., the population variance $\text{Var}(\theta)=\sum_{\tilde s} p_{\tilde s} \theta^2_{\tilde s}-\langle \theta\rangle^2$ divided by the sample size $M$.  
Intuitively, this quantity implies that the larger the sample size $M$, the smaller the variance of the sampling distribution of $\theta_M$. 
For the $\mathcal{N}$ and $\Delta p$ estimators we have
\begin{align}
    \text{Var}_M\,(\mathcal{N})&=\frac{1}{M}\left(\sum_{\tilde s}p_{\tilde s} A^2_{\tilde s} -\mathcal{N}^2 \right) \approx \frac{1}{M}\sum_{\tilde s}p_{\tilde s} A^2_{\tilde s}
    \\
    \text{Var}_M\,(\Delta p)&=\frac{1}{M}(1-\Delta p^2) \approx \frac{1}{M},
\end{align}
where on the rightmost sides we used the fact that both $\mathcal{N}$ and $\Delta p$ decay exponentially with system size.  Comparing to Eq.~\eqref{scaling}, we see here that accurately determining both the numerator and denominator of $r(A)$ requires a number of trials $M$ that grows exponentially with $N$.  The relative error $\sqrt{\text{Var}_M\,(\Delta p)}/\Delta p$ in determining $\Delta p$, for instance, becomes smaller than one for $M \gtrsim e^{2\eta_{\text{sra}}(u)N}$; similar reasoning applies to $\mathcal{N}$.  

We are primarily interested in the number of trials required to reliably estimate $r(A)$ itself. The corresponding ratio estimator reads
\begin{equation}  
 r_M(A)=\frac{\mathcal{N}_M}{\Delta p_M},
\end{equation}
while to O$(M^{-1})$ its variance is \cite{RatioEstimator} 
\begin{multline}
    \text{Var}_M[r(A)]\approx \frac{1}{\Delta p^2}\left[ \text{Var}_M(\mathcal{N})+ r(A)^2 \text{Var}_M(\Delta p )\right.\\
    \left.- 2 r(A)\text{Covar}_M(\mathcal{N},\Delta p)\right].
\end{multline}
Evaluating the terms in brackets and neglecting contributions that are exponentially small in system size yields
\begin{equation}
    \text{Var}_M[r(A)]\approx \frac{1}{M\Delta p^2}\{\text{Var}(A) + [r(A)-\langle A \rangle]^2\}. 
    \label{VarM2}
\end{equation}
The dominant remaining system-size dependence appears through $\Delta p$ in the denominator.  Consequently, accurate extraction of the symmetry-resolved average ratio $r(A)$ requires a number of trials satisfying \footnote{Equation~\eqref{eq:Msra} provides the leading $N$ dependence needed to ensure that $\sqrt{{\rm Var}_M[r(A)]}/r(A)$ becomes smaller than one for the cases of interest.  This conclusion follows from the fact that $r(A)$ decays at most as a power-law in system size, whereas the factor in braces in Eq.~\eqref{VarM2} is at most O(1).} 
\begin{equation}\label{eq:Msra}
    M_{\rm sra} \gtrsim \frac{1}{\Delta p^2} \sim e^{2 \eta_{\text{sra}}(u)N}
\end{equation}
(which is the same criterion for separately determining the numerator and denominator).  

To diagnose a potential advantage of this approach with respect to post-selection, we next estimate the minimum sample size required 
to obtain the target measurement outcome $\tilde s_{\rm uni}$ with high likelihood.  
The probability to measure this string, 
\begin{equation}
  p_{\textrm{uni}}= \langle \psi_U | \tilde{s}_{\text{uni}}\rangle\langle \tilde{s}_{\text{uni}}|\psi_U\rangle \sim e^{-\eta_{\text{uni}}(u)N},
\end{equation}
also decreases exponentially with system size, as expected from the fact that it arises from the overlap of two very different many-body wave functions. Importantly, the function $\eta_{\rm uni}(u)$, unlike $\eta_{\rm sra}(u)$, generically does \emph{not} vanish as $u \rightarrow 0$: At $u = 0$ exponential decay with $N$ persists due to nontrivial overlap between $\ket{\tilde s_{\rm uni}}$ and the initial ancilla wavefunction, except in the extreme limit $h_{\rm anc}/J_{\rm anc} \rightarrow \infty$.   
Since the probability of not measuring $\tilde s_{\rm uni}$ after $M$ trials is $(1–p_{\text{uni}})^M$, the probability of finding this measurement outcome at least once is $p_{\text{success}}=1–(1–p_{\text{uni}})^M$. The number of trials required for post-selecting the uniform measurement outcome with high success probability $p_{\text{success}} = 1-\varepsilon$ (ideally $1$) accordingly satisfies 
\begin{equation} \label{eq:M_uni}
M_{\text{uni}}=\frac{\log(\varepsilon)}{\log(1-p_{\text{uni}})}\sim \frac{1}{p_{\text{uni}}}\sim e^{\eta_{\text{uni}}(u)N}.   
\end{equation}

Both the symmetry-resolved average and brute-force post-selection approaches thus require an exponentially large (in system size) number of measurements specified by Eqs.~\eqref{eq:Msra} and \eqref{eq:M_uni}, respectively.  It is crucial to observe, however, that the scaling with $N$ is tunable via the choice of entangling gate and ancilla initialization in a manner that differs for the two methods. 
On very general grounds, since $\eta_{\rm sra}(u = 0)$ vanishes whereas $\eta_{\rm uni}(u = 0)$ is positive, there always exists a window of sufficiently small $u$ for which symmetry-resolved averages can be probed more efficiently compared to post-selection.  To be more quantitative, Appendix~\ref{app:exp_decay} provides numerical evidence that for small $u$ these functions typically behave as
\begin{equation}
     \eta_{\text{sra}}(u)\approx c_{\rm sra} u^{\zeta},~~~\eta_{\text{uni}}(u)\approx \Upsilon + c_{\rm uni}u^{\zeta}.
     \label{eta_small_u}
\end{equation}
Here $c_{\rm sra}, c_{\rm uni}$ are positive constants, $\zeta$ is a case-dependent exponent that we extract (see Fig.~\ref{fig:scaling}), and $\Upsilon$ determines the probability of finding the uniform measurement outcome at $u = 0$ (see Appendix~\ref{app:Gaussian} and Fig.~\ref{fig:Upsilon}).  Equation~\eqref{eta_small_u} implies that when $2c_{\rm sra} > c_{\rm uni}$---which we find holds in practice---$M_{\rm sra}$ grows with system size exponentially more slowly compared to $M_{\rm uni}$ for $u$ between 0 and $u_* \equiv \left(\frac{\Upsilon}{2c_{\rm sra} - c_{\rm uni}}\right)^{1/\zeta}$.  When $u$ increases just beyond $u_*$, post-selection begins to become more efficient than symmetry-resolved averages.  Intuitively, as the ancilla correlation length increases, the probability for obtaining the uniform measurement outcome decreases, thereby enhancing $u_*$ and broadening the window in which symmetry-resolved averages are advantageous.  In case IV with critical ancilla, we find that $\Delta p$ does not decay monotonically to zero with $N$, but rather changes sign along the way.  Equation~\eqref{eta_small_u} does not capture such non-monotonic behavior; similar conclusions nevertheless hold also in that case as we will see.

\begin{figure}[t!]
\centering
\includegraphics[width = \columnwidth]{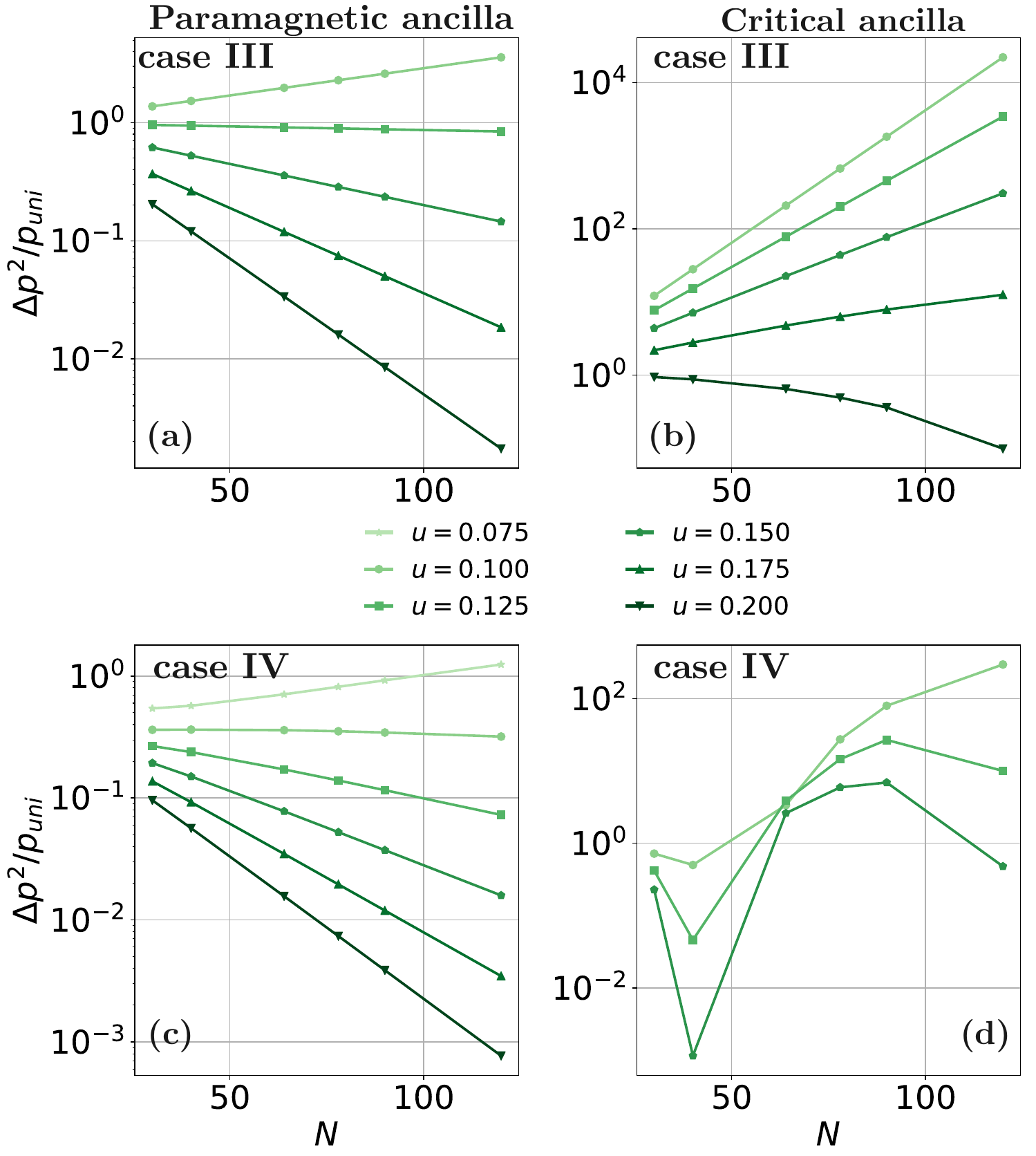}
\caption{(Color online).  \textbf{Comparison between symmetry-resolved averages and post-selection.} The vertical axis captures the ratio $M_{\rm uni}/M_{\rm sra} \sim \Delta p^2/p_{\rm uni}$, where $M_{\rm uni}$ characterizes the number of trials needed for post-selection that targets the uniform measurement outcome and $M_{\rm sra}$ characterizes the number of trials required for evaluation of symmetry-resolved averages with order one variance.   
 All panels are consistent with symmetry-resolved averages providing more favorable scaling with system size $N$ over a window of small $u$---as argued on general grounds in the main text.  Data were obtained using finite DMRG with bond dimension $\chi=800$ and periodic boundary conditions; case IV results used $C = -1$.}
 \label{fig_ratio1}
\end{figure}

We validate the preceding picture by numerically analyzing the ratio $M_{\rm uni}/M_{\rm sra} \sim \Delta p^2/p_{\rm uni}$, in particular by simulating the system-size dependence of $\Delta p^2/p_{\rm uni}$ for different $u$ values in cases III and IV.  (When $u N \ll 1$ the unitary entangling gates are sufficiently close to the identity that they do not induce appreciable decay of either $\Delta p$ or $p_{\rm uni}$; hence we restrict the range of $u$ such that the $N$ values accessible in our simulations include regimes with $uN \gtrsim 1$. With this constraint we also avoid possible artificial phenomena appearing as a result of scaling $u$ with system size.) Growth of $\Delta p^2/p_{\rm uni}$ with $N$ indicates more favorable scaling for symmetry-resolved averages, while decay with $N$ indicates an advantage for post-selection.  Figure~\ref{fig_ratio1} presents our results.  For paramagentic ancilla (left panels), data for cases III and IV are consistent with post-selection becoming favorable for $u \gtrsim 0.1$. 
For critical ancilla (right panels), the data show that symmetry-resolved averages can remain advantageous out to larger values of $u$---consistent with the intuition above---especially in case III.  Non-monotonic behavior evident in panel (d) arises because of the aforementioned sign changes in $\Delta p$ arising in case IV with critical ancilla.  

\begin{figure}[t!]
\centering
\includegraphics[width = \columnwidth]{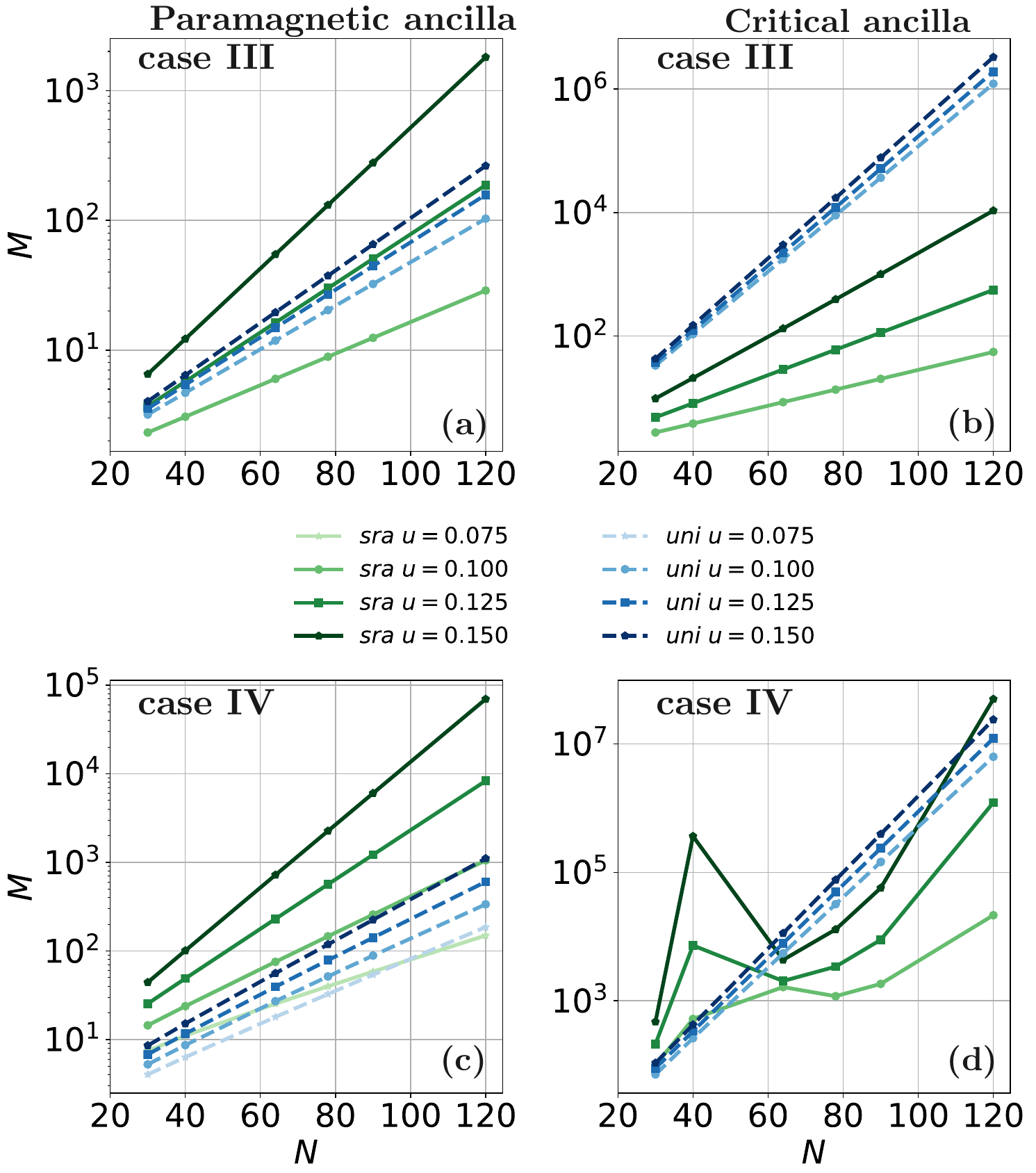}
\caption{(Color online).  \textbf{Scaling of $M_{\text{uni}}$ and $M_{\text{sra}}$ with system size $N$.}  The number of trials needed for post-selection of the uniform measurement outcome and to extract symmetry-resolved averages are here estimated by $M_{\rm uni} = 1/p_{\rm uni}$ and $M_{\rm sra} = 1/(\Delta p)^2$, respectively.  Both approaches offer complementary regimes of experimental viability even at large systems with $N \sim {\rm O}(100)$, as evidenced by a number of required trials of O$(10^3)$ or smaller.  Data were obtained identically as in Fig.~\ref{fig_ratio1}.
 }
 \label{fig_Ms}
\end{figure}

Assessing practicality of either scheme also requires  quantifying the separate values $M_{\rm sra}$ and $M_{\rm uni}$ (as opposed to just their ratio) for experimentally reasonable system sizes.  Relevant $N$ values will certainly be platform dependent, as will the number of trials that one can feasibly conduct on laboratory timescales.  For concreteness, we will focus on $N\sim{\rm O}(100)$---which is relevant for present-day hardware---and postulate that trials up to O$(10^3)$ are accessible.  Furthermore, we will simply take $M_{\rm sra} = 1/(\Delta p)^2$ and $M_{\rm uni} = 1/p_{\rm uni}$ to roughly evaluate the trials required for symmetry-resolved averages and for post-selection, respectively; Fig.~\ref{fig_Ms} displays the $N$ dependence of these quantities for select $u$ values.  Remarkably, all four panels explored in the figure reveal regimes for which symmetry-resolved averages satisfy the experimental plausibility criteria laid out above. With paramagnetic ancilla, post-selection also enjoys regimes that require a surprisingly moderate number of trials even out to fairly large system sizes---ultimately because ancilla measurement outcomes obey a highly biased, controllable distribution.  For reference, had all measurement outcomes been equally likely, one would obtain $M_{\rm uni} = 2^N \sim 10^{24}$ in an $N = 80$ system!  Figure~\ref{fig_Ms} additionally reveals that $M_{\rm uni}$ increases relatively slowly with $u$ (compared to $M_{\rm sra}$), extending the experimentally plausible regime for post-selection to larger $u$'s that display correspondingly stronger signatures of measurement-altered criticality.    

\section{Discussion and outlook}
\label{sec:discussion}

We analyzed the initialize-entangle-measure-probe protocol summarized in Fig.~\ref{fig:protocol} to investigate how measurements impact correlations in 1D Ising quantum critical points.  Specifically, we developed a perturbative formalism that allowed us to analytically study the outcome of our protocol applied with the four classes of unitaries and projective ancilla measurements listed in Table~\ref{tab.unitaries}.  Within this approach, long-distance correlations of microscopic spin operators were related to  correlations of low-energy fields evaluated with respect to the usual Ising CFT action perturbed by a `defect line'.  The detailed structure of the defect line depends on the choice of entangling unitary, the initial ancilla state, and the outcome of ancilla measurements.  We argued that, with $\tilde Z$-basis ancilla measurements, this formalism applies to general measurement outcomes; with $\tilde X$-basis ancilla measurements, however, well-behaved defect-line actions emerge only for a restricted set of (high-probability) measurement outcomes.  In the latter context, we hope that future work can develop a more complete analytic theory capable of treating  arbitrary measurement outcomes and assessing their probabilities for general ancilla initializations.

Various predictions follow from this framework---most of which we supported with numerical simulations.  We recapitulate our main findings here (see also Fig.~\ref{fig:protocol}(e)): 

{\bf Case I: unitary $U_j = e^{iu(X_j-\langle X\rangle)\tilde X_j}$, $\tilde Z$-basis measurements.} Non-perturbative CFT results \cite{NT2011} allow one to formally compute the coarse-grained two-point spin correlation function $\langle Z_j Z_{j'}\rangle_{\tilde s}$ for general measurement outcomes $\tilde s$.  For a uniform measurement outcome---which occurs with highest probability---the two-point function exhibits power-law decay with a measurement-induced change in the scaling dimension.  Our formulation also captured subtle changes in correlations that arise with measurement outcomes featuring a domain wall. 
The agreement we found between analytical and numerical results here represents a highly nontrivial check for the validity of our approach.

{\bf Case II: unitary $U_j = e^{iu(Z_j-C)\tilde X_j}$, $\tilde Z$-basis measurements.} With $C \neq 0$ the defect-line action includes a longitudinal-field term that explicitly breaks the $\mathbb{Z}_2$ symmetry enjoyed by the critical chain prior to measurement.  Correspondingly, the one-point function $\langle Z_j\rangle_{\tilde s}$ becomes non-zero, with a spatial profile dependent on the measurement outcome. Averaging $\langle Z_j\rangle_{\tilde s}$ over measurement outcomes yields a vanishing one-point function as required on general grounds.  By contrast, averaging $\langle Z_j\rangle_{\tilde s}^2$ retains memory of the measurements and yields a non-zero result that, based on our exact diagonalization results, appears to survive in the thermodynamic limit. 

{\bf Case III: unitary $U_j = e^{iu(X_j-\langle X\rangle)\tilde Z_j}$, $\tilde X$-basis measurements.}  Just as for case I, the uniform string measurement outcome occurs with highest probability and yields a two-point function $\langle Z_j Z_{j'}\rangle_{\tilde s}$ with modified scaling dimension.  

{\bf Case IV: unitary $U_j = e^{iu(Z_j-C)\tilde Z_j}$, $\tilde X$-basis measurements.}  As for case II, explicit breaking of $\mathbb{Z}_2$ symmetry induced by $C \neq 0$ yields a non-zero one-point function $\langle Z_j\rangle_{\tilde s}$ for the nearly uniform measurement outcomes amenable to our perturbative formalism.  Taking $C = 0$ restores $\mathbb{Z}_2$ symmetry for the critical chain.  Here, when the ancilla are \emph{also} critical, the defect-line action hosts a long-range power-law decaying interaction among CFT spin-fields that, based on iDMRG simulations, appears to qualitatively alter $\langle Z_j Z_{j'}\rangle_{\tilde s}$ correlations (for uniform or nearly uniform measurement outcomes).  That is, on short distances the correlations decay faster than power-law, though we argued that on longer-distances power-law correlations are likely to re-emerge.
Further substantiating this scenario, possibly drawing connections to previous work on long-range-interacting Ising chains \cite{Tagliacozzo,Vodola_2016}, raises an interesting open problem.

In the cases with $\tilde X$-basis ancilla measurements, we further proposed a new post-selection-free method for detecting non-trivial effects of measurements on Ising quantum criticality.   
Here we exploited the fact that $\tilde X$-basis measurement outcomes factorize into two symmetry sectors depending on the 
value of the generator $\tilde G = \prod_j \tilde X_j \in \pm 1$ of the global $\mathbb{Z}_2$ symmetry for the ancilla.  Although $\tilde G$ is a non-local operator, its eigenvalue for any outcome $\ket{\tilde s}$ follows trivially given measurements of $\tilde X_j$ for each ancilla site; 
one can, in turn, average critical-chain observables over measurement outcomes separately within each symmetry sector. We found analytically that the \emph{difference} in averages between the two sectors (normalized by the difference in probability for accessing the sectors) encodes measurement effects on Ising criticality that survive in the thermodynamic limit.  Strikingly, such ratios  evaluated for one- and two-point spin correlations mimic the behavior predicted for uniform post-selection outcomes by our perturbative defect-line framework, as recovered also in our simulations.  The practical catch is that the ratio involves a numerator and denominator that individually decay to zero as the system size increases.

To address feasibility given this catch, we quantified the number of trials needed to meaningfully extract symmetry-resolved averages and contrasted with the alternative technique of post-selecting for the uniform measurement outcome.  We established very generally that symmetry-resolved averages require exponentially fewer trials compared to post-selection over a window of small entangling-gate strength $u$ that widens as the ancilla become more correlated.  Moreover, within this window symmetry-resolved averages necessitate a strikingly modest number of protocol runs---estimated to be on the scale of $10^2$ to $10^3$ for critical chains composed of O$(100)$ spins.  Another important message of this work, however, is that post-selection poses a far less daunting challenge compared to the situation for other measurement-induced phenomena.  With paramagnetic ancilla in particular, we found that in larger-$u$ regimes where post-selection outperforms symmetry-resolved averages, the uniform measurement outcome can emerge with high probability after a similarly modest number of protocol runs in systems with as large as O$(100)$ spins.  Apparent viability of post-selection here originates from the fact that measurement outcomes are far from random, but rather follow a highly biased distribution that one can control via $u$ and the initial ancilla configuration. 
 It would be valuable to understand the effects of measurement errors and decoherence on both symmetry-resolved averages and post-selection to further address their suitability for experimental application.  More broadly, might  factorization into symmetry sectors as exploited for symmetry-resolved averages prove fruitful for detecting measurement-induced phenomena in other contexts?

In much of this work, correlations in the initialized ancilla state played an important role. 
Cases I and II, for instance, become completely trivial if the ancilla are initialized into the (product-state) ground state of Eq.~\eqref{ancilla} at $h_{\rm anc}/J_{\rm anc} \rightarrow \infty$.  In this extreme case $\ket{\psi_a}$ is an eigenstate of the $U_j$'s used in our protocol for cases I and II.  Hence those unitaries do not actually entangle the critical chain with the ancilla, and measurements of the latter do not affect the former.  Case IV highlights a more striking example, where once again critical ancilla can produce a defect-line action exhibiting inherently long-range interaction among CFT fields, mediating physics qualitatively different from what we found with paramagnetic ancilla.  

Outside of this last example, we invariably concluded that the defect-line action could be approximated by a single term linear in either the $\sigma$ or $\varepsilon$ field (depending on the protocol details under consideration).  Microscopically, we showed that for the post-measurement state $\ket{\psi_{\tilde s}} = \frac{1}{\sqrt{\mathcal{N}}}U' e^{-H_m/2}\ket{\psi_c}$ [Eq.~\eqref{psi_m_goal}], the important non-unitary $e^{-H_m/2}$ part generically does not factorize into a product of operators acting at individual sites $j$ due to the $V_{jk}$ term in Eq.~\eqref{H_m}.  An approximate factorized form,
\begin{equation}
    \ket{\psi_{\tilde{s}}}\approx \frac{1}{\sqrt{p_{\tilde{s}}}}\prod_j M_j \ket{\psi_c},
    \label{psi_factorized}
\end{equation}
nevertheless captures the leading defect-line action linear in $\sigma$ or $\varepsilon$ that we typically obtained.  When $m_j$ is non-zero, the factorized measurement operators $M_j$ follow by simply setting $V_{jk} = 0$ in $H_m$; with $m_j = 0$, one instead modifies the $V_{jk}$ term by `fusing' the constituent microscopic operators (mimicking the CFT-field fusion rules) to arrive at a factorizable form.  Intuitively, the more highly entangled the ancilla are, the worse this approximation becomes---culminating in its complete breakdown in case IV with $C = 0$ when the ancilla are critical.  The breakdown is anticipated to be especially stark in the case of symmetry-resolved averages due to the very slowly decaying longe-range interaction generated by measurement. It would be interesting to quantify the accuracy of Eq.~\eqref{psi_factorized} from a microscopic viewpoint, e.g., by studying the operator space entanglement of $M_{\tilde{s}}$ as a function of the ancilla wavefunction.  

Chains of laser-excited Rydberg atoms trapped in optical tweezer arrays comprise a promising experimental platform for  measurement-altered Ising criticality.  A single Rydberg chain effectively realizes an antiferromagnetic spin model with power-law-decaying Ising interactions, supplemented by both transverse \emph{and} longitudinal fields---though the latter can be tuned to zero by choosing an appropriate detuning from resonance.  The phase diagram hosts a readily accessible Ising quantum phase transition (among other more exotic critical points) \cite{FSS} that is well-understood also at the lattice level in this setting \cite{Slagle2021}.  Moreover, a second Rydberg chain could furnish the ancilla degrees of freedom in our protocol.  Devising concrete implementations of the requisite unitaries and ancilla measurements in this venue poses a nontrivial problem for future work.  Additionally, the pursuit of measurement-altered criticality in Rydberg arrays highlights several fundamental open questions---including the impact of antiferromagnetic Ising interactions, a non-zero longitudinal field,  integrability-breaking perturbations, etc. 
Erasure conversion developed for Rydberg arrays in Ref~\cite{scholl2023} is a promising tool for probing measurement-altered quantum criticality in this arena.  Recent work in a quite different setting has also shown the possibility of creating the ground state of a critical transverse-field Ising chain using a quantum computer \cite{Haghshenas23}, highlighting tantalizing prospects for realization also in digital quantum hardware.

Finally, many other variations on the present work would be interesting to explore.  Extension to strongly interacting CFTs---e.g., tricritical Ising, parafermionic, etc.---is particularly intriguing given their rich field content, and correspondingly rich set of possible measurement-induced defect-line actions.  Measurements could also be performed in various alternative ways that add a new twist into the problem; for instance, one could contemplate joint measurements of operators on the critical and ancilla chains, or measure different quantities in different regions of space.
We hope that the approach used here will prove useful for addressing such problems in the future.

\emph{Note added:} While finishing this work we became aware of Refs.~\onlinecite{EhudMeasurementIsing,JianMeasurementIsing}, which examined the effects of measurement on Ising criticality from a perspective largely complementary to ours.

\begin{acknowledgments}

It is a pleasure to acknowledge enlightening conversations with Ehud Altman, Mario Collura, Adolfo Del Campo, Manuel Endres, Matthew Fisher, Sam Garratt, Tim Hsieh, Dan Mao, Chao-Ming Jian, Sheng-Hsuan Lin, Olexei Motrunich, Kelly Shane, Xhek Turkeshi, Ettore Vicari, Omar Wani, and Zack Weinstein. Tensor network calculations were performed using the TeNPy Library~\cite{tenpy}. The U.S. Department of
Energy, Office of Science, National Quantum Information Science Research Centers, Quantum Science Center supported the construction and numerical analysis of the protocol for probing measurement-altered Ising criticality. Additional support was provided by the National Science Foundation through grant
DMR-1848336 (RM); the Caltech Institute for Quantum
Information and Matter, an NSF Physics Frontiers Center with support of the Gordon and Betty Moore Foundation through Grant GBMF1250; and the Walter Burke
Institute for Theoretical Physics at Caltech.

\end{acknowledgments}

\appendix

\onecolumngrid
\pagebreak[4]

\section{Post-measurement wavefunction}\label{app:post}

Here we provide technical details leading to the post-measurement state specified in Eqs.~\eqref{psi_m_goal}, \eqref{H_U}, and \eqref{H_m}.
We start by observing that the ancilla measurement basis is always orthogonal to $\tilde O_j$; hence if measurement projects ancilla site $j$ to $\ket{\tilde s_j}$, then $\tilde O_j$ simply flips that measured spin: $\tilde O_j\ket{\tilde s_j} = \ket{-\tilde s_j}$.
For the unitary in Eq.~\eqref{Uj_general}, we can then use properties of Pauli operators to express the post-measurement state defined in Eq.~\eqref{psi_m_def} as
\begin{align}
  \ket{\psi_{\tilde s}} &= \frac{1}{\sqrt{p_{\tilde s}}}\left\{\left[\prod_j\Big(C_j\bra{\tilde s_j} + S_j \bra{-\tilde s_j}\Big)\right]\ket{\psi_a}\right\}
   \ket{\psi_c}.
  \label{psi_m_1}
\end{align}
Equation~\eqref{psi_m_1} uses the short-hand notation
\begin{align}\label{eq:cossin}
  C_j = \cos[u(O_j-\theta)],~~~~
  S_j = i \sin[u(O_j-\theta)],
\end{align}
which satisfy standard trigonometric identities even including the Pauli operators in the arguments.

Observe that multiplying all elements in the product from Eq.~\eqref{psi_m_1} yields a sum of terms with anywhere from $N_f = 0$ to $N$ flipped ancilla spins ($N$ is the total number of ancilla sites), and that these flipped spins can occur at arbitrary sites $i_1<i_2<\cdots<i_{N_f}$.  Let $\ket{\tilde s(i_1,\cdots, i_{N_f})}$ be the state with $N_f$ flipped ancilla spins at these sites, and let $F(i_1,\cdots , i_{N_f})$ denote a product of $C_j$'s for the unflipped sites and $S_j$'s for the flipped sites.  For example, 
\begin{align}
    \ket{\tilde{s}(i_1,i_2)} &= \tilde{O}_{i_1} \tilde{O}_{i_2} \ket{\tilde s}\nonumber,
\\
  F(i_1,i_2) &= C_1\cdots C_{i_1-1}S_{i_1}C_{i_1+1}\cdots C_{i_2-1}S_{i_2}C_{i_2+1}\cdots .
\end{align}
We can then explicitly write
\begin{align}
  \ket{\psi_{\tilde s}} = \frac{1}{\sqrt{p_{\tilde s}}}\sum_{N_f = 0}^N \; \sum_{i_1<i_2<\cdots<i_{N_f}} \!\! \bra{ \tilde s(i_1,\cdots, i_{N_f}) } \ket{\psi_a}
  F(i_1,\cdots, i_{N_f})\ket{\psi_c}
  \label{psi_m_2}
\end{align}
which, en route to the form in Eq.~\eqref{psi_m_goal}, can be trivially re-expressed as
\begin{align}
  \ket{\psi_{\tilde s}} = \frac{1}{\sqrt{p_{\tilde s}}}\exp\mathopen\bigg\{\ln\mathopen\bigg[\bra{\tilde s}\psi_a \rangle F_0
  +\sum_i \bra{\tilde s(i)}\psi_a\rangle F(i)
  + \sum_{i_1<i_2}\bra{\tilde s(i_1,i_2)}\psi_a\rangle F(i_1,i_2)  
  + \cdots\bigg]\bigg\}\ket{\psi_c}
  \label{psi_m_2again}
\end{align}
with $F_0 \equiv C_1 \cdots C_N$.

Suppose now that $\bra{\tilde s}\psi_a \rangle$ is non-zero.  In this case we can factor out the first term in the log from Eq.~\eqref{psi_m_2again} to obtain, after some manipulation and absorbing an $\tilde s$-dependent 
constant into a new normalization factor $\mathcal{N}$,
\begin{align}
  \ket{\psi_{\tilde s}} &= \frac{1}{\sqrt{\mathcal{N}}} e^{-H_{\rm temp}/2}\ket{\psi_c},
  \label{psitemp}
  \\
  H_{\rm temp} &= -2\ln\bigg[1 + \sum_i a(i)T_i 
  \nonumber + \sum_{i_1<i_2} a(i_1,i_2)T_{i_1}T_{i_2} + \cdots\bigg]
  - 2\sum_i \ln(C_i).
\end{align}
In the second and third lines we introduced the quantities
\begin{align}
    a(i_1,\cdots, i_{N_f}) = \frac{\bra{\tilde s(i_1\cdots i_{N_f})}\psi_a\rangle}{\bra{\tilde s}\psi_a \rangle},~~~
    T_i = C_i^{-1}S_i,\label{eq:aj}
\end{align}
which are a generalization of Eq.~\eqref{eq:ajk} of the main text.
At this point one can expand $H_{\rm temp}$ to the desired order in $u \ll 1$.  To proceed it is convenient to decompose $H_{\rm temp}$ via 
\begin{equation}\label{Htemp}
    H_{\rm temp} = -2i H' + H_m,
\end{equation} 
where $H', H_m$ are commuting Hermitian operators; $U'= e^{iH'}$ is the unitary transformation from Eq.~\eqref{psi_m_goal} while $H_m$ contains the crucial non-unitary effects from measurement.  
Upon absorbing constants into the normalization $\mathcal{N}$, to $O(u^2)$ one obtains Eqs.~\eqref{H_U} and~\eqref{H_m} in the main text.

\section{Gaussian overlaps}\label{app:Gaussian}

Throughout the manuscript, we are interested in the evaluation of correlation functions like the ones appearing in Eq.~\eqref{eq:aj}. To establish the main ideas, consider the case in which the unitary involves $\tilde{X}$, we measure in the $\tilde Z$-basis, and we want to evaluate 
\begin{equation}\label{eq:goal}
a(i_1,\cdots,i_{N_F})= \frac{\bra{\tilde{s}}\prod_{j=1}^{N_F}\tilde{X}_{i_j}\ket{\psi_a}}{\braket{\tilde s}{\psi_a}}.
\end{equation}
Here $\tilde{s}$ is an arbitrary string outcome and $N_F$ denotes the number of flipped spins generated by $\tilde X_i$ operators in the product above.  The ket $\ket{\psi_a}$ is the ground state of the transverse-field Ising model in Eq.~\eqref{ancilla} or, equivalently, of its fermionic representation obtained via the Jordan-Wigner transformation.  We can collect the Majorana operators appearing in the latter form of the Hamiltonian into a vector $\Vec{\gamma}$ defined by
\begin{equation}
    \Vec{\gamma}=\begin{pmatrix}
\gamma_{A,1}\\
\gamma_{A,2}\\
\vdots \\
\gamma_{A,N}\\
\gamma_{B,1}\\
\vdots\\
\gamma_{B,N}
\\
\end{pmatrix},
\end{equation}
where $N$ is the system size and $\{\Vec{\gamma}_i,\Vec{\gamma}_j\}=2\delta_{ij}$. To simplify notation, in this appendix we suppress tildes on the Majorana fermion operators for the ancilla, and also replace $h_{\rm anc} \rightarrow h$ and $J_{\rm anc} \rightarrow 1$. 
The Hamiltonian is quadratic in the fermionic operators, and any fermionic Gaussian state can be described through the covariance matrix 
\begin{equation}
    \Gamma_{jk}=\frac{i}{2}\langle [\gamma_j, \gamma_k] \rangle,
\end{equation}
with $[ \gamma_j, \gamma_k ]$ the commutator of the two Majorana operators $\gamma_j$ and $\gamma_k$. From the definition, we observe that $\Gamma$ is a real and skew-symmetric matrix.
For the transverse field Ising chain, the covariance matrix is known analytically \cite{correlator}.
In particular it has a Toeplitz structure given by
\begin{equation}
   \Gamma_{jk}=\frac{1}{N}\sum_{k\in \Omega_{\mathrm{gs}} }e^{-\frac{2\pi i k}{N}(j-k)}\begin{pmatrix}
       & 0 & \frac{h-\cos(\frac{2\pi k}{N})-i\sin(\frac{2\pi k}{N})}{\sqrt{(h-\cos(\frac{2\pi k}{N}))^2+\sin(\frac{2\pi k}{N})^2}}\\
       &-\frac{h-\cos(\frac{2\pi k}{N})-i\sin(\frac{2\pi k}{N})}{\sqrt{(h-\cos(\frac{2\pi k}{N}))^2+\sin(\frac{2\pi k}{N})^2}} & 0
   \end{pmatrix},
\end{equation}
where $\Omega_{\mathrm{gs}}$ is the set of occupied momenta in the ground state. 
In the limit $N\to \infty$, the Eq. above simplifies as
\begin{equation}
   \Gamma_{jk}=\frac{1}{2\pi}\int_0^{2\pi}d\phi e^{-i \phi (j-k)}\begin{pmatrix}
       & 0 & \frac{h-\cos(\phi)-i\sin(\phi)}{\sqrt{(h-\cos(\phi)^2+\sin(\phi)^2}}\\
       &-\frac{h-\cos(\frac{2\pi k}{N})-i\sin(\phi)}{\sqrt{(h-\cos(\phi)^2+\sin(\phi)^2}} & 0
   \end{pmatrix},
\end{equation}
We can also use an alternative approach to determine the covariance matrix. First, rewrite Eq.~\eqref{eq:majorana} as $H=\frac{1}{2}\sum_{j,k}h_{jk} \gamma_j \gamma_k$ with $h$ a matrix encoding the free-fermion Hamiltonian.  
We proceed by finding the fermionic transformation $U$ that diagonalizes $h$; in this diagonal basis, the correlation matrix associated to the ground state, $\Gamma_{\mathrm{diag}}$, is simply obtained by substituting $-1$ ($+1$) for any positive (negative) eigenvalue~\cite{Surace_2022}. To obtain $\Gamma$, we just need to move back to the original basis, i.e., $\Gamma=U^{\dagger}\Gamma_{\mathrm{diag}} U$.

When we measure in the $\tilde Z$-basis, $\ket{\Tilde{s}}$ is not a Gaussian state, but noticing that 
$ \bra{\Tilde{s}}\Tilde{X}_j\ket{\psi_a}=\bra{\Tilde{s}}\Tilde{X}_j\tilde{G}\ket{\psi_a}= \bra{-\Tilde{s}}\Tilde{X}_j\ket{\psi_a}$,
we find
\begin{equation}
    \bra{\Tilde{s}}\Tilde{X}_j\ket{\psi_a}=\frac{1}{\sqrt{2}}\bra{\psi_+}\Tilde{X}_j\ket{\psi_a}, \qquad \ket{\psi_+}=\frac{\ket{\Tilde{s}}+\ket{-\Tilde{s}}}{\sqrt{2}}.
\end{equation}
The advantage of using the cat state $\ket{\psi_+} $ is that now it is Gaussian and corresponds to the ground state of a quadratic Hamiltonian
\begin{equation}
    H_{\Tilde{s}}=-\sum_j \Tilde{s}_j\Tilde{s}_{j+1}\Tilde{Z}_j\Tilde{Z}_{j+1}
\end{equation}
that, after a Jordan-Wigner transformation, reads
\begin{equation}\label{eq:Htilde}
    H_{\Tilde{s}}=i\sum_j \Tilde{s}_j\Tilde{s}_{j+1}\gamma_{A,j+1}\gamma_{B,j}\equiv \frac{1}{2} \sum_{jk}h^{\Tilde{s}}_{jk} \gamma_j\gamma_k.
\end{equation}
By applying the procedure described above, we can find the covariance matrix $\Gamma_{\Tilde{s}}$ describing the Gaussian ground state of Eq.~\eqref{eq:Htilde}. Once we know the covariance matrix both for the ground state of the ancilla and for the Hamiltonian $ H_{\Tilde{s}}$, we can apply a result found in Ref.~\onlinecite{Bravyi-08}: the absolute value of the inner product $\langle \Tilde{s}| \psi_a\rangle $ is
\begin{equation}\label{eq:bravyi}
    |\langle \Tilde{s}| \psi_a\rangle |=\sqrt{2^{-N-1}\Pf(\Gamma+\Gamma_{\Tilde{s}})},
\end{equation}
where $\Pf$ is the Pfaffian of $\Gamma+\Gamma_{\Tilde{s}}$. 

Evaluation of Eq.~\eqref{eq:goal} follows straightforwardly from the above results since we know explicitly the action of $\prod_j\Tilde{X}_{i_j}$ on $\ket{\Tilde{s}}$. For instance, if $N_F=1$, $\Tilde{X}_j\ket{\Tilde{s}}=\ket{\Tilde{s}(j)}$, i.e., the action of $\tilde X_j$ simply flips the spin at site $j$. We therefore have to also compute the covariance matrix for the Hamiltonian associated with the outcome $\Tilde{s}(j)$, and then extract the corresponding overlap with the ancilla ground state using Eq.~\eqref{eq:bravyi} with $\tilde s \rightarrow \tilde s(j)$. 
This procedure yields 
\begin{equation}
\label{eq:aj_gaussian}    a(j)=\sqrt{\frac{\Pf(\Gamma+\Gamma_{ \Tilde{s}(j) })}{\Pf(\Gamma+\Gamma_{\Tilde{s}})}}, \qquad a(j,k)=\sqrt{\frac{\Pf(\Gamma+\Gamma_{\Tilde{s}(j,k)  })}{\Pf(\Gamma+\Gamma_{\Tilde{s}})}};
\end{equation}
other $a(i_1,\cdots,i_{N_F})$ coefficients from Eq.~\eqref{eq:goal} follow similarly.
Importantly, the absolute value in Eq.~\eqref{eq:bravyi} is superfluous for the model we consider in this manuscript due to the stoquasticity of the transverse-field Ising model.

\begin{figure}[t!]
\centering
\includegraphics[width=0.4\linewidth]{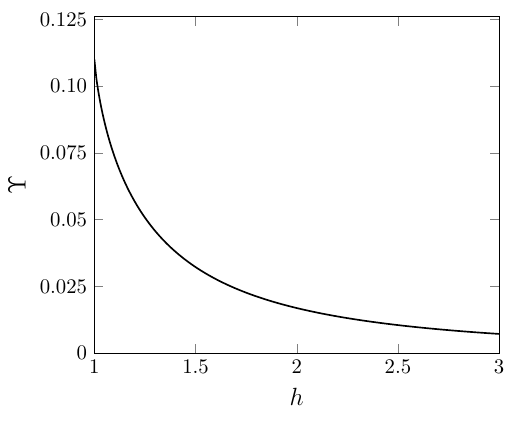}
\caption{\textbf{Coefficient $\Upsilon$ in Eq. \eqref{eq:p0_Upsilon}.} The plot shows the behavior of the coefficient describing the exponential growth with system size of the probability $p^{(0)}_{\rm uni}$ for obtaining the uniform measurement outcome at $u = 0$.} 
\label{fig:Upsilon}
\end{figure}

When we measure in the $\tilde X$-basis, $\ket{\Tilde{s}}$ is automatically a Gaussian state, corresponding to the ground state of the quadratic Hamiltonian
\begin{equation}
    H_{\Tilde{s}}=-\sum_j \Tilde{s}_j\Tilde{X}_j = -i\sum_j \Tilde{s}_j\gamma_{A,j}\gamma_{B,j}.
\end{equation}
Therefore, computing the corresponding covariance matrix, we can again apply Eq.~\eqref{eq:bravyi} to evaluate 
\begin{equation}\label{eq:bravyi2}
    \langle \Tilde{s} |\psi_a\rangle =\sqrt{2^{-N}\Pf(\Gamma+\Gamma_{\Tilde{s}})},
\end{equation}
as well as correlation functions like $\frac{\bra{\tilde s}\prod_{j=1}^{N_F}\Tilde{Z}_{i_j}|\psi_a\rangle}{\bra{\tilde s}\psi_a \rangle}$.
We remark here that if $\ket{\tilde s}$ is the uniform measurement outcome, $\Gamma+\Gamma_{\Tilde{s}}$ is still a block-Toeplitz matrix because the system preserves its translational invariance. Indeed, for large $N$, it can be written as  
\begin{equation}
   (\Gamma+\Gamma_{\tilde s})_{jk}=\frac{1}{2\pi}\int_0^{2\pi}d\phi e^{-i \phi (j-k)}\mathcal{G}(\phi,h),\quad \mathcal{G}(\phi,h)=\begin{pmatrix}
       & 0 & \frac{h-\cos(\phi)-i\sin(\phi)}{\sqrt{(h-\cos(\phi)^2+\sin(\phi)^2}}+1\\
       &-\frac{h-\cos(\frac{2\pi k}{N})-i\sin(\phi)}{\sqrt{(h-\cos(\phi)^2+\sin(\phi)^2}}-1 & 0
        \end{pmatrix}
\end{equation}
In order to evaluate $p^{(0)}_{\tilde s}$ in Eq.~\eqref{p0} for a uniform measurement outcome, we need to evaluate the Pfaffian (therefore the determinant) of $\Gamma+\Gamma_{\tilde s}$ using Eq.~\eqref{eq:bravyi2}. One of the main results on the theory of block Toeplitz determinants is the Widom-Szeg\"o theorem~\cite{szego}. According to it, the determinant of a block Toeplitz matrix, like $\Gamma+\Gamma_{\tilde s}$, with symbol
$\mathcal{G}(\phi,h)$, behaves for large $N$ as 
\begin{equation}
   \log \mathrm{det}[\Gamma+\Gamma_{\tilde s}]\sim N\int_0^{2\pi}\frac{d\phi}{2\pi}\log \det \mathcal{G}(\phi,h)= N\int_0^{2\pi}\frac{d\phi}{\pi}\log\left[\frac{\sqrt{h^2-2 h \cos (\phi )+1}+h-\cos (\phi )}{\sqrt{h^2-2 h \cos (\phi )+1}}\right].
\end{equation}
This results allows us to compute explicitly the probability $p^{(0)}_{\rm uni}$ for finding the uniform measurement outcome at $u = 0$:
\begin{equation}\label{eq:p0_Upsilon}
    p^{(0)}_{\rm uni} \sim e^{-\Upsilon N}, \quad \Upsilon=\log 2-\int_0^{2\pi}\frac{d\phi}{2\pi}\log\left[\frac{\sqrt{h^2-2 h \cos (\phi )+1}+h-\cos (\phi )}{\sqrt{h^2-2 h \cos (\phi )+1}}\right].
\end{equation}
In particular the integral above can be explicitly solved for $h=1$, i.e., when the ancilla chain becomes critical, and we obtain
\begin{equation}
    \Upsilon=\log 2-2\mathrm{Catalan}/\pi,
\end{equation}
where Catalan's constant is $\simeq 0.92$.  As Fig.~\ref{fig:Upsilon} shows, $\Upsilon$ monotonically decreases to zero for $h\geq 1$.

\section{Effective action formalism for ratio $r(A)$ }
\label{AppB}

In this appendix, we show how the formalism developed in Sec.~\ref{perturbative} and Appendix~\ref{app:post} allows us to rewrite $r(A)$ from Eq.~\eqref{ratio} in a more illuminating form that immediately allows us to take advantage of the effective action formalism exploited throughout this paper. Let us start from the case where $[A,U]=0$.
The numerator of Eq.~\eqref{ratio} can be rewritten as 
\begin{equation} \label{eq:product}   \langle \psi_c|A\langle \psi_a|\prod_j[C^r_j+S^r_j \tilde Z_j]|\psi_a\rangle |\psi_c\rangle, 
\end{equation}
where
\begin{align}
  C^r_j = \cos[2u(O_j-\theta)],~~~~
  S^r_j = i \sin[2u(O_j-\theta)].
  \label{CrSr}
\end{align}
Notice the extra factor of $2$ in front of $u$ compared to Eq.~\eqref{eq:cossin}. The form of Eq.~\eqref{eq:product} resembles Eq.~\eqref{psi_m_1} and is therefore amenable to similar 
manipulations that led to Eqs.~\eqref{psitemp} and \eqref{Htemp}.  Following these steps, we re-express the ratio in Eq.~\eqref{ratio} as in Eq. \eqref{rA_nice}, where 
now
\begin{align}
  H^r_m =& -\ln\bigg[1 + \sum_{i_1<i_2} \langle\psi_a|\tilde Z_{i_1} \tilde Z_{i_2}|\psi_a\rangle T^r_{i_1}T^r_{i_2} + \cdots\bigg] - \sum_i \ln(C^r_i), \label{eq:H_mr}
\end{align}
with $T^r_i=(C^r_i)^{-1} S^r_i$.  Compared to Eq.~\eqref{Htemp}, the argument of the log contains only terms with an even number of $T^r_i$ operators.  This difference results from the fact that $\tilde{G}\ket{\psi_a}=\ket{\psi_a}$, implying that only multi-point ancilla correlators with an even number of $\tilde{Z}$ operators give non-trivial contributions. It turn, $\bra{\psi_c}A e^{-H^r_m}\ket{\psi_c}$ does not contain a unitary contribution, even though $\bra{\psi_a}\bra{\psi_c} AU^2\ket{\psi_c}\ket{\psi_a}$ naively does. This conclusion can be understood by noticing that $\bra{\psi_a}\bra{\psi_c}  AU^2\ket{\psi_c}\ket{\psi_a}=\bra{\psi_a}\bra{\psi_c}  A\textrm{Re}(U^2)\ket{\psi_c}\ket{\psi_a}$ with $\textrm{Re}(U^2)$ a non-unitary operator. Upon expanding $H^r_m$ to $O(u^2)$ and simplifying all constant terms between the numerator and denominator, we obtain Eq.~\eqref{eq:H_m2} in the main text.

So far, we derived Eq.~\eqref{rA_nice} assuming that $[A,U]=0$. When $[A,U] \neq 0$ we obtain a modified form of $H_m$ as follows.  Consider case III from Table~\ref{tab.unitaries} and suppose that we are interested in observables $A= Z_j Z_{j'}$ with $j \neq j'$.  We start by noticing that $ U^2A_U=AU^2 +[U,A]U$ which leads to 
\begin{equation}
 \langle \psi_a|\langle \psi_c|U^2 A_U  |\psi_c\rangle|\psi_a\rangle=\langle \psi_a|\langle \psi_c|A U^2|\psi_c\rangle|\psi_a\rangle+\langle \psi_a|\langle \psi_c|[U,Z_jZ_{j^\prime}]U|\psi_c\rangle|\psi_a\rangle.
\end{equation}
Computing the commutator explicitly and using the notation $U_{\neq j}\equiv \prod_{k\neq j}e^{iu(X_k-\langle X\rangle)\tilde Z_k}$ then yields
\begin{align}
 \label{eq:product2}
&\langle \psi_a|\langle \psi_c|U^2 A_U  |\psi_c\rangle|\psi_a\rangle=\langle \psi_a|\langle \psi_c|A U^2|\psi_c\rangle|\psi_a\rangle-2i\sin(u)\langle \psi_a|\langle \psi_c|Y_jZ_{j'}[i\cos(u\,\langle X\rangle)\tilde Z_{j}+\sin(u\,\langle X\rangle )]U_{\neq j}U |\psi_c\rangle|\psi_a\rangle \nonumber \\&-2i\sin(u)\langle \psi_a |\langle \psi_c|Z_jY_{j'}[i\cos(u\,\langle X\rangle)\tilde Z_{j'}+\sin(u\,\langle X\rangle)]U_{\neq j^\prime}U |\psi_c\rangle|\psi_a\rangle \nonumber \\&+4\sin(u)^2\langle \psi_a |\langle \psi_c|Y_jY_{j'}[i\cos(u\,\langle X\rangle)\tilde Z_{j}+\sin(u\,\langle X\rangle)][i\cos(u\,\langle X\rangle)\tilde Z_{j'}+\sin(u\,\langle X\rangle)]U_{\neq j,j^\prime}U |\psi_c\rangle|\psi_a\rangle. 
\end{align}

Let us focus on the second term on the right side, which can be recast into the form of Eqs.~\eqref{eq:product} in the main text but with $C_k^r = \sin(2u\avg{X}-uX_j)$, $S_k^r = i\cos(2u\avg{X}-uX_j)$ when $k=j$, and given by Eq.~\eqref{CrSr} otherwise. Hence, one recovers the Hamiltonian $H^r_m$ in Eq.~\eqref{eq:H_mr} although with local modifications for terms involving the $j$-th site. We denote the resulting Hamiltonian as $H^{(j)}_m$. All together, this procedure allows the second term on the right side to be compactly expressed as (up to constants that cancel when normalized by the denominator in $r(A)$), 
\begin{align}
-2i\sin(u)\bra{\psi_c}Y_jZ_{j'} e^{-H^{(j)}_m}\ket{\psi_c}.
\end{align} 
Compared to $H_m^r$ in Eq.~\eqref{eq:H_m2}, $H_m^{(j)}$ and $H_m^{(j')}$ are very similar and differ only by some corrections involving sites $j$ or $j'$. Importantly, both $\bra{\psi_c}Y_jZ_{j'} e^{-H^{(j)}_m}\ket{\psi_c}$ and $\bra{\psi_c}Z_jY_{j'} e^{-H^{(j')}_m}\ket{\psi_c}$ contain $Y$ operators, which map to a CFT operator with larger scaling dimension than
that for $Z$. 
One can find an effective action also for the fourth term in Eq. \eqref{eq:product2}, but since it contains operators $Y_j Y_{j'}$, its contribution will be even more subleading with respect to the previous ones.
Therefore, as already mentioned in Sec.~\ref{sec:evenodd}, we can neglect their subleading contributions at large distances.

\section{Dependence of $r(A)$ on system size and scaling form of $\Delta p, p_{\rm uni}$} \label{app:exp_decay}
\begin{figure}[t!]
\centering
\includegraphics[width=0.7\linewidth]{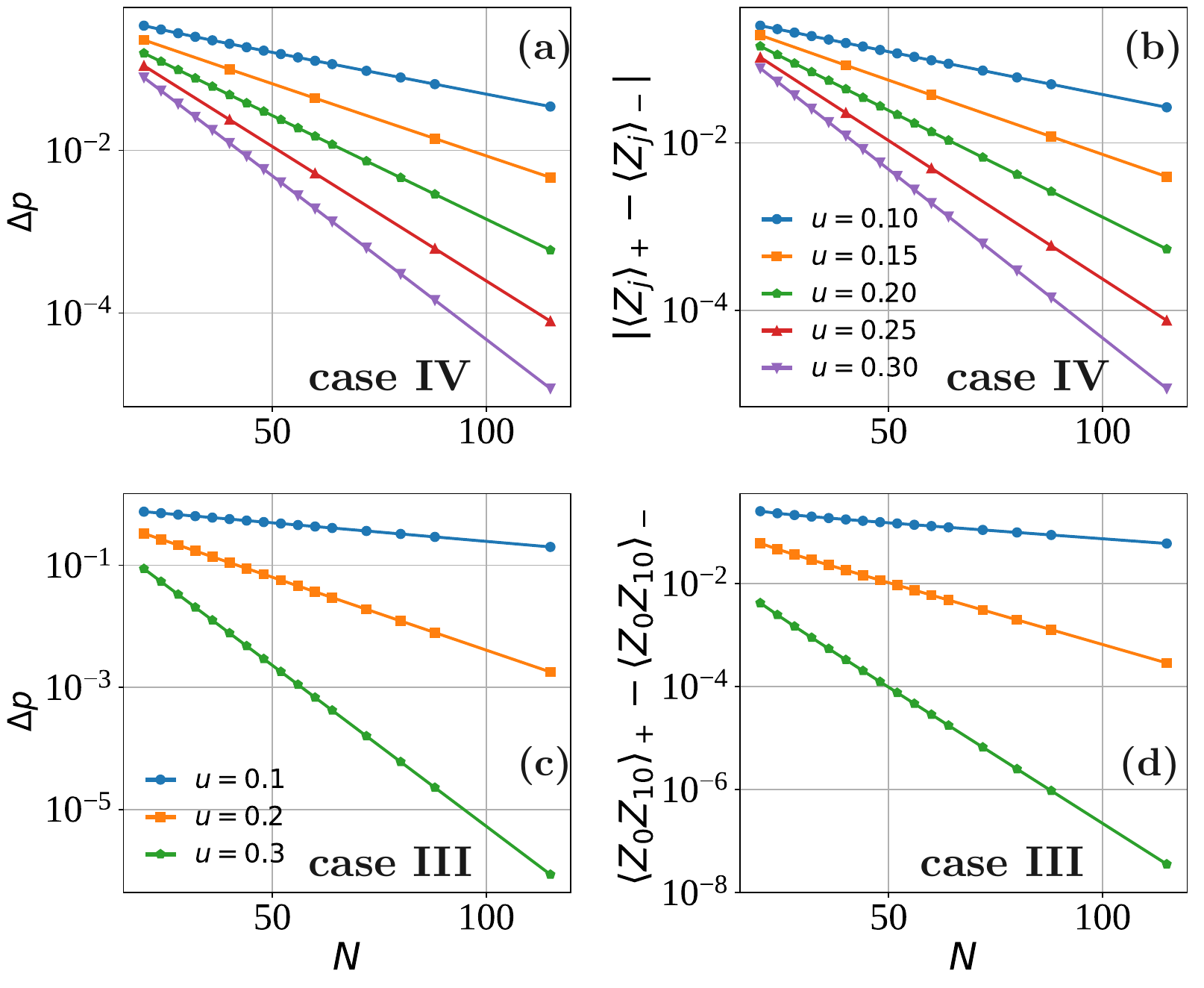}
\caption{(Color online).  \textbf{Exponential decay of symmetry-resolved differences $\avg{A}_+-\avg{A}_-$ with system size.} (a) Probability difference $\Delta p$ corresponding to $A=\mathds{1}$. (b) One-point expectation value corresponding to $A=Z_j$. The result does not depend on the evaluated site $j$ due to translation invariance. (c) Two-point correlator $A=Z_0Z_{10}$ evaluated at a distance $10$ from the reference site.  Data were obtained using (finite) DMRG with periodic boundary conditions for case IV (with $C=-1$) in panels (a,b), and case III in panels (c,d) assuming paramagnetic ancilla.} 
\label{fig:exp_decay}
\end{figure}

In Sec.~\ref{sec:evenoddA}, we argued that the symmetry-resolved differences $\avg{A}_+-\avg{A}_-=\langle \psi_c |\langle \psi_a| U^2 A_U |\psi_a\rangle| \psi_c\rangle$, with $A$ corresponding to correlators of local operators, vanish in the thermodynamic limit $N\to \infty$. Figure~\ref{fig:exp_decay} numerically shows that with paramagnetic ancilla these quantities decay exponentially with system size for the values of $u$ considered in the main text. In particular, panels (a,b) respectively show numerical results for case IV (at $C=-1$) with $A=\mathds{1}$, i.e., the probability difference $\Delta p = p_+-p_-$, and $A=Z_j$. Panels (c,d) respectively show results in case III for $\Delta p$ and $A=Z_0Z_{10}$.

In Sec.~\ref{sec:evenoddB}, when comparing symmetry-resolved averages and post-selection, we used the scaling forms $\Delta p \sim e^{-\eta_{\text{sra}}(u)N}$ and $p_{\rm uni}\sim  e^{-\eta_{\text{uni}}(u)N}$.  
Here we numerically establish that, for small $u$, 
 these quantities in most cases conform well to the more precise scaling behavior $\Delta p \sim e^{-c_{\rm sra} u^{\zeta}N}$ and $p_{\rm uni}/p^{(0)}_{\rm uni}\sim  e^{-c_{\rm uni} u^{\zeta}N}$ with $c_{\rm sra},c_{\rm uni}$ constants, $\zeta$ an exponent that depends on the entangling gate $U$ and on the initial ancilla wavefunction, and $p^{(0)}_{\rm uni} \sim e^{-\Upsilon N}$ given in Eq.~\eqref{eq:p0_Upsilon}.  
 Equivalently, the prefactors determining the scaling with $N$ obey $\eta_{\text{sra}}(u)\approx c_{\rm sra} u^{\zeta}$ and $\eta_{\text{uni}}(u)\approx \Upsilon + c_{\rm uni}u^{\zeta}$ at small $u$.

\begin{figure}[t!]
\centering
\includegraphics[width=1.0\linewidth]{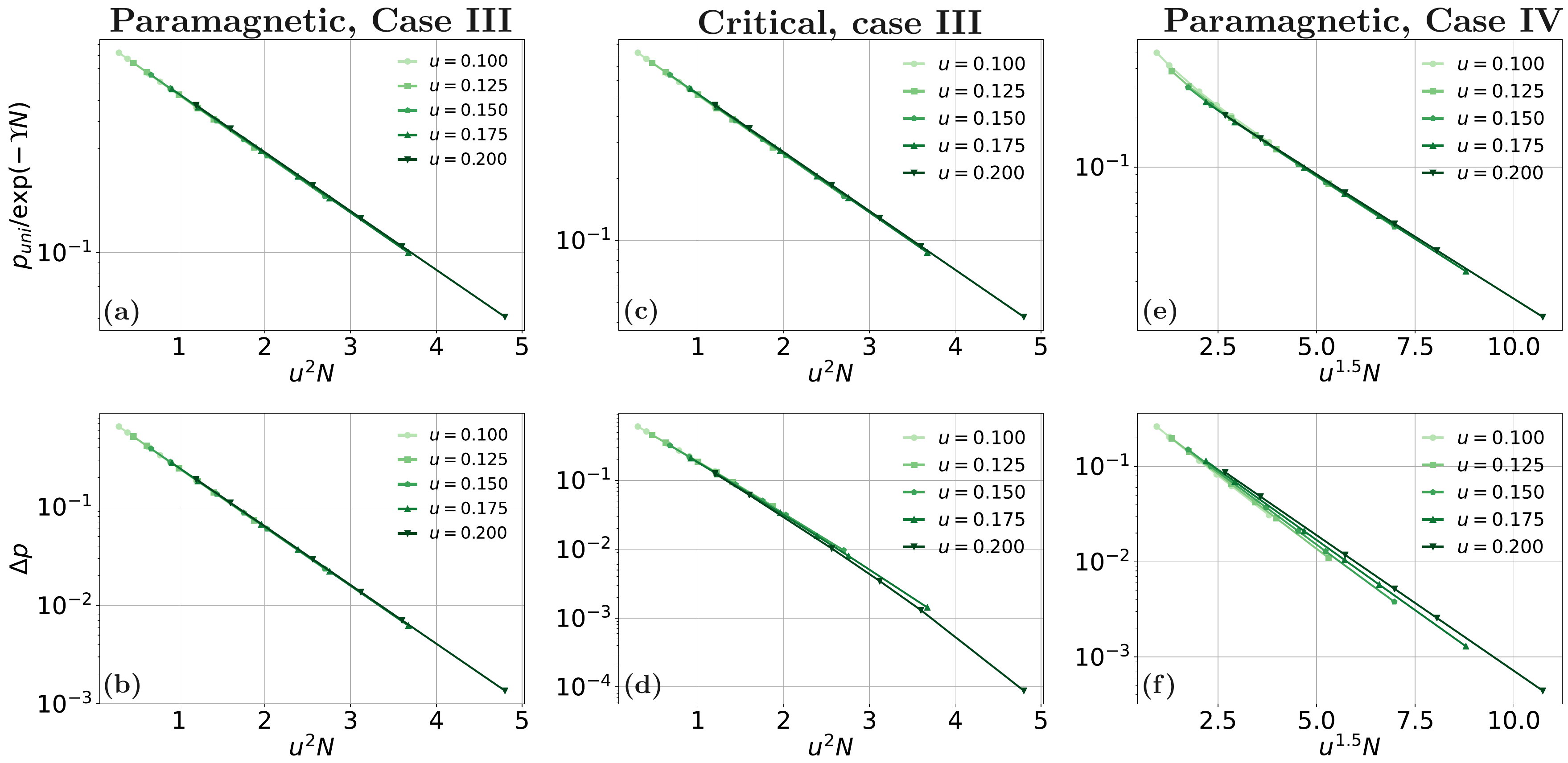}
\caption{(Color online). \textbf{Scaling of $\Delta p$ and $p_{\text{uni}}$.} Upper panels show that the ratio $p_{\rm uni}/p_{\rm uni}^{(0)}$ exhibits excellent data collapse when plotted versus $u^\zeta N$, with $\zeta$ exponents specified in the horizontal axes.  Lower panels show similar data collapse for $\Delta p$.   These results are consistent with scaling behavior $\Delta p\sim  e^{-c_{\rm sra} u^{\zeta}N}$ and $p_{\rm uni}/p^{(0)}_{\rm uni}\sim  e^{-c_{\rm uni} u^{\zeta}N}$. 
Data were obtained using finite DMRG with bond dimension $\chi=800$ and periodic boundary conditions.  Panels (e,f) for case IV correspond to $C = -1$.  } 
\label{fig:scaling}
\end{figure}

In agreement with the expressions above, Fig.~\ref{fig:scaling} illustrates data collapse of both $p_{\rm uni}/e^{-\Upsilon N}$ (upper row) and $\Delta p$ 
 (lower row) when plotted versus $u^\zeta N$, with exponents $\zeta$ indicated on the horizontal axes.  The range of $u$ and $N$ considered here are the same as those examined in the figures of Sec.~\ref{sec:evenoddB}. Panels (a,b) and (c,d) correspond to case III of Table~\ref{tab.unitaries} with paramagnetic ancilla and critical ancilla, respectively. Panels (e) and (f) correspond to case IV with $C=-1$ and paramagnetic ancilla.  The non-ideal data collapse in (f) possibly originates from neglected subleading corrections in $u$ (but in any event does not matter for the conclusions drawn in Sec.~\ref{sec:evenoddB}).  
In case IV with critical ancilla, we find good data collapse for  $p_{\text{uni}}/e^{-\Upsilon N}$ with $\zeta=0.25$ (data not shown); data collapse with our ansatz does not arise for $\Delta p$, however. 
The reason is that $\Delta p$ undergoes sign changes in this case as a function of $N$ for fixed $u$, thus no longer displaying monotonic behavior. 

\twocolumngrid

\begin{thebibliography}{100}
\makeatletter
\providecommand \@ifxundefined [1]{%
 \@ifx{#1\undefined}
}%
\providecommand \@ifnum [1]{%
 \ifnum #1\expandafter \@firstoftwo
 \else \expandafter \@secondoftwo
 \fi
}%
\providecommand \@ifx [1]{%
 \ifx #1\expandafter \@firstoftwo
 \else \expandafter \@secondoftwo
 \fi
}%
\providecommand \natexlab [1]{#1}%
\providecommand \enquote  [1]{``#1''}%
\providecommand \bibnamefont  [1]{#1}%
\providecommand \bibfnamefont [1]{#1}%
\providecommand \citenamefont [1]{#1}%
\providecommand \href@noop [0]{\@secondoftwo}%
\providecommand \href [0]{\begingroup \@sanitize@url \@href}%
\providecommand \@href[1]{\@@startlink{#1}\@@href}%
\providecommand \@@href[1]{\endgroup#1\@@endlink}%
\providecommand \@sanitize@url [0]{\catcode `\\12\catcode `\$12\catcode
  `\&12\catcode `\#12\catcode `\^12\catcode `\_12\catcode `\%12\relax}%
\providecommand \@@startlink[1]{}%
\providecommand \@@endlink[0]{}%
\providecommand \url  [0]{\begingroup\@sanitize@url \@url }%
\providecommand \@url [1]{\endgroup\@href {#1}{\urlprefix }}%
\providecommand \urlprefix  [0]{URL }%
\providecommand \Eprint [0]{\href }%
\providecommand \doibase [0]{https://doi.org/}%
\providecommand \selectlanguage [0]{\@gobble}%
\providecommand \bibinfo  [0]{\@secondoftwo}%
\providecommand \bibfield  [0]{\@secondoftwo}%
\providecommand \translation [1]{[#1]}%
\providecommand \BibitemOpen [0]{}%
\providecommand \bibitemStop [0]{}%
\providecommand \bibitemNoStop [0]{.\EOS\space}%
\providecommand \EOS [0]{\spacefactor3000\relax}%
\providecommand \BibitemShut  [1]{\csname bibitem#1\endcsname}%
\let\auto@bib@innerbib\@empty

\bibitem [{\citenamefont {Li}\ \emph {et~al.}(2018)\citenamefont {Li},
  \citenamefont {Chen},\ and\ \citenamefont {Fisher}}]{Li2018}%
  \BibitemOpen
  \bibfield  {author} {\bibinfo {author} {\bibfnamefont {Y.}~\bibnamefont
  {Li}}, \bibinfo {author} {\bibfnamefont {X.}~\bibnamefont {Chen}},\ and\
  \bibinfo {author} {\bibfnamefont {M.~P.~A.}\ \bibnamefont {Fisher}},\
  }\bibfield  {title} {\bibinfo {title} {Quantum zeno effect and the many-body
  entanglement transition},\ }\href
  {https://doi.org/10.1103/PhysRevB.98.205136} {\bibfield  {journal} {\bibinfo
  {journal} {Phys. Rev. B}\ }\textbf {\bibinfo {volume} {98}},\ \bibinfo
  {pages} {205136} (\bibinfo {year} {2018})}\BibitemShut {NoStop}%
\bibitem [{\citenamefont {Skinner}\ \emph {et~al.}(2019)\citenamefont
  {Skinner}, \citenamefont {Ruhman},\ and\ \citenamefont
  {Nahum}}]{skinner2019}%
  \BibitemOpen
  \bibfield  {author} {\bibinfo {author} {\bibfnamefont {B.}~\bibnamefont
  {Skinner}}, \bibinfo {author} {\bibfnamefont {J.}~\bibnamefont {Ruhman}},\
  and\ \bibinfo {author} {\bibfnamefont {A.}~\bibnamefont {Nahum}},\ }\bibfield
   {title} {\bibinfo {title} {Measurement-induced phase transitions in the
  dynamics of entanglement},\ }\href
  {https://doi.org/10.1103/PhysRevX.9.031009} {\bibfield  {journal} {\bibinfo
  {journal} {Phys. Rev. X}\ }\textbf {\bibinfo {volume} {9}},\ \bibinfo {pages}
  {031009} (\bibinfo {year} {2019})}\BibitemShut {NoStop}%
\bibitem [{\citenamefont {Li}\ \emph {et~al.}(2019)\citenamefont {Li},
  \citenamefont {Chen},\ and\ \citenamefont {Fisher}}]{Li2019}%
  \BibitemOpen
  \bibfield  {author} {\bibinfo {author} {\bibfnamefont {Y.}~\bibnamefont
  {Li}}, \bibinfo {author} {\bibfnamefont {X.}~\bibnamefont {Chen}},\ and\
  \bibinfo {author} {\bibfnamefont {M.~P.~A.}\ \bibnamefont {Fisher}},\
  }\bibfield  {title} {\bibinfo {title} {Measurement-driven entanglement
  transition in hybrid quantum circuits},\ }\href
  {https://doi.org/10.1103/PhysRevB.100.134306} {\bibfield  {journal} {\bibinfo
   {journal} {Phys. Rev. B}\ }\textbf {\bibinfo {volume} {100}},\ \bibinfo
  {pages} {134306} (\bibinfo {year} {2019})}\BibitemShut {NoStop}%
\bibitem [{\citenamefont {Chan}\ \emph {et~al.}(2019)\citenamefont {Chan},
  \citenamefont {Nandkishore}, \citenamefont {Pretko},\ and\ \citenamefont
  {Smith}}]{Chan2019}%
  \BibitemOpen
  \bibfield  {author} {\bibinfo {author} {\bibfnamefont {A.}~\bibnamefont
  {Chan}}, \bibinfo {author} {\bibfnamefont {R.~M.}\ \bibnamefont
  {Nandkishore}}, \bibinfo {author} {\bibfnamefont {M.}~\bibnamefont
  {Pretko}},\ and\ \bibinfo {author} {\bibfnamefont {G.}~\bibnamefont
  {Smith}},\ }\bibfield  {title} {\bibinfo {title} {Unitary-projective
  entanglement dynamics},\ }\href {https://doi.org/10.1103/PhysRevB.99.224307}
  {\bibfield  {journal} {\bibinfo  {journal} {Phys. Rev. B}\ }\textbf {\bibinfo
  {volume} {99}},\ \bibinfo {pages} {224307} (\bibinfo {year}
  {2019})}\BibitemShut {NoStop}%
\bibitem [{\citenamefont {Gullans}\ and\ \citenamefont
  {Huse}(2020{\natexlab{a}})}]{Gullans2020}%
  \BibitemOpen
  \bibfield  {author} {\bibinfo {author} {\bibfnamefont {M.~J.}\ \bibnamefont
  {Gullans}}\ and\ \bibinfo {author} {\bibfnamefont {D.~A.}\ \bibnamefont
  {Huse}},\ }\bibfield  {title} {\bibinfo {title} {Dynamical purification phase
  transition induced by quantum measurements},\ }\href
  {https://doi.org/10.1103/PhysRevX.10.041020} {\bibfield  {journal} {\bibinfo
  {journal} {Phys. Rev. X}\ }\textbf {\bibinfo {volume} {10}},\ \bibinfo
  {pages} {041020} (\bibinfo {year} {2020}{\natexlab{a}})}\BibitemShut
  {NoStop}%
\bibitem [{\citenamefont {Choi}\ \emph {et~al.}(2020)\citenamefont {Choi},
  \citenamefont {Bao}, \citenamefont {Qi},\ and\ \citenamefont
  {Altman}}]{choi2020}%
  \BibitemOpen
  \bibfield  {author} {\bibinfo {author} {\bibfnamefont {S.}~\bibnamefont
  {Choi}}, \bibinfo {author} {\bibfnamefont {Y.}~\bibnamefont {Bao}}, \bibinfo
  {author} {\bibfnamefont {X.-L.}\ \bibnamefont {Qi}},\ and\ \bibinfo {author}
  {\bibfnamefont {E.}~\bibnamefont {Altman}},\ }\bibfield  {title} {\bibinfo
  {title} {Quantum error correction in scrambling dynamics and
  measurement-induced phase transition},\ }\href
  {https://doi.org/10.1103/PhysRevLett.125.030505} {\bibfield  {journal}
  {\bibinfo  {journal} {Phys. Rev. Lett.}\ }\textbf {\bibinfo {volume} {125}},\
  \bibinfo {pages} {030505} (\bibinfo {year} {2020})}\BibitemShut {NoStop}%
\bibitem [{\citenamefont {Jian}\ \emph {et~al.}(2020)\citenamefont {Jian},
  \citenamefont {You}, \citenamefont {Vasseur},\ and\ \citenamefont
  {Ludwig}}]{Vasseur2020}%
  \BibitemOpen
  \bibfield  {author} {\bibinfo {author} {\bibfnamefont {C.-M.}\ \bibnamefont
  {Jian}}, \bibinfo {author} {\bibfnamefont {Y.-Z.}\ \bibnamefont {You}},
  \bibinfo {author} {\bibfnamefont {R.}~\bibnamefont {Vasseur}},\ and\ \bibinfo
  {author} {\bibfnamefont {A.~W.~W.}\ \bibnamefont {Ludwig}},\ }\bibfield
  {title} {\bibinfo {title} {Measurement-induced criticality in random quantum
  circuits},\ }\href {https://doi.org/10.1103/PhysRevB.101.104302} {\bibfield
  {journal} {\bibinfo  {journal} {Phys. Rev. B}\ }\textbf {\bibinfo {volume}
  {101}},\ \bibinfo {pages} {104302} (\bibinfo {year} {2020})}\BibitemShut
  {NoStop}%
\bibitem [{\citenamefont {Alberton}\ \emph {et~al.}(2021)\citenamefont
  {Alberton}, \citenamefont {Buchhold},\ and\ \citenamefont
  {Diehl}}]{Alberton2021}%
  \BibitemOpen
  \bibfield  {author} {\bibinfo {author} {\bibfnamefont {O.}~\bibnamefont
  {Alberton}}, \bibinfo {author} {\bibfnamefont {M.}~\bibnamefont {Buchhold}},\
  and\ \bibinfo {author} {\bibfnamefont {S.}~\bibnamefont {Diehl}},\ }\bibfield
   {title} {\bibinfo {title} {Entanglement transition in a monitored
  free-fermion chain: From extended criticality to area law},\ }\href
  {https://doi.org/10.1103/PhysRevLett.126.170602} {\bibfield  {journal}
  {\bibinfo  {journal} {Phys. Rev. Lett.}\ }\textbf {\bibinfo {volume} {126}},\
  \bibinfo {pages} {170602} (\bibinfo {year} {2021})}\BibitemShut {NoStop}%
\bibitem [{\citenamefont {Biella}\ and\ \citenamefont
  {Schir{\'{o}}}(2021)}]{Biella_2021}%
  \BibitemOpen
  \bibfield  {author} {\bibinfo {author} {\bibfnamefont {A.}~\bibnamefont
  {Biella}}\ and\ \bibinfo {author} {\bibfnamefont {M.}~\bibnamefont
  {Schir{\'{o}}}},\ }\bibfield  {title} {\bibinfo {title} {Many-body quantum
  zeno effect and measurement-induced subradiance transition},\ }\href
  {https://doi.org/10.22331/q-2021-08-19-528} {\bibfield  {journal} {\bibinfo
  {journal} {Quantum}\ }\textbf {\bibinfo {volume} {5}},\ \bibinfo {pages}
  {528} (\bibinfo {year} {2021})}\BibitemShut {NoStop}%
\bibitem [{\citenamefont {Turkeshi}\ \emph
  {et~al.}(2022{\natexlab{a}})\citenamefont {Turkeshi}, \citenamefont
  {Piroli},\ and\ \citenamefont {Schir\'o}}]{Turkeshi2022}%
  \BibitemOpen
  \bibfield  {author} {\bibinfo {author} {\bibfnamefont {X.}~\bibnamefont
  {Turkeshi}}, \bibinfo {author} {\bibfnamefont {L.}~\bibnamefont {Piroli}},\
  and\ \bibinfo {author} {\bibfnamefont {M.}~\bibnamefont {Schir\'o}},\
  }\bibfield  {title} {\bibinfo {title} {Enhanced entanglement negativity in
  boundary-driven monitored fermionic chains},\ }\href
  {https://doi.org/10.1103/PhysRevB.106.024304} {\bibfield  {journal} {\bibinfo
   {journal} {Phys. Rev. B}\ }\textbf {\bibinfo {volume} {106}},\ \bibinfo
  {pages} {024304} (\bibinfo {year} {2022}{\natexlab{a}})}\BibitemShut
  {NoStop}%
\bibitem [{\citenamefont {Cao}\ \emph {et~al.}(2019)\citenamefont {Cao},
  \citenamefont {Tilloy},\ and\ \citenamefont {Luca}}]{deluca}%
  \BibitemOpen
  \bibfield  {author} {\bibinfo {author} {\bibfnamefont {X.}~\bibnamefont
  {Cao}}, \bibinfo {author} {\bibfnamefont {A.}~\bibnamefont {Tilloy}},\ and\
  \bibinfo {author} {\bibfnamefont {A.~D.}\ \bibnamefont {Luca}},\ }\bibfield
  {title} {\bibinfo {title} {{Entanglement in a fermion chain under continuous
  monitoring}},\ }\href {https://doi.org/10.21468/SciPostPhys.7.2.024}
  {\bibfield  {journal} {\bibinfo  {journal} {SciPost Phys.}\ }\textbf
  {\bibinfo {volume} {7}},\ \bibinfo {pages} {024} (\bibinfo {year}
  {2019})}\BibitemShut {NoStop}%
\bibitem [{\citenamefont {Gullans}\ and\ \citenamefont
  {Huse}(2020{\natexlab{b}})}]{Gullans_2020}%
  \BibitemOpen
  \bibfield  {author} {\bibinfo {author} {\bibfnamefont {M.~J.}\ \bibnamefont
  {Gullans}}\ and\ \bibinfo {author} {\bibfnamefont {D.~A.}\ \bibnamefont
  {Huse}},\ }\bibfield  {title} {\bibinfo {title} {Scalable probes of
  measurement-induced criticality},\ }\href
  {https://doi.org/10.1103/physrevlett.125.070606} {\bibfield  {journal}
  {\bibinfo  {journal} {Phys. Rev. Lett.}\ }\textbf {\bibinfo {volume} {125}},\
  \bibinfo {pages} {070606} (\bibinfo {year} {2020}{\natexlab{b}})}\BibitemShut
  {NoStop}%
\bibitem [{\citenamefont {Bao}\ \emph {et~al.}(2020)\citenamefont {Bao},
  \citenamefont {Choi},\ and\ \citenamefont {Altman}}]{bao2020}%
  \BibitemOpen
  \bibfield  {author} {\bibinfo {author} {\bibfnamefont {Y.}~\bibnamefont
  {Bao}}, \bibinfo {author} {\bibfnamefont {S.}~\bibnamefont {Choi}},\ and\
  \bibinfo {author} {\bibfnamefont {E.}~\bibnamefont {Altman}},\ }\bibfield
  {title} {\bibinfo {title} {Theory of the phase transition in random unitary
  circuits with measurements},\ }\href
  {https://doi.org/10.1103/PhysRevB.101.104301} {\bibfield  {journal} {\bibinfo
   {journal} {Phys. Rev. B}\ }\textbf {\bibinfo {volume} {101}},\ \bibinfo
  {pages} {104301} (\bibinfo {year} {2020})}\BibitemShut {NoStop}%
\bibitem [{\citenamefont {Boorman}\ \emph {et~al.}(2022)\citenamefont
  {Boorman}, \citenamefont {Szyniszewski}, \citenamefont {Schomerus},\ and\
  \citenamefont {Romito}}]{Boorman}%
  \BibitemOpen
  \bibfield  {author} {\bibinfo {author} {\bibfnamefont {T.}~\bibnamefont
  {Boorman}}, \bibinfo {author} {\bibfnamefont {M.}~\bibnamefont
  {Szyniszewski}}, \bibinfo {author} {\bibfnamefont {H.}~\bibnamefont
  {Schomerus}},\ and\ \bibinfo {author} {\bibfnamefont {A.}~\bibnamefont
  {Romito}},\ }\bibfield  {title} {\bibinfo {title} {Diagnostics of
  entanglement dynamics in noisy and disordered spin chains via the
  measurement-induced steady-state entanglement transition},\ }\href
  {https://doi.org/10.1103/PhysRevB.105.144202} {\bibfield  {journal} {\bibinfo
   {journal} {Phys. Rev. B}\ }\textbf {\bibinfo {volume} {105}},\ \bibinfo
  {pages} {144202} (\bibinfo {year} {2022})}\BibitemShut {NoStop}%
\bibitem [{\citenamefont {Fan}\ \emph {et~al.}(2021)\citenamefont {Fan},
  \citenamefont {Vijay}, \citenamefont {Vishwanath},\ and\ \citenamefont
  {You}}]{fan2021}%
  \BibitemOpen
  \bibfield  {author} {\bibinfo {author} {\bibfnamefont {R.}~\bibnamefont
  {Fan}}, \bibinfo {author} {\bibfnamefont {S.}~\bibnamefont {Vijay}}, \bibinfo
  {author} {\bibfnamefont {A.}~\bibnamefont {Vishwanath}},\ and\ \bibinfo
  {author} {\bibfnamefont {Y.-Z.}\ \bibnamefont {You}},\ }\bibfield  {title}
  {\bibinfo {title} {Self-organized error correction in random unitary circuits
  with measurement},\ }\href {https://doi.org/10.1103/PhysRevB.103.174309}
  {\bibfield  {journal} {\bibinfo  {journal} {Phys. Rev. B}\ }\textbf {\bibinfo
  {volume} {103}},\ \bibinfo {pages} {174309} (\bibinfo {year}
  {2021})}\BibitemShut {NoStop}%
\bibitem [{\citenamefont {Bentsen}\ \emph {et~al.}(2021)\citenamefont
  {Bentsen}, \citenamefont {Sahu},\ and\ \citenamefont {Swingle}}]{swingle}%
  \BibitemOpen
  \bibfield  {author} {\bibinfo {author} {\bibfnamefont {G.~S.}\ \bibnamefont
  {Bentsen}}, \bibinfo {author} {\bibfnamefont {S.}~\bibnamefont {Sahu}},\ and\
  \bibinfo {author} {\bibfnamefont {B.}~\bibnamefont {Swingle}},\ }\bibfield
  {title} {\bibinfo {title} {Measurement-induced purification in large-$n$
  hybrid brownian circuits},\ }\href
  {https://doi.org/10.1103/PhysRevB.104.094304} {\bibfield  {journal} {\bibinfo
   {journal} {Phys. Rev. B}\ }\textbf {\bibinfo {volume} {104}},\ \bibinfo
  {pages} {094304} (\bibinfo {year} {2021})}\BibitemShut {NoStop}%
\bibitem [{\citenamefont {Li}\ \emph {et~al.}(2021)\citenamefont {Li},
  \citenamefont {Chen}, \citenamefont {Ludwig},\ and\ \citenamefont
  {Fisher}}]{Li2021}%
  \BibitemOpen
  \bibfield  {author} {\bibinfo {author} {\bibfnamefont {Y.}~\bibnamefont
  {Li}}, \bibinfo {author} {\bibfnamefont {X.}~\bibnamefont {Chen}}, \bibinfo
  {author} {\bibfnamefont {A.~W.~W.}\ \bibnamefont {Ludwig}},\ and\ \bibinfo
  {author} {\bibfnamefont {M.~P.~A.}\ \bibnamefont {Fisher}},\ }\bibfield
  {title} {\bibinfo {title} {Conformal invariance and quantum nonlocality in
  critical hybrid circuits},\ }\href
  {https://doi.org/10.1103/PhysRevB.104.104305} {\bibfield  {journal} {\bibinfo
   {journal} {Phys. Rev. B}\ }\textbf {\bibinfo {volume} {104}},\ \bibinfo
  {pages} {104305} (\bibinfo {year} {2021})}\BibitemShut {NoStop}%
\bibitem [{\citenamefont {Friedman}\ \emph {et~al.}(2022)\citenamefont
  {Friedman}, \citenamefont {Yin}, \citenamefont {Hong},\ and\ \citenamefont
  {Lucas}}]{Friedman}%
  \BibitemOpen
  \bibfield  {author} {\bibinfo {author} {\bibfnamefont {A.~J.}\ \bibnamefont
  {Friedman}}, \bibinfo {author} {\bibfnamefont {C.}~\bibnamefont {Yin}},
  \bibinfo {author} {\bibfnamefont {Y.}~\bibnamefont {Hong}},\ and\ \bibinfo
  {author} {\bibfnamefont {A.}~\bibnamefont {Lucas}},\ }\href@noop {} {\bibinfo
  {title} {Locality and error correction in quantum dynamics with measurement}}
  (\bibinfo {year} {2022}),\ \Eprint {https://arxiv.org/abs/2206.09929}
  {arXiv:2206.09929 [quantum-ph]} \BibitemShut {NoStop}%
\bibitem [{\citenamefont {Turkeshi}\ \emph {et~al.}(2020)\citenamefont
  {Turkeshi}, \citenamefont {Fazio},\ and\ \citenamefont
  {Dalmonte}}]{Turkeshi1}%
  \BibitemOpen
  \bibfield  {author} {\bibinfo {author} {\bibfnamefont {X.}~\bibnamefont
  {Turkeshi}}, \bibinfo {author} {\bibfnamefont {R.}~\bibnamefont {Fazio}},\
  and\ \bibinfo {author} {\bibfnamefont {M.}~\bibnamefont {Dalmonte}},\
  }\bibfield  {title} {\bibinfo {title} {Measurement-induced criticality in
  $(2+1)$-dimensional hybrid quantum circuits},\ }\href
  {https://doi.org/10.1103/PhysRevB.102.014315} {\bibfield  {journal} {\bibinfo
   {journal} {Phys. Rev. B}\ }\textbf {\bibinfo {volume} {102}},\ \bibinfo
  {pages} {014315} (\bibinfo {year} {2020})}\BibitemShut {NoStop}%
\bibitem [{\citenamefont {Turkeshi}\ \emph {et~al.}(2021)\citenamefont
  {Turkeshi}, \citenamefont {Biella}, \citenamefont {Fazio}, \citenamefont
  {Dalmonte},\ and\ \citenamefont {Schir\'o}}]{Turkeshi2}%
  \BibitemOpen
  \bibfield  {author} {\bibinfo {author} {\bibfnamefont {X.}~\bibnamefont
  {Turkeshi}}, \bibinfo {author} {\bibfnamefont {A.}~\bibnamefont {Biella}},
  \bibinfo {author} {\bibfnamefont {R.}~\bibnamefont {Fazio}}, \bibinfo
  {author} {\bibfnamefont {M.}~\bibnamefont {Dalmonte}},\ and\ \bibinfo
  {author} {\bibfnamefont {M.}~\bibnamefont {Schir\'o}},\ }\bibfield  {title}
  {\bibinfo {title} {Measurement-induced entanglement transitions in the
  quantum ising chain: From infinite to zero clicks},\ }\href
  {https://doi.org/10.1103/PhysRevB.103.224210} {\bibfield  {journal} {\bibinfo
   {journal} {Phys. Rev. B}\ }\textbf {\bibinfo {volume} {103}},\ \bibinfo
  {pages} {224210} (\bibinfo {year} {2021})}\BibitemShut {NoStop}%
\bibitem [{\citenamefont {M\"uller}\ \emph {et~al.}(2022)\citenamefont
  {M\"uller}, \citenamefont {Diehl},\ and\ \citenamefont {Buchhold}}]{muller}%
  \BibitemOpen
  \bibfield  {author} {\bibinfo {author} {\bibfnamefont {T.}~\bibnamefont
  {M\"uller}}, \bibinfo {author} {\bibfnamefont {S.}~\bibnamefont {Diehl}},\
  and\ \bibinfo {author} {\bibfnamefont {M.}~\bibnamefont {Buchhold}},\
  }\bibfield  {title} {\bibinfo {title} {Measurement-induced dark state phase
  transitions in long-ranged fermion systems},\ }\href
  {https://doi.org/10.1103/PhysRevLett.128.010605} {\bibfield  {journal}
  {\bibinfo  {journal} {Phys. Rev. Lett.}\ }\textbf {\bibinfo {volume} {128}},\
  \bibinfo {pages} {010605} (\bibinfo {year} {2022})}\BibitemShut {NoStop}%
\bibitem [{\citenamefont {Lavasani}\ \emph {et~al.}(2021)\citenamefont
  {Lavasani}, \citenamefont {Y.},\ and\ \citenamefont
  {Maissam~Barkeshli}}]{Lavasani}%
  \BibitemOpen
  \bibfield  {author} {\bibinfo {author} {\bibfnamefont {A.}~\bibnamefont
  {Lavasani}}, \bibinfo {author} {\bibfnamefont {A.}~\bibnamefont {Y.}},\ and\
  \bibinfo {author} {\bibfnamefont {M.}~\bibnamefont {Maissam~Barkeshli}},\
  }\bibfield  {title} {\bibinfo {title} {Measurement-induced topological
  entanglement transitions in symmetric random quantum circuits},\ }\href
  {https://www.nature.com/articles/s41567-022-01619-7} {\bibfield  {journal}
  {\bibinfo  {journal} {Nat.Phys.}\ }\textbf {\bibinfo {volume} {17}},\
  \bibinfo {pages} {342–347} (\bibinfo {year} {2021})}\BibitemShut {NoStop}%
\bibitem [{\citenamefont {Sang}\ and\ \citenamefont {Hsieh}(2021)}]{Hsieh3}%
  \BibitemOpen
  \bibfield  {author} {\bibinfo {author} {\bibfnamefont {S.}~\bibnamefont
  {Sang}}\ and\ \bibinfo {author} {\bibfnamefont {T.~H.}\ \bibnamefont
  {Hsieh}},\ }\bibfield  {title} {\bibinfo {title} {Measurement-protected
  quantum phases},\ }\href {https://doi.org/10.1103/PhysRevResearch.3.023200}
  {\bibfield  {journal} {\bibinfo  {journal} {Phys. Rev. Res.}\ }\textbf
  {\bibinfo {volume} {3}},\ \bibinfo {pages} {023200} (\bibinfo {year}
  {2021})}\BibitemShut {NoStop}%
\bibitem [{\citenamefont {Bao}\ \emph {et~al.}(2021)\citenamefont {Bao},
  \citenamefont {Choi},\ and\ \citenamefont {Altman}}]{BAO2021}%
  \BibitemOpen
  \bibfield  {author} {\bibinfo {author} {\bibfnamefont {Y.}~\bibnamefont
  {Bao}}, \bibinfo {author} {\bibfnamefont {S.}~\bibnamefont {Choi}},\ and\
  \bibinfo {author} {\bibfnamefont {E.}~\bibnamefont {Altman}},\ }\bibfield
  {title} {\bibinfo {title} {Symmetry enriched phases of quantum circuits},\
  }\href {https://doi.org/https://doi.org/10.1016/j.aop.2021.168618} {\bibfield
   {journal} {\bibinfo  {journal} {Ann. of Phys.}\ }\textbf {\bibinfo {volume}
  {435}},\ \bibinfo {pages} {168618} (\bibinfo {year} {2021})}\BibitemShut
  {NoStop}%
\bibitem [{\citenamefont {Van~Regemortel}\ \emph {et~al.}(2021)\citenamefont
  {Van~Regemortel}, \citenamefont {Cian}, \citenamefont {Seif}, \citenamefont
  {Dehghani},\ and\ \citenamefont {Hafezi}}]{Regemortel}%
  \BibitemOpen
  \bibfield  {author} {\bibinfo {author} {\bibfnamefont {M.}~\bibnamefont
  {Van~Regemortel}}, \bibinfo {author} {\bibfnamefont {Z.-P.}\ \bibnamefont
  {Cian}}, \bibinfo {author} {\bibfnamefont {A.}~\bibnamefont {Seif}}, \bibinfo
  {author} {\bibfnamefont {H.}~\bibnamefont {Dehghani}},\ and\ \bibinfo
  {author} {\bibfnamefont {M.}~\bibnamefont {Hafezi}},\ }\bibfield  {title}
  {\bibinfo {title} {Entanglement entropy scaling transition under competing
  monitoring protocols},\ }\href
  {https://doi.org/10.1103/PhysRevLett.126.123604} {\bibfield  {journal}
  {\bibinfo  {journal} {Phys. Rev. Lett.}\ }\textbf {\bibinfo {volume} {126}},\
  \bibinfo {pages} {123604} (\bibinfo {year} {2021})}\BibitemShut {NoStop}%
\bibitem [{\citenamefont {Ippoliti}\ \emph {et~al.}(2021)\citenamefont
  {Ippoliti}, \citenamefont {Gullans}, \citenamefont {Gopalakrishnan},
  \citenamefont {Huse},\ and\ \citenamefont {Khemani}}]{ippoliti_2021}%
  \BibitemOpen
  \bibfield  {author} {\bibinfo {author} {\bibfnamefont {M.}~\bibnamefont
  {Ippoliti}}, \bibinfo {author} {\bibfnamefont {M.~J.}\ \bibnamefont
  {Gullans}}, \bibinfo {author} {\bibfnamefont {S.}~\bibnamefont
  {Gopalakrishnan}}, \bibinfo {author} {\bibfnamefont {D.~A.}\ \bibnamefont
  {Huse}},\ and\ \bibinfo {author} {\bibfnamefont {V.}~\bibnamefont
  {Khemani}},\ }\bibfield  {title} {\bibinfo {title} {Entanglement phase
  transitions in measurement-only dynamics},\ }\href
  {https://doi.org/10.1103/PhysRevX.11.011030} {\bibfield  {journal} {\bibinfo
  {journal} {Phys. Rev. X}\ }\textbf {\bibinfo {volume} {11}},\ \bibinfo
  {pages} {011030} (\bibinfo {year} {2021})}\BibitemShut {NoStop}%
\bibitem [{\citenamefont {Turkeshi}\ \emph
  {et~al.}(2022{\natexlab{b}})\citenamefont {Turkeshi}, \citenamefont
  {Dalmonte}, \citenamefont {Fazio},\ and\ \citenamefont {Schir\`o}}]{xhek1}%
  \BibitemOpen
  \bibfield  {author} {\bibinfo {author} {\bibfnamefont {X.}~\bibnamefont
  {Turkeshi}}, \bibinfo {author} {\bibfnamefont {M.}~\bibnamefont {Dalmonte}},
  \bibinfo {author} {\bibfnamefont {R.}~\bibnamefont {Fazio}},\ and\ \bibinfo
  {author} {\bibfnamefont {M.}~\bibnamefont {Schir\`o}},\ }\bibfield  {title}
  {\bibinfo {title} {Entanglement transitions from stochastic resetting of
  non-hermitian quasiparticles},\ }\href
  {https://doi.org/10.1103/PhysRevB.105.L241114} {\bibfield  {journal}
  {\bibinfo  {journal} {Phys. Rev. B}\ }\textbf {\bibinfo {volume} {105}},\
  \bibinfo {pages} {L241114} (\bibinfo {year}
  {2022}{\natexlab{b}})}\BibitemShut {NoStop}%
\bibitem [{\citenamefont {Turkeshi}\ and\ \citenamefont
  {Schir\'o}(2023)}]{xhek2}%
  \BibitemOpen
  \bibfield  {author} {\bibinfo {author} {\bibfnamefont {X.}~\bibnamefont
  {Turkeshi}}\ and\ \bibinfo {author} {\bibfnamefont {M.}~\bibnamefont
  {Schir\'o}},\ }\bibfield  {title} {\bibinfo {title} {Entanglement and
  correlation spreading in non-hermitian spin chains},\ }\href
  {https://doi.org/10.1103/PhysRevB.107.L020403} {\bibfield  {journal}
  {\bibinfo  {journal} {Phys. Rev. B}\ }\textbf {\bibinfo {volume} {107}},\
  \bibinfo {pages} {L020403} (\bibinfo {year} {2023})}\BibitemShut {NoStop}%
\bibitem [{\citenamefont {Sierant}\ \emph {et~al.}(2022)\citenamefont
  {Sierant}, \citenamefont {Chiriacò}, \citenamefont {Surace}, \citenamefont
  {Sharma}, \citenamefont {Turkeshi}, \citenamefont {Dalmonte}, \citenamefont
  {Fazio},\ and\ \citenamefont {Pagano}}]{federica}%
  \BibitemOpen
  \bibfield  {author} {\bibinfo {author} {\bibfnamefont {P.}~\bibnamefont
  {Sierant}}, \bibinfo {author} {\bibfnamefont {G.}~\bibnamefont {Chiriacò}},
  \bibinfo {author} {\bibfnamefont {F.~M.}\ \bibnamefont {Surace}}, \bibinfo
  {author} {\bibfnamefont {S.}~\bibnamefont {Sharma}}, \bibinfo {author}
  {\bibfnamefont {X.}~\bibnamefont {Turkeshi}}, \bibinfo {author}
  {\bibfnamefont {M.}~\bibnamefont {Dalmonte}}, \bibinfo {author}
  {\bibfnamefont {R.}~\bibnamefont {Fazio}},\ and\ \bibinfo {author}
  {\bibfnamefont {G.}~\bibnamefont {Pagano}},\ }\bibfield  {title} {\bibinfo
  {title} {Dissipative floquet dynamics: from steady state to measurement
  induced criticality in trapped-ion chains},\ }\href
  {https://doi.org/10.22331/q-2022-02-02-638} {\bibfield  {journal} {\bibinfo
  {journal} {Quantum}\ }\textbf {\bibinfo {volume} {6}},\ \bibinfo {pages}
  {638} (\bibinfo {year} {2022})}\BibitemShut {NoStop}%
\bibitem [{\citenamefont {Coppola}\ \emph {et~al.}(2022)\citenamefont
  {Coppola}, \citenamefont {Tirrito}, \citenamefont {Karevski},\ and\
  \citenamefont {Collura}}]{mario2}%
  \BibitemOpen
  \bibfield  {author} {\bibinfo {author} {\bibfnamefont {M.}~\bibnamefont
  {Coppola}}, \bibinfo {author} {\bibfnamefont {E.}~\bibnamefont {Tirrito}},
  \bibinfo {author} {\bibfnamefont {D.}~\bibnamefont {Karevski}},\ and\
  \bibinfo {author} {\bibfnamefont {M.}~\bibnamefont {Collura}},\ }\bibfield
  {title} {\bibinfo {title} {Growth of entanglement entropy under local
  projective measurements},\ }\href
  {https://doi.org/10.1103/PhysRevB.105.094303} {\bibfield  {journal} {\bibinfo
   {journal} {Phys. Rev. B}\ }\textbf {\bibinfo {volume} {105}},\ \bibinfo
  {pages} {094303} (\bibinfo {year} {2022})}\BibitemShut {NoStop}%
\bibitem [{\citenamefont {Kelly}\ \emph {et~al.}(2022)\citenamefont {Kelly},
  \citenamefont {Poschinger}, \citenamefont {Schmidt-Kaler}, \citenamefont
  {Fisher},\ and\ \citenamefont {Marino}}]{shane1}%
  \BibitemOpen
  \bibfield  {author} {\bibinfo {author} {\bibfnamefont {S.~P.}\ \bibnamefont
  {Kelly}}, \bibinfo {author} {\bibfnamefont {U.}~\bibnamefont {Poschinger}},
  \bibinfo {author} {\bibfnamefont {F.}~\bibnamefont {Schmidt-Kaler}}, \bibinfo
  {author} {\bibfnamefont {M.~P.~A.}\ \bibnamefont {Fisher}},\ and\ \bibinfo
  {author} {\bibfnamefont {J.}~\bibnamefont {Marino}},\ }\href@noop {}
  {\bibinfo {title} {Coherence requirements for quantum communication from
  hybrid circuit dynamics}} (\bibinfo {year} {2022}),\ \Eprint
  {https://arxiv.org/abs/2210.11547} {arXiv:2210.11547 [quant-ph]} \BibitemShut
  {NoStop}%
\bibitem [{\citenamefont {Piroli}\ \emph {et~al.}(2021)\citenamefont {Piroli},
  \citenamefont {Styliaris},\ and\ \citenamefont {Cirac}}]{piroli2021}%
  \BibitemOpen
  \bibfield  {author} {\bibinfo {author} {\bibfnamefont {L.}~\bibnamefont
  {Piroli}}, \bibinfo {author} {\bibfnamefont {G.}~\bibnamefont {Styliaris}},\
  and\ \bibinfo {author} {\bibfnamefont {J.~I.}\ \bibnamefont {Cirac}},\
  }\bibfield  {title} {\bibinfo {title} {Quantum circuits assisted by local
  operations and classical communication: Transformations and phases of
  matter},\ }\href {https://doi.org/10.1103/PhysRevLett.127.220503} {\bibfield
  {journal} {\bibinfo  {journal} {Phys. Rev. Lett.}\ }\textbf {\bibinfo
  {volume} {127}},\ \bibinfo {pages} {220503} (\bibinfo {year}
  {2021})}\BibitemShut {NoStop}%
\bibitem [{\citenamefont {Verresen}\ \emph {et~al.}(2021)\citenamefont
  {Verresen}, \citenamefont {Tantivasadakarn},\ and\ \citenamefont
  {Vishwanath}}]{Verresen}%
  \BibitemOpen
  \bibfield  {author} {\bibinfo {author} {\bibfnamefont {R.}~\bibnamefont
  {Verresen}}, \bibinfo {author} {\bibfnamefont {N.}~\bibnamefont
  {Tantivasadakarn}},\ and\ \bibinfo {author} {\bibfnamefont {A.}~\bibnamefont
  {Vishwanath}},\ }\href@noop {} {\bibinfo {title} {Efficiently preparing
  schroedinger's cat, fractons and non-abelian topological order in quantum
  devices}} (\bibinfo {year} {2021}),\ \Eprint
  {https://arxiv.org/abs/2112.03061} {arXiv:2112.03061 [quant-ph]} \BibitemShut
  {NoStop}%
\bibitem [{\citenamefont {Tantivasadakarn}\ \emph {et~al.}(2021)\citenamefont
  {Tantivasadakarn}, \citenamefont {Thorngren}, \citenamefont {Vishwanath},\
  and\ \citenamefont {Verresen}}]{nat21}%
  \BibitemOpen
  \bibfield  {author} {\bibinfo {author} {\bibfnamefont {N.}~\bibnamefont
  {Tantivasadakarn}}, \bibinfo {author} {\bibfnamefont {R.}~\bibnamefont
  {Thorngren}}, \bibinfo {author} {\bibfnamefont {A.}~\bibnamefont
  {Vishwanath}},\ and\ \bibinfo {author} {\bibfnamefont {R.}~\bibnamefont
  {Verresen}},\ }\href@noop {} {\bibinfo {title} {Long-range entanglement from
  measuring symmetry-protected topological phases}} (\bibinfo {year} {2021}),\
  \Eprint {https://arxiv.org/abs/2112.01519} {arXiv:2112.01519 [quant-ph]}
  \BibitemShut {NoStop}%
\bibitem [{\citenamefont {Tantivasadakarn}\ \emph {et~al.}(2022)\citenamefont
  {Tantivasadakarn}, \citenamefont {Vishwanath},\ and\ \citenamefont
  {Verresen}}]{nat22}%
  \BibitemOpen
  \bibfield  {author} {\bibinfo {author} {\bibfnamefont {N.}~\bibnamefont
  {Tantivasadakarn}}, \bibinfo {author} {\bibfnamefont {A.}~\bibnamefont
  {Vishwanath}},\ and\ \bibinfo {author} {\bibfnamefont {R.}~\bibnamefont
  {Verresen}},\ }\href@noop {} {\bibinfo {title} {A hierarchy of topological
  order from finite-depth unitaries, measurement and feedforward}} (\bibinfo
  {year} {2022}),\ \Eprint {https://arxiv.org/abs/2209.06202} {arXiv:2209.06202
  [quant-ph]} \BibitemShut {NoStop}%
\bibitem [{\citenamefont {Lu}\ \emph {et~al.}(2022)\citenamefont {Lu},
  \citenamefont {Lessa}, \citenamefont {Kim},\ and\ \citenamefont
  {Hsieh}}]{Lu22}%
  \BibitemOpen
  \bibfield  {author} {\bibinfo {author} {\bibfnamefont {T.-C.}\ \bibnamefont
  {Lu}}, \bibinfo {author} {\bibfnamefont {L.~A.}\ \bibnamefont {Lessa}},
  \bibinfo {author} {\bibfnamefont {I.~H.}\ \bibnamefont {Kim}},\ and\ \bibinfo
  {author} {\bibfnamefont {T.~H.}\ \bibnamefont {Hsieh}},\ }\href@noop {}
  {\bibinfo {title} {Measurement as a shortcut to long-range entangled quantum
  matter}} (\bibinfo {year} {2022}),\ \Eprint
  {https://arxiv.org/abs/2206.13527} {arXiv:2206.13527 [quant-ph]} \BibitemShut
  {NoStop}%
\bibitem [{\citenamefont {Bravyi}\ \emph {et~al.}(2022)\citenamefont {Bravyi},
  \citenamefont {Kim}, \citenamefont {Kliesch},\ and\ \citenamefont
  {Koenig}}]{bravyi_2022}%
  \BibitemOpen
  \bibfield  {author} {\bibinfo {author} {\bibfnamefont {S.}~\bibnamefont
  {Bravyi}}, \bibinfo {author} {\bibfnamefont {I.}~\bibnamefont {Kim}},
  \bibinfo {author} {\bibfnamefont {A.}~\bibnamefont {Kliesch}},\ and\ \bibinfo
  {author} {\bibfnamefont {R.}~\bibnamefont {Koenig}},\ }\href@noop {}
  {\bibinfo {title} {Adaptive constant-depth circuits for manipulating
  non-abelian anyons}} (\bibinfo {year} {2022}),\ \Eprint
  {https://arxiv.org/abs/2205.01933} {arXiv:2205.01933 [quant-ph]} \BibitemShut
  {NoStop}%
\bibitem [{\citenamefont {Zhu}\ \emph {et~al.}(2022)\citenamefont {Zhu},
  \citenamefont {Tantivasadakarn}, \citenamefont {Vishwanath}, \citenamefont
  {Trebst},\ and\ \citenamefont {Verresen}}]{Zhu22}%
  \BibitemOpen
  \bibfield  {author} {\bibinfo {author} {\bibfnamefont {G.-Y.}\ \bibnamefont
  {Zhu}}, \bibinfo {author} {\bibfnamefont {N.}~\bibnamefont
  {Tantivasadakarn}}, \bibinfo {author} {\bibfnamefont {A.}~\bibnamefont
  {Vishwanath}}, \bibinfo {author} {\bibfnamefont {S.}~\bibnamefont {Trebst}},\
  and\ \bibinfo {author} {\bibfnamefont {R.}~\bibnamefont {Verresen}},\
  }\href@noop {} {\bibinfo {title} {Nishimori's cat: stable long-range
  entanglement from finite-depth unitaries and weak measurements}} (\bibinfo
  {year} {2022}),\ \Eprint {https://arxiv.org/abs/2208.11136} {arXiv:2208.11136
  [quant-ph]} \BibitemShut {NoStop}%
\bibitem [{\citenamefont {Lee}\ \emph {et~al.}(2022)\citenamefont {Lee},
  \citenamefont {Ji}, \citenamefont {Bi},\ and\ \citenamefont
  {Fisher}}]{leeji2022}%
  \BibitemOpen
  \bibfield  {author} {\bibinfo {author} {\bibfnamefont {J.~Y.}\ \bibnamefont
  {Lee}}, \bibinfo {author} {\bibfnamefont {W.}~\bibnamefont {Ji}}, \bibinfo
  {author} {\bibfnamefont {Z.}~\bibnamefont {Bi}},\ and\ \bibinfo {author}
  {\bibfnamefont {M.~P.~A.}\ \bibnamefont {Fisher}},\ }\href@noop {} {\bibinfo
  {title} {Decoding measurement-prepared quantum phases and transitions: from
  ising model to gauge theory, and beyond}} (\bibinfo {year} {2022}),\ \Eprint
  {https://arxiv.org/abs/2208.11699} {arXiv:2208.11699 [cond-mat.str-el]}
  \BibitemShut {NoStop}%
\bibitem [{\citenamefont {Garc\'{\i}a-Pintos}\ \emph
  {et~al.}(2019)\citenamefont {Garc\'{\i}a-Pintos}, \citenamefont {Tielas},\
  and\ \citenamefont {del Campo}}]{delcampo}%
  \BibitemOpen
  \bibfield  {author} {\bibinfo {author} {\bibfnamefont {L.~P.}\ \bibnamefont
  {Garc\'{\i}a-Pintos}}, \bibinfo {author} {\bibfnamefont {D.}~\bibnamefont
  {Tielas}},\ and\ \bibinfo {author} {\bibfnamefont {A.}~\bibnamefont {del
  Campo}},\ }\bibfield  {title} {\bibinfo {title} {Spontaneous symmetry
  breaking induced by quantum monitoring},\ }\href
  {https://doi.org/10.1103/PhysRevLett.123.090403} {\bibfield  {journal}
  {\bibinfo  {journal} {Phys. Rev. Lett.}\ }\textbf {\bibinfo {volume} {123}},\
  \bibinfo {pages} {090403} (\bibinfo {year} {2019})}\BibitemShut {NoStop}%
\bibitem [{\citenamefont {Noel}\ \emph {et~al.}(2022)\citenamefont {Noel},
  \citenamefont {Niroula}, \citenamefont {Zhu}, \citenamefont {Egan},
  \citenamefont {Biswas}, \citenamefont {Cetina}, \citenamefont {Gorshkov},
  \citenamefont {Gullans}, \citenamefont {Huse},\ and\ \citenamefont
  {Monroe}}]{exp1}%
  \BibitemOpen
  \bibfield  {author} {\bibinfo {author} {\bibfnamefont {C.}~\bibnamefont
  {Noel}}, \bibinfo {author} {\bibfnamefont {P.}~\bibnamefont {Niroula}},
  \bibinfo {author} {\bibfnamefont {A.}~\bibnamefont {Zhu}, \bibfnamefont
  {Dand~Risinger}}, \bibinfo {author} {\bibfnamefont {L.}~\bibnamefont {Egan}},
  \bibinfo {author} {\bibfnamefont {D.}~\bibnamefont {Biswas}}, \bibinfo
  {author} {\bibfnamefont {M.}~\bibnamefont {Cetina}}, \bibinfo {author}
  {\bibfnamefont {A.~V.}\ \bibnamefont {Gorshkov}}, \bibinfo {author}
  {\bibfnamefont {M.~J.}\ \bibnamefont {Gullans}}, \bibinfo {author}
  {\bibfnamefont {D.~A.}\ \bibnamefont {Huse}},\ and\ \bibinfo {author}
  {\bibfnamefont {C.}~\bibnamefont {Monroe}},\ }\bibfield  {title} {\bibinfo
  {title} {Measurement-induced quantum phases realized in a trapped-ion quantum
  computer},\ }\href {https://doi.org/10.1038/s41567-022-01619-7} {\bibfield
  {journal} {\bibinfo  {journal} {Nat. Phys.}\ }\textbf {\bibinfo {volume}
  {18}},\ \bibinfo {pages} {760–764} (\bibinfo {year} {2022})}\BibitemShut
  {NoStop}%
\bibitem [{\citenamefont {Koh}\ \emph {et~al.}(2022)\citenamefont {Koh},
  \citenamefont {Sun}, \citenamefont {Motta},\ and\ \citenamefont
  {Minnich}}]{exp2}%
  \BibitemOpen
  \bibfield  {author} {\bibinfo {author} {\bibfnamefont {J.~M.}\ \bibnamefont
  {Koh}}, \bibinfo {author} {\bibfnamefont {S.-N.}\ \bibnamefont {Sun}},
  \bibinfo {author} {\bibfnamefont {M.}~\bibnamefont {Motta}},\ and\ \bibinfo
  {author} {\bibfnamefont {A.~J.}\ \bibnamefont {Minnich}},\ }\href@noop {}
  {\bibinfo {title} {Experimental realization of a measurement-induced
  entanglement phase transition on a superconducting quantum processor}}
  (\bibinfo {year} {2022}),\ \Eprint {https://arxiv.org/abs/2203.04338}
  {arXiv:2203.04338 [quant-ph]} \BibitemShut {NoStop}%
\bibitem [{\citenamefont {Iqbal}\ \emph {et~al.}(2023)\citenamefont {Iqbal},
  \citenamefont {Tantivasadakarn}, \citenamefont {Gatterman}, \citenamefont
  {Gerber}, \citenamefont {Gilmore}, \citenamefont {Gresh}, \citenamefont
  {Hankin}, \citenamefont {Hewitt}, \citenamefont {Horst}, \citenamefont
  {Matheny}, \citenamefont {Mengle}, \citenamefont {Neyenhuis}, \citenamefont
  {Vishwanath}, \citenamefont {Foss-Feig}, \citenamefont {Verresen},\ and\
  \citenamefont {Dreyer}}]{Iqbal22}%
  \BibitemOpen
  \bibfield  {author} {\bibinfo {author} {\bibfnamefont {M.}~\bibnamefont
  {Iqbal}}, \bibinfo {author} {\bibfnamefont {N.}~\bibnamefont
  {Tantivasadakarn}}, \bibinfo {author} {\bibfnamefont {T.~M.}\ \bibnamefont
  {Gatterman}}, \bibinfo {author} {\bibfnamefont {J.~A.}\ \bibnamefont
  {Gerber}}, \bibinfo {author} {\bibfnamefont {K.}~\bibnamefont {Gilmore}},
  \bibinfo {author} {\bibfnamefont {D.}~\bibnamefont {Gresh}}, \bibinfo
  {author} {\bibfnamefont {A.}~\bibnamefont {Hankin}}, \bibinfo {author}
  {\bibfnamefont {N.}~\bibnamefont {Hewitt}}, \bibinfo {author} {\bibfnamefont
  {C.~V.}\ \bibnamefont {Horst}}, \bibinfo {author} {\bibfnamefont
  {M.}~\bibnamefont {Matheny}}, \bibinfo {author} {\bibfnamefont
  {T.}~\bibnamefont {Mengle}}, \bibinfo {author} {\bibfnamefont
  {B.}~\bibnamefont {Neyenhuis}}, \bibinfo {author} {\bibfnamefont
  {A.}~\bibnamefont {Vishwanath}}, \bibinfo {author} {\bibfnamefont
  {M.}~\bibnamefont {Foss-Feig}}, \bibinfo {author} {\bibfnamefont
  {R.}~\bibnamefont {Verresen}},\ and\ \bibinfo {author} {\bibfnamefont
  {H.}~\bibnamefont {Dreyer}},\ }\href@noop {} {\bibinfo {title} {Topological
  order from measurements and feed-forward on a trapped ion quantum computer}}
  (\bibinfo {year} {2023}),\ \Eprint {https://arxiv.org/abs/2302.01917}
  {arXiv:2302.01917 [quant-ph]} \BibitemShut {NoStop}%
\bibitem [{\citenamefont {Weinstein}\ \emph {et~al.}(2022)\citenamefont
  {Weinstein}, \citenamefont {Kelly}, \citenamefont {Marino},\ and\
  \citenamefont {Altman}}]{shane2}%
  \BibitemOpen
  \bibfield  {author} {\bibinfo {author} {\bibfnamefont {Z.}~\bibnamefont
  {Weinstein}}, \bibinfo {author} {\bibfnamefont {S.~P.}\ \bibnamefont
  {Kelly}}, \bibinfo {author} {\bibfnamefont {J.}~\bibnamefont {Marino}},\ and\
  \bibinfo {author} {\bibfnamefont {E.}~\bibnamefont {Altman}},\ }\href@noop {}
  {\bibinfo {title} {Scrambling transition in a radiative random unitary
  circuit}} (\bibinfo {year} {2022}),\ \Eprint
  {https://arxiv.org/abs/2210.14242} {arXiv:2210.14242 [quant-ph]} \BibitemShut
  {NoStop}%
\bibitem [{\citenamefont {Dehghani}\ \emph {et~al.}(2022)\citenamefont
  {Dehghani}, \citenamefont {Lavasani}, \citenamefont {Hafezi},\ and\
  \citenamefont {Gullans}}]{mach_learning}%
  \BibitemOpen
  \bibfield  {author} {\bibinfo {author} {\bibfnamefont {H.}~\bibnamefont
  {Dehghani}}, \bibinfo {author} {\bibfnamefont {A.}~\bibnamefont {Lavasani}},
  \bibinfo {author} {\bibfnamefont {M.}~\bibnamefont {Hafezi}},\ and\ \bibinfo
  {author} {\bibfnamefont {M.~J.}\ \bibnamefont {Gullans}},\ }\href@noop {}
  {\bibinfo {title} {Neural-network decoders for measurement induced phase
  transitions}} (\bibinfo {year} {2022}),\ \Eprint
  {https://arxiv.org/abs/2204.10904} {arXiv:2204.10904 [quant-ph]} \BibitemShut
  {NoStop}%
\bibitem [{\citenamefont {Turkeshi}(2022)}]{xhek_machine}%
  \BibitemOpen
  \bibfield  {author} {\bibinfo {author} {\bibfnamefont {X.}~\bibnamefont
  {Turkeshi}},\ }\bibfield  {title} {\bibinfo {title} {Measurement-induced
  criticality as a data-structure transition},\ }\href
  {https://doi.org/10.1103/PhysRevB.106.144313} {\bibfield  {journal} {\bibinfo
   {journal} {Phys. Rev. B}\ }\textbf {\bibinfo {volume} {106}},\ \bibinfo
  {pages} {144313} (\bibinfo {year} {2022})}\BibitemShut {NoStop}%
\bibitem [{\citenamefont {Ippoliti}\ and\ \citenamefont
  {Khemani}(2021)}]{ippolitiPRL}%
  \BibitemOpen
  \bibfield  {author} {\bibinfo {author} {\bibfnamefont {M.}~\bibnamefont
  {Ippoliti}}\ and\ \bibinfo {author} {\bibfnamefont {V.}~\bibnamefont
  {Khemani}},\ }\bibfield  {title} {\bibinfo {title} {Postselection-free
  entanglement dynamics via spacetime duality},\ }\href
  {https://doi.org/10.1103/PhysRevLett.126.060501} {\bibfield  {journal}
  {\bibinfo  {journal} {Phys. Rev. Lett.}\ }\textbf {\bibinfo {volume} {126}},\
  \bibinfo {pages} {060501} (\bibinfo {year} {2021})}\BibitemShut {NoStop}%
\bibitem [{\citenamefont {Ippoliti}\ \emph {et~al.}(2022)\citenamefont
  {Ippoliti}, \citenamefont {Rakovszky},\ and\ \citenamefont
  {Khemani}}]{fractal}%
  \BibitemOpen
  \bibfield  {author} {\bibinfo {author} {\bibfnamefont {M.}~\bibnamefont
  {Ippoliti}}, \bibinfo {author} {\bibfnamefont {T.}~\bibnamefont
  {Rakovszky}},\ and\ \bibinfo {author} {\bibfnamefont {V.}~\bibnamefont
  {Khemani}},\ }\bibfield  {title} {\bibinfo {title} {Fractal, logarithmic, and
  volume-law entangled nonthermal steady states via spacetime duality},\ }\href
  {https://doi.org/10.1103/PhysRevX.12.011045} {\bibfield  {journal} {\bibinfo
  {journal} {Phys. Rev. X}\ }\textbf {\bibinfo {volume} {12}},\ \bibinfo
  {pages} {011045} (\bibinfo {year} {2022})}\BibitemShut {NoStop}%
\bibitem [{\citenamefont {Lu}\ and\ \citenamefont {Grover}(2021)}]{grover2021}%
  \BibitemOpen
  \bibfield  {author} {\bibinfo {author} {\bibfnamefont {T.-C.}\ \bibnamefont
  {Lu}}\ and\ \bibinfo {author} {\bibfnamefont {T.}~\bibnamefont {Grover}},\
  }\bibfield  {title} {\bibinfo {title} {Spacetime duality between localization
  transitions and measurement-induced transitions},\ }\href
  {https://doi.org/10.1103/PRXQuantum.2.040319} {\bibfield  {journal} {\bibinfo
   {journal} {PRX Quantum}\ }\textbf {\bibinfo {volume} {2}},\ \bibinfo {pages}
  {040319} (\bibinfo {year} {2021})}\BibitemShut {NoStop}%
\bibitem [{\citenamefont {Li}\ \emph {et~al.}(2022)\citenamefont {Li},
  \citenamefont {Zou}, \citenamefont {Glorioso}, \citenamefont {Altman},\ and\
  \citenamefont {Fisher}}]{CrossEntropy}%
  \BibitemOpen
  \bibfield  {author} {\bibinfo {author} {\bibfnamefont {Y.}~\bibnamefont
  {Li}}, \bibinfo {author} {\bibfnamefont {Y.}~\bibnamefont {Zou}}, \bibinfo
  {author} {\bibfnamefont {P.}~\bibnamefont {Glorioso}}, \bibinfo {author}
  {\bibfnamefont {E.}~\bibnamefont {Altman}},\ and\ \bibinfo {author}
  {\bibfnamefont {M.~P.~A.}\ \bibnamefont {Fisher}},\ }\href@noop {} {\bibinfo
  {title} {Cross entropy benchmark for measurement-induced phase transitions}}
  (\bibinfo {year} {2022}),\ \Eprint {https://arxiv.org/abs/2209.00609}
  {arXiv:2209.00609 [quant-ph]} \BibitemShut {NoStop}%
\bibitem[{\citenamefont{Garratt et~al.}(2023)\citenamefont{Garratt, Weinstein,
  and Altman}}]{AltmanMeasurementLL}
\bibinfo{author}{\bibfnamefont{S.~J.} \bibnamefont{Garratt}},
  \bibinfo{author}{\bibfnamefont{Z.}~\bibnamefont{Weinstein}},
  \bibnamefont{and} \bibinfo{author}{\bibfnamefont{E.}~\bibnamefont{Altman}},
  \href{https://link.aps.org/doi/10.1103/PhysRevX.13.021026}{\bibinfo{journal}{Phys. Rev. X} \textbf{\bibinfo{volume}{13}},
  \bibinfo{pages}{021026} (\bibinfo{year}{2023})}.

\bibitem[{\citenamefont{Kane and Fisher}(1992)}]{KaneFisher}
\bibinfo{author}{\bibfnamefont{C.~L.} \bibnamefont{Kane}} \bibnamefont{and}
  \bibinfo{author}{\bibfnamefont{M.~P.~A.} \bibnamefont{Fisher}},
 \href{https://link.aps.org/doi/10.1103/PhysRevLett.68.1220}{ \bibinfo{journal}{Phys. Rev. Lett.} \textbf{\bibinfo{volume}{68}},
  \bibinfo{pages}{1220} (\bibinfo{year}{1992})}.

\bibitem [{\citenamefont {Tirrito}\ \emph {et~al.}(2022)\citenamefont
  {Tirrito}, \citenamefont {Santini}, \citenamefont {Fazio},\ and\
  \citenamefont {Collura}}]{mario1}%
  \BibitemOpen
  \bibfield  {author} {\bibinfo {author} {\bibfnamefont {E.}~\bibnamefont
  {Tirrito}}, \bibinfo {author} {\bibfnamefont {A.}~\bibnamefont {Santini}},
  \bibinfo {author} {\bibfnamefont {R.}~\bibnamefont {Fazio}},\ and\ \bibinfo
  {author} {\bibfnamefont {M.}~\bibnamefont {Collura}},\ }\href@noop {}
  {\bibinfo {title} {Full counting statistics as probe of measurement-induced
  transitions in the quantum ising chain}} (\bibinfo {year} {2022}),\ \Eprint
  {https://arxiv.org/abs/2212.09405} {arXiv:2212.09405 [cond-mat.stat-mech]}
  \BibitemShut {NoStop}%
\bibitem [{\citenamefont {Rossini}\ and\ \citenamefont
  {Vicari}(2020)}]{vicari}%
  \BibitemOpen
  \bibfield  {author} {\bibinfo {author} {\bibfnamefont {D.}~\bibnamefont
  {Rossini}}\ and\ \bibinfo {author} {\bibfnamefont {E.}~\bibnamefont
  {Vicari}},\ }\bibfield  {title} {\bibinfo {title} {Measurement-induced
  dynamics of many-body systems at quantum criticality},\ }\href
  {https://doi.org/10.1103/PhysRevB.102.035119} {\bibfield  {journal} {\bibinfo
   {journal} {Phys. Rev. B}\ }\textbf {\bibinfo {volume} {102}},\ \bibinfo
  {pages} {035119} (\bibinfo {year} {2020})}\BibitemShut {NoStop}%
\bibitem [{\citenamefont {Lin}\ \emph {et~al.}(2023)\citenamefont {Lin},
  \citenamefont {Ye}, \citenamefont {Zou}, \citenamefont {Sang},\ and\
  \citenamefont {Hsieh}}]{Hsieh1}%
  \BibitemOpen
  \bibfield  {author} {\bibinfo {author} {\bibfnamefont {C.-J.}\ \bibnamefont
  {Lin}}, \bibinfo {author} {\bibfnamefont {W.}~\bibnamefont {Ye}}, \bibinfo
  {author} {\bibfnamefont {Y.}~\bibnamefont {Zou}}, \bibinfo {author}
  {\bibfnamefont {S.}~\bibnamefont {Sang}},\ and\ \bibinfo {author}
  {\bibfnamefont {T.~H.}\ \bibnamefont {Hsieh}},\ }\bibfield  {title} {\bibinfo
  {title} {Probing sign structure using measurement-induced entanglement},\
  }\href {https://doi.org/10.22331/q-2023-02-02-910} {\bibfield  {journal}
  {\bibinfo  {journal} {Quantum}\ }\textbf {\bibinfo {volume} {7}},\ \bibinfo
  {pages} {910} (\bibinfo {year} {2023})}\BibitemShut {NoStop}%
\bibitem [{\citenamefont {Sun}\ \emph {et~al.}(2023)\citenamefont {Sun},
  \citenamefont {Yao},\ and\ \citenamefont {Jian}}]{entropyLuttinger}%
  \BibitemOpen
  \bibfield  {author} {\bibinfo {author} {\bibfnamefont {X.}~\bibnamefont
  {Sun}}, \bibinfo {author} {\bibfnamefont {H.}~\bibnamefont {Yao}},\ and\
  \bibinfo {author} {\bibfnamefont {S.-K.}\ \bibnamefont {Jian}},\ }\href@noop
  {} {\bibinfo {title} {{New critical states induced by measurement}}}
  (\bibinfo {year} {2023}),\ \Eprint {https://arxiv.org/abs/2301.11337}
  {arXiv:2301.11337 [quant-ph]}. 

\bibitem [{\citenamefont {Lee}\ \emph {et~al.}(2023)\citenamefont {Lee},
  \citenamefont {Jian},\ and\ \citenamefont {Xu}}]{lee2023}%
  \BibitemOpen
  \bibfield  {author} {\bibinfo {author} {\bibfnamefont {J.~Y.}\ \bibnamefont
  {Lee}}, \bibinfo {author} {\bibfnamefont {C.-M.}\ \bibnamefont {Jian}},\ and\
  \bibinfo {author} {\bibfnamefont {C.}~\bibnamefont {Xu}},\ }\href@noop {}
  {\bibinfo {title} {Quantum criticality under decoherence or weak
  measurement}} (\bibinfo {year} {2023}),\ \Eprint
  {https://arxiv.org/abs/2301.05238} {arXiv:2301.05238 [cond-mat.stat-mech]}.

\bibitem [{\citenamefont {Bao}\ \emph {et~al.}(2023)\citenamefont {Bao},
  \citenamefont {Fan}, \citenamefont {Vishwanath},\ and\ \citenamefont
  {Altman}}]{topological1}%
  \BibitemOpen
  \bibfield  {author} {\bibinfo {author} {\bibfnamefont {Y.}~\bibnamefont
  {Bao}}, \bibinfo {author} {\bibfnamefont {R.}~\bibnamefont {Fan}}, \bibinfo
  {author} {\bibfnamefont {A.}~\bibnamefont {Vishwanath}},\ and\ \bibinfo
  {author} {\bibfnamefont {E.}~\bibnamefont {Altman}},\ }\href@noop {}
  {\bibinfo {title} {Mixed-state topological order and the errorfield double
  formulation of decoherence-induced transitions}} (\bibinfo {year} {2023}),\
  \Eprint {https://arxiv.org/abs/2301.05687} {arXiv:2301.05687 [quant-ph]}.

\bibitem [{\citenamefont {Fan}\ \emph {et~al.}(2023)\citenamefont {Fan},
  \citenamefont {Bao}, \citenamefont {Altman},\ and\ \citenamefont
  {Vishwanath}}]{topological2}%
  \BibitemOpen
  \bibfield  {author} {\bibinfo {author} {\bibfnamefont {R.}~\bibnamefont
  {Fan}}, \bibinfo {author} {\bibfnamefont {Y.}~\bibnamefont {Bao}}, \bibinfo
  {author} {\bibfnamefont {E.}~\bibnamefont {Altman}},\ and\ \bibinfo {author}
  {\bibfnamefont {A.}~\bibnamefont {Vishwanath}},\ }\href@noop {} {\bibinfo
  {title} {Diagnostics of mixed-state topological order and breakdown of
  quantum memory}} (\bibinfo {year} {2023}),\ \Eprint
  {https://arxiv.org/abs/2301.05689} {arXiv:2301.05689 [quant-ph]}.
\bibitem [{\citenamefont {Zou}\ \emph {et~al.}(2023)\citenamefont {Zou},
  \citenamefont {Sang},\ and\ \citenamefont {Hsieh}}]{Hsieh2}%
  \BibitemOpen
  \bibfield  {author} {\bibinfo {author} {\bibfnamefont {Y.}~\bibnamefont
  {Zou}}, \bibinfo {author} {\bibfnamefont {S.}~\bibnamefont {Sang}},\ and\
  \bibinfo {author} {\bibfnamefont {T.~H.}\ \bibnamefont {Hsieh}},\ }\href@noop
  {} {\bibinfo {title} {Channeling quantum criticality}} (\bibinfo {year}
  {2023}),\ \Eprint {https://arxiv.org/abs/2301.07141} {arXiv:2301.07141}.
\bibitem [{\citenamefont {Slagle}\ \emph {et~al.}(2021)\citenamefont {Slagle},
  \citenamefont {Aasen}, \citenamefont {Pichler}, \citenamefont {Mong},
  \citenamefont {Fendley}, \citenamefont {Chen}, \citenamefont {Endres},\ and\
  \citenamefont {Alicea}}]{Slagle2021}%
  \BibitemOpen
  \bibfield  {author} {\bibinfo {author} {\bibfnamefont {K.}~\bibnamefont
  {Slagle}}, \bibinfo {author} {\bibfnamefont {D.}~\bibnamefont {Aasen}},
  \bibinfo {author} {\bibfnamefont {H.}~\bibnamefont {Pichler}}, \bibinfo
  {author} {\bibfnamefont {R.~S.~K.}\ \bibnamefont {Mong}}, \bibinfo {author}
  {\bibfnamefont {P.}~\bibnamefont {Fendley}}, \bibinfo {author} {\bibfnamefont
  {X.}~\bibnamefont {Chen}}, \bibinfo {author} {\bibfnamefont {M.}~\bibnamefont
  {Endres}},\ and\ \bibinfo {author} {\bibfnamefont {J.}~\bibnamefont
  {Alicea}},\ }\bibfield  {title} {\bibinfo {title} {Microscopic
  characterization of ising conformal field theory in rydberg chains},\ }\href
  {https://doi.org/10.1103/PhysRevB.104.235109} {\bibfield  {journal} {\bibinfo
   {journal} {Phys. Rev. B}\ }\textbf {\bibinfo {volume} {104}},\ \bibinfo
  {pages} {235109} (\bibinfo {year} {2021})}

\bibitem[{\citenamefont{Scholl et~al.}(2023)\citenamefont{Scholl, Shaw, Tsai,
  Finkelstein, Choi, and Endres}}]{scholl2023}
\bibinfo{author}{\bibfnamefont{P.}~\bibnamefont{Scholl}},
  \bibinfo{author}{\bibfnamefont{A.~L.} \bibnamefont{Shaw}},
  \bibinfo{author}{\bibfnamefont{R.~B.-S.} \bibnamefont{Tsai}},
  \bibinfo{author}{\bibfnamefont{R.}~\bibnamefont{Finkelstein}},
  \bibinfo{author}{\bibfnamefont{J.}~\bibnamefont{Choi}}, \bibnamefont{and}
  \bibinfo{author}{\bibfnamefont{M.}~\bibnamefont{Endres}},
  \emph{\bibinfo{title}{Erasure conversion in a high-fidelity rydberg quantum
  simulator}} (\bibinfo{year}{2023}), \Eprint {https://arxiv.org/abs/2305.03406}{arXiv:2305.03406 [quantum-ph]}.

\bibitem[{\citenamefont{Haghshenas et~al.}(2023)\citenamefont{Haghshenas,
  Chertkov, DeCross, Gatterman, Gerber, Gilmore, Gresh, Hewitt, Horst, Matheny
  et~al.}}]{Haghshenas23}
\bibinfo{author}{\bibfnamefont{R.}~\bibnamefont{Haghshenas}},
  \bibinfo{author}{\bibfnamefont{E.}~\bibnamefont{Chertkov}},
  \bibinfo{author}{\bibfnamefont{M.}~\bibnamefont{DeCross}},
  \bibinfo{author}{\bibfnamefont{T.~M.} \bibnamefont{Gatterman}},
  \bibinfo{author}{\bibfnamefont{J.~A.} \bibnamefont{Gerber}},
  \bibinfo{author}{\bibfnamefont{K.}~\bibnamefont{Gilmore}},
  \bibinfo{author}{\bibfnamefont{D.}~\bibnamefont{Gresh}},
  \bibinfo{author}{\bibfnamefont{N.}~\bibnamefont{Hewitt}},
  \bibinfo{author}{\bibfnamefont{C.~V.} \bibnamefont{Horst}},
  \bibinfo{author}{\bibfnamefont{M.}~\bibnamefont{Matheny}},
  \bibnamefont{et~al.}, \bibinfo{title}{Probing critical states of matter
  on a digital quantum computer}, \Eprint {https://arxiv.org/abs/2305.01650}{arXiv:2305.01650 [quantum-ph]}.

\bibitem [{\citenamefont {Di~Francesco}\ \emph {et~al.}(1997)\citenamefont
  {Di~Francesco}, \citenamefont {Mathieu},\ and\ \citenamefont
  {Senechal}}]{yellowbook}%
  \BibitemOpen
  \bibfield  {author} {\bibinfo {author} {\bibfnamefont {P.}~\bibnamefont
  {Di~Francesco}}, \bibinfo {author} {\bibfnamefont {P.}~\bibnamefont
  {Mathieu}},\ and\ \bibinfo {author} {\bibfnamefont {D.}~\bibnamefont
  {Senechal}},\ }\href@noop {} {\emph {\bibinfo {title} {Conformal Field
  Theory}}}\ (\bibinfo {year} {1997})\BibitemShut {NoStop}%
\bibitem [{Note1()}]{Note1}%
  \BibitemOpen
  \bibinfo {note} {Notice that $\left \langle i\gamma _R\gamma _L \right
  \rangle =0$ when evaluated in the critical Ising CFT}\BibitemShut {NoStop}%
\bibitem [{Note2()}]{Note2}%
  \BibitemOpen
  \bibinfo {note} {Otherwise the measurement translates into a control unitary
  on the top chain depending on the measurement outcome of the bottom
  chain.}\BibitemShut {Stop}%
\bibitem [{\citenamefont {Nielsen}\ and\ \citenamefont
  {Chuang}(2011)}]{nielsen2002quantum}%
  \BibitemOpen
  \bibfield  {author} {\bibinfo {author} {\bibfnamefont {M.~A.}\ \bibnamefont
  {Nielsen}}\ and\ \bibinfo {author} {\bibfnamefont {I.~L.}\ \bibnamefont
  {Chuang}},\ }\href@noop {} {\emph {\bibinfo {title} {Quantum Computation and
  Quantum Information: 10th Anniversary Edition}}},\ \bibinfo {edition} {10th}\
  ed.\ (\bibinfo  {publisher} {Cambridge University Press},\ \bibinfo {address}
  {USA},\ \bibinfo {year} {2011})\BibitemShut {NoStop}%
\bibitem [{\citenamefont {Hauschild}\ and\ \citenamefont
  {Pollmann}(2018)}]{tenpy}%
  \BibitemOpen
  \bibfield  {author} {\bibinfo {author} {\bibfnamefont {J.}~\bibnamefont
  {Hauschild}}\ and\ \bibinfo {author} {\bibfnamefont {F.}~\bibnamefont
  {Pollmann}},\ }\bibfield  {title} {\bibinfo {title} {{Efficient numerical
  simulations with Tensor Networks: Tensor Network Python (TeNPy)}},\ }\href
  {https://doi.org/10.21468/SciPostPhysLectNotes.5} {\bibfield  {journal}
  {\bibinfo  {journal} {SciPost Phys. Lect. Notes}\ ,\ \bibinfo {pages} {5}}
  (\bibinfo {year} {2018})},\ \bibinfo {note} {code available from
  \url{https://github.com/tenpy/tenpy}},\ \Eprint
  {https://arxiv.org/abs/1805.00055} {arXiv:1805.00055} \BibitemShut {NoStop}%
\bibitem [{\citenamefont {White}(1992)}]{dmrg}%
  \BibitemOpen
  \bibfield  {author} {\bibinfo {author} {\bibfnamefont {S.~R.}\ \bibnamefont
  {White}},\ }\bibfield  {title} {\bibinfo {title} {Density matrix formulation
  for quantum renormalization groups},\ }\href
  {https://doi.org/10.1103/PhysRevLett.69.2863} {\bibfield  {journal} {\bibinfo
   {journal} {Phys. Rev. Lett.}\ }\textbf {\bibinfo {volume} {69}},\ \bibinfo
  {pages} {2863} (\bibinfo {year} {1992})}\BibitemShut {NoStop}%
\bibitem [{\citenamefont {McCulloch}(2008)}]{iDMRG}%
  \BibitemOpen
  \bibfield  {author} {\bibinfo {author} {\bibfnamefont {I.~P.}\ \bibnamefont
  {McCulloch}},\ }\href@noop {} {\bibinfo {title} {Infinite size density matrix
  renormalization group, revisited}} (\bibinfo {year} {2008}),\ \Eprint
  {https://arxiv.org/abs/0804.2509} {arXiv:0804.2509 [cond-mat.str-el]}
  \BibitemShut {NoStop}%
\bibitem [{\citenamefont {Bravyi}\ \emph {et~al.}(2008)\citenamefont {Bravyi},
  \citenamefont {Divincenzo}, \citenamefont {Oliveira},\ and\ \citenamefont
  {Terhal}}]{Bravyi-08}%
  \BibitemOpen
  \bibfield  {author} {\bibinfo {author} {\bibfnamefont {S.}~\bibnamefont
  {Bravyi}}, \bibinfo {author} {\bibfnamefont {D.~P.}\ \bibnamefont
  {Divincenzo}}, \bibinfo {author} {\bibfnamefont {R.}~\bibnamefont
  {Oliveira}},\ and\ \bibinfo {author} {\bibfnamefont {B.~M.}\ \bibnamefont
  {Terhal}},\ }\bibfield  {title} {\bibinfo {title} {The complexity of
  stoquastic local hamiltonian problems},\ }\href
  {https://dl.acm.org/doi/10.5555/2011772.2011773} {\bibfield  {journal}
  {\bibinfo  {journal} {Quantum Info. Comput.}\ }\textbf {\bibinfo {volume}
  {8}},\ \bibinfo {pages} {361–385} (\bibinfo {year} {2008})}\BibitemShut
  {NoStop}%
\bibitem [{Note3()}]{Note3}%
  \BibitemOpen
  \bibinfo {note} {The high probability of this measurement outcome becomes
  intuitive in the $h_{\protect \rm anc}/J_{\protect \rm anc} \gg 1$
  regime.}\BibitemShut {Stop}%
\bibitem [{Note4()}]{Note4}%
  \BibitemOpen
  \bibinfo {note} {Explicitly, after dropping terms that vanish by
  time-reversal symmetry, one finds $\left \langle (U^{\prime })^{ \dagger
  }Z_jZ_{j^\prime }U^\prime \right \rangle _{\protect \tilde {s}}=\cos
  (2ua(j))\cos (2ua(j^\prime ))\left \langle Z_jZ_{j^\prime } \right \rangle
  _{\protect \tilde {s}}+\sin (2ua(j))\sin (2ua(j^\prime ))\left \langle
  Y_jY_{j^\prime } \right \rangle _{\protect \tilde {s}}$. Given that $(i)$
  $Y_j$ maps to a CFT operator with larger scaling dimension than that for
  $Z_j$ ($Y_j\sim i\partial _{\tau }\sigma $, $\Delta _{\partial _{\tau }\sigma
  }=9/8$~\cite {PFEUTY}) and $(ii)$ our perturbative expansion focuses on the
  $u \ll 1$ regime, the $U'$ unitary can be safely neglected.}\BibitemShut
  {Stop}%
\bibitem [{\citenamefont {McCoy}\ and\ \citenamefont
  {Perk}(1980)}]{mccoy_1980}%
  \BibitemOpen
  \bibfield  {author} {\bibinfo {author} {\bibfnamefont {B.~M.}\ \bibnamefont
  {McCoy}}\ and\ \bibinfo {author} {\bibfnamefont {J.~H.~H.}\ \bibnamefont
  {Perk}},\ }\bibfield  {title} {\bibinfo {title} {Spin correlation functions
  of an ising model with continuous exponents},\ }\href
  {https://doi.org/10.1103/PhysRevLett.44.840} {\bibfield  {journal} {\bibinfo
  {journal} {Phys. Rev. Lett.}\ }\textbf {\bibinfo {volume} {144}},\ \bibinfo
  {pages} {840} (\bibinfo {year} {1980})}\BibitemShut {NoStop}%
\bibitem [{\citenamefont {Cabra}\ and\ \citenamefont
  {Na{\'o}}(1994)}]{CABRA_1994}%
  \BibitemOpen
  \bibfield  {author} {\bibinfo {author} {\bibfnamefont {D.}~\bibnamefont
  {Cabra}}\ and\ \bibinfo {author} {\bibfnamefont {C.}~\bibnamefont
  {Na{\'o}}},\ }\bibfield  {title} {\bibinfo {title} {2d ising model with a
  defect line},\ }\href {https://doi.org/10.1142/s0217732394001969} {\bibfield
  {journal} {\bibinfo  {journal} {Mod. Phys. Lett. A}\ }\textbf {\bibinfo
  {volume} {09}},\ \bibinfo {pages} {2017} (\bibinfo {year}
  {1994})}\BibitemShut {NoStop}%
\bibitem [{\citenamefont {Na{\'{o} }n}\ and\ \citenamefont
  {Trobo}(2011)}]{NT2011}%
  \BibitemOpen
  \bibfield  {author} {\bibinfo {author} {\bibfnamefont {C.}~\bibnamefont
  {Na{\'{o} }n}}\ and\ \bibinfo {author} {\bibfnamefont {M.}~\bibnamefont
  {Trobo}},\ }\bibfield  {title} {\bibinfo {title} {The spin correlation
  function in 2d statistical mechanics models with inhomogeneous line
  defects},\ }\href {https://doi.org/10.1088/1742-5468/2011/02/p02021}
  {\bibfield  {journal} {\bibinfo  {journal} {J. Stat. Mech.}\ }\textbf
  {\bibinfo {volume} {2011}},\ \bibinfo {pages} {P02021} (\bibinfo {year}
  {2011})}\BibitemShut {NoStop}%
\bibitem [{Note5()}]{Note5}%
  \BibitemOpen
  \bibinfo {note} {Notice that for $C=0$, $\ket {\psi _{\protect \tilde {s}}}$
  is an eigenstate of $G$, even though $U$ breaks this symmetry
  explicitly.}\BibitemShut {Stop}%
\bibitem [{Note6()}]{Note6}%
  \BibitemOpen
  \bibinfo {note} {We show connected correlations, as for finite bond dimension
  $\langle Z\rangle _{\protect \tilde {s}}$ gives non-zero values.}\BibitemShut
  {Stop}%
\bibitem [{\citenamefont {Koffel}\ \emph {et~al.}(2012)\citenamefont {Koffel},
  \citenamefont {Lewenstein},\ and\ \citenamefont {Tagliacozzo}}]{Tagliacozzo}%
  \BibitemOpen
  \bibfield  {author} {\bibinfo {author} {\bibfnamefont {T.}~\bibnamefont
  {Koffel}}, \bibinfo {author} {\bibfnamefont {M.}~\bibnamefont {Lewenstein}},\
  and\ \bibinfo {author} {\bibfnamefont {L.}~\bibnamefont {Tagliacozzo}},\
  }\bibfield  {title} {\bibinfo {title} {Entanglement entropy for the
  long-range ising chain in a transverse field},\ }\href
  {https://doi.org/10.1103/PhysRevLett.109.267203} {\bibfield  {journal}
  {\bibinfo  {journal} {Phys. Rev. Lett.}\ }\textbf {\bibinfo {volume} {109}},\
  \bibinfo {pages} {267203} (\bibinfo {year} {2012})}\BibitemShut {NoStop}%
\bibitem [{\citenamefont {Vodola}\ \emph {et~al.}(2015)\citenamefont {Vodola},
  \citenamefont {Lepori}, \citenamefont {Ercolessi},\ and\ \citenamefont
  {Pupillo}}]{Vodola_2016}%
  \BibitemOpen
  \bibfield  {author} {\bibinfo {author} {\bibfnamefont {D.}~\bibnamefont
  {Vodola}}, \bibinfo {author} {\bibfnamefont {L.}~\bibnamefont {Lepori}},
  \bibinfo {author} {\bibfnamefont {E.}~\bibnamefont {Ercolessi}},\ and\
  \bibinfo {author} {\bibfnamefont {G.}~\bibnamefont {Pupillo}},\ }\bibfield
  {title} {\bibinfo {title} {Long-range ising and kitaev models: phases,
  correlations and edge modes},\ }\href
  {https://doi.org/10.1088/1367-2630/18/1/015001} {\bibfield  {journal}
  {\bibinfo  {journal} {New Jour. of Phys.}\ }\textbf {\bibinfo {volume}
  {18}},\ \bibinfo {pages} {015001} (\bibinfo {year} {2015})}.

\bibitem[{\citenamefont{Alicea et~al.}(in preparation)\citenamefont{Alicea,
  Liu, Murciano, and Sala}}]{future_work}
\bibinfo{author}{\bibfnamefont{J.}~\bibnamefont{Alicea}},
  \bibinfo{author}{\bibfnamefont{Y.}~\bibnamefont{Liu}},
  \bibinfo{author}{\bibfnamefont{S.}~\bibnamefont{Murciano}}, \bibnamefont{and}
  \bibinfo{author}{\bibfnamefont{P.}~\bibnamefont{Sala}} (\bibinfo{year}{in
  preparation}).
\bibitem [{Note7()}]{Note7}%
  \BibitemOpen
  \bibinfo {note} {We are very grateful to Zack Weinstein for suggesting this
  approach.}
\bibitem[{\citenamefont{van Kempen and van Vliet}(2000)}]{RatioEstimator}
\bibinfo{author}{\bibfnamefont{G.}~\bibnamefont{van Kempen}} \bibnamefont{and}
  \bibinfo{author}{\bibfnamefont{L.}~\bibnamefont{van Vliet}},
  \href{https://onlinelibrary.wiley.com/doi/abs/10.1002/%28SICI%291097-0320%2820000401%2939%3A4%3C300%3A%3AAID-CYTO8%3E3.0.CO%3B2-O}{\bibinfo{journal}{Cytometry} \textbf{\bibinfo{volume}{39}},
  \bibinfo{pages}{300} (\bibinfo{year}{2000})}.
  
\bibitem [{Note8()}]{Note8}%
  \BibitemOpen
  \bibinfo {note}   {Equation\protect \nobreakspace  {}\protect \textup  {\hbox {\mathsurround \z@ \protect \normalfont  (\ignorespaces \ref {eq:Msra}\unskip \@@italiccorr )}} provides the leading $N$ dependence needed to ensure that $\protect \sqrt  {{\protect \rm  Var}_M[r(A)]}/r(A)$ becomes smaller than one for the cases of interest. This conclusion follows from the fact that $r(A)$ decays at most as a power-law in system size, whereas the factor in braces in Eq.\protect \nobreakspace  {}\protect \textup  {\hbox {\mathsurround \z@ \protect \normalfont  (\ignorespaces \ref {VarM2}\unskip \@@italiccorr )}} is at most O(1).}

\bibitem [{\citenamefont {Fendley}\ \emph {et~al.}(2004)\citenamefont
  {Fendley}, \citenamefont {Sengupta},\ and\ \citenamefont {Sachdev}}]{FSS}%
  \BibitemOpen
  \bibfield  {author} {\bibinfo {author} {\bibfnamefont {P.}~\bibnamefont
  {Fendley}}, \bibinfo {author} {\bibfnamefont {K.}~\bibnamefont {Sengupta}},\
  and\ \bibinfo {author} {\bibfnamefont {S.}~\bibnamefont {Sachdev}},\
  }\bibfield  {title} {\bibinfo {title} {Competing density-wave orders in a
  one-dimensional hard-boson model},\ }\href
  {https://doi.org/10.1103/PhysRevB.69.075106} {\bibfield  {journal} {\bibinfo
  {journal} {Phys. Rev. B}\ }\textbf {\bibinfo {volume} {69}},\ \bibinfo
  {pages} {075106} (\bibinfo {year} {2004})}\BibitemShut {NoStop}%
\bibitem[{\citenamefont{Weinstein et~al.}(2023)\citenamefont{Weinstein, Sajith,
  Altman, and Garratt}}]{EhudMeasurementIsing}
\bibinfo{author}{\bibfnamefont{Z.}~\bibnamefont{Weinstein}},
  \bibinfo{author}{\bibfnamefont{R.}~\bibnamefont{Sajith}},
  \bibinfo{author}{\bibfnamefont{E.}~\bibnamefont{Altman}}, \bibnamefont{and}
  \bibinfo{author}{\bibfnamefont{S.~J.} \bibnamefont{Garratt}},
  \href{https://doi.org/10.1103\%2Fphysrevb.107.245132}{ \bibinfo{journal}{Phys. Rev. B} \textbf{\bibinfo{volume}{107}},
  \bibinfo{pages}{245132} (\bibinfo{year}{2023})}.
\bibitem [{\citenamefont {Yang}\ \emph {et~al.}(2023)\citenamefont {Yang},
  \citenamefont {Mao},\ and\ \citenamefont {Jian}}]{JianMeasurementIsing}%
  \BibitemOpen
  \bibfield  {author} {\bibinfo {author} {\bibfnamefont {Z.}~\bibnamefont
  {Yang}}, \bibinfo {author} {\bibfnamefont {D.}~\bibnamefont {Mao}},\ and\
  \bibinfo {author} {\bibfnamefont {C.-M.}\ \bibnamefont {Jian}},\ }\href@noop
  {} {\bibinfo {title} {Entanglement in one-dimensional critical state after
  measurements}} (\bibinfo {year} {2023}),\ \Eprint
  {https://arxiv.org/abs/2301.08255} {arXiv:2301.08255 [quant-ph]} \BibitemShut
  {NoStop}%
\bibitem [{\citenamefont {Latorre}\ \emph {et~al.}(2004)\citenamefont
  {Latorre}, \citenamefont {Rico},\ and\ \citenamefont {Vidal}}]{correlator}%
  \BibitemOpen
  \bibfield  {author} {\bibinfo {author} {\bibfnamefont {J.~I.}\ \bibnamefont
  {Latorre}}, \bibinfo {author} {\bibfnamefont {E.}~\bibnamefont {Rico}},\ and\
  \bibinfo {author} {\bibfnamefont {G.}~\bibnamefont {Vidal}},\ }\bibfield
  {title} {\bibinfo {title} {Ground state entanglement in quantum spin
  chains},\ }\href {https://arxiv.org/abs/quant-ph/0304098} {\bibfield
  {journal} {\bibinfo  {journal} {Quantum Info. Comput.}\ }\textbf {\bibinfo
  {volume} {4}},\ \bibinfo {pages} {48–92} (\bibinfo {year}
  {2004})}\BibitemShut {NoStop}%
\bibitem [{\citenamefont {Surace}\ and\ \citenamefont
  {Tagliacozzo}(2022)}]{Surace_2022}%
  \BibitemOpen
  \bibfield  {author} {\bibinfo {author} {\bibfnamefont {J.}~\bibnamefont
  {Surace}}\ and\ \bibinfo {author} {\bibfnamefont {L.}~\bibnamefont
  {Tagliacozzo}},\ }\bibfield  {title} {\bibinfo {title} {Fermionic gaussian
  states: an introduction to numerical approaches},\ }\href
  {https://doi.org/10.21468/scipostphyslectnotes.54} {\bibfield  {journal}
  {\bibinfo  {journal} {SciPost Phys. Lect. Notes}\ }\textbf {\bibinfo {volume}
  {54}} (\bibinfo {year} {2022})}.

\bibitem[{\citenamefont{Widom}(1974)}]{szego}
\bibinfo{author}{\bibfnamefont{H.}~\bibnamefont{Widom}}, 
  \href{https://www.sciencedirect.com/science/article/pii/0001870874900723}{\bibinfo{journal}{Adv.
  in Math.} \textbf{\bibinfo{volume}{13}}, \bibinfo{pages}{284}
  (\bibinfo{year}{1974})}.

\bibitem[{\citenamefont{Pfeuty}(1970)}]{PFEUTY}
\bibinfo{author}{\bibfnamefont{P.}~\bibnamefont{Pfeuty}},
  \href{https://www.sciencedirect.com/science/article/pii/0003491670902708}{ \bibinfo{journal}{Annals of Physics} \textbf{\bibinfo{volume}{57}},
  \bibinfo{pages}{79} (\bibinfo{year}{1970})}.

\end{thebibliography}

\end{document}